\documentclass[aps,twocolumn,preprintnumbers,amsmath,amssymb,floatfix,longbibliography]{revtex4-1}
\usepackage{graphicx}
\usepackage{graphics}
\usepackage{psfrag}
\usepackage{hyperref}
\usepackage{multirow}
\usepackage[sort&compress]{natbib}
\usepackage{amssymb}
\usepackage{amsmath}
\usepackage{amsthm}
\usepackage{bbold}
\usepackage{bbm}
\usepackage{enumerate}
\usepackage{esint} 

\newcommand{\vare}{\varepsilon}
\newcommand{\sgn}{\mbox{sgn}}
\newcommand{\rmi}{{\rm i}}

\newcommand{\eps}{\epsilon}

\usepackage{xcolor}

\begin{document}

\hypersetup{pdftitle={title}}
\title{Optical response of Luttinger semimetals in the normal and superconducting states}

\author{Igor Boettcher}
\email{iboettch@umd.edu}
\affiliation{Joint Quantum Institute, University of Maryland, College Park, MD 20742, USA}

\begin{abstract}
We investigate the optical response properties of three-dimensional Luttinger semimetals with the Fermi energy close to a quadratic band touching point. In particular, in order to address recent experiments on the spectroscopy of Pyrochlore Iridates and half-Heusler superconductors, we derive expressions for the optical conductivity in both the normal and general superconducting states in the linear response regime within the random phase approximation. The response functions can be decomposed into contributions from intraband and interband transitions, the latter comprising a genuine signature of the quadratic band touching point. We demonstrate the importance of interband transitions in the optical response in the normal state both in the homogeneous and quasi-static limit. Our analysis reveals a factorization property of the homogeneous conductivity in the spatially anisotropic case and the divergence of the conductivity for strong spatial anisotropy. In the quasi-static limit, the response is dominated by interband transitions and significantly different from systems with a single parabolic band. As an applications of the formalism in the superconducting state we compute the optical conductivity and superfluid density for the s-wave singlet superconducting case for both finite and vanishing chemical potential.
\end{abstract}

\maketitle

\section{Introduction}

Ignited by recent advances in growth and characterization of novel classes of spin-orbit coupled materials, the study of many-body physics in three-dimensional Luttinger semimetals with the Fermi energy close to an inverted quadratic band touching point (QBT) is part of the forefront of both theoretical and experimental research on quantum materials. Already in the noninteracting case these systems are highly compelling, as applying strain or quantum confinement can induce a topological insulator state, which furthermore is robust against weak perturbations \cite{RevModPhys.83.1057}. An even richer manifold of possible macroscopic phases emerges when considering the effects of long-range or sufficiently strong short-range interactions. Some of the currently most actively investigated platforms for exploring interactions in QBT systems are Pyrochlore Iridates \cite{WKrempa} and half-Heusler superconductors \cite{SmidmanReview,ChadovNatMater}. In particular, two recent measurements of their intriguing conductance properties constitute the motivation for the present work \cite{2017NatCo...8.2097C,Kimeaao4513}.

What makes the study of many-body physics and interactions in Luttinger semimetals so fascinating can be attributed to two main features. Firstly, as realized by Abrikosov, the long-range Coulomb repulsion between electrons at the QBT point induces a non-Fermi liquid (NFL) phase of the system \cite{abrikosov,abrben,moon}. Although the ultimate stability of this phase is currently still debated, as emergent strong short-range interactions may eventually drive the system into a topological Mott insulator state \cite{PhysRevLett.113.106401,PhysRevB.93.165109,PhysRevB.95.075101,PhysRevB.97.125121}, it is fairly certain that correlation functions will show anomalous scaling over some extended range of experimental parameters such as temperature, momentum, and frequency. Secondly, since the electrons occupying the QBT point carry an effective spin of 3/2, many novel and often tensorial order parameters can be constructed close to the touching point \cite{PhysRevB.85.045124,PhysRevX.4.041027,PhysRevB.92.045117,PhysRevB.92.035137,PhysRevB.93.205138,PhysRevLett.116.137001,PhysRevLett.116.177001,PhysRevB.95.085120,PhysRevB.95.075149,PhysRevLett.118.127001,PhysRevB.95.144503,PhysRevB.96.144514,PhysRevB.96.214514,PhysRevLett.120.057002,PhysRevX.8.011029,PhysRevB.97.205402,Mandal,PhysRevB.98.104514,PhysRevX.8.041039,2017arXiv170807825R,2018arXiv181104046S,2018arXiv181112415S}. Fortunately both magnetic and superconducting orders of this type are, respectively, covered by the Pyrochlore Iridates and half-Heusler compounds in experiment.

Pyrochlore Iridates, having structural formula $R$$_2$Ir$_2$O$_7$ (denoted $R$-227 for short) with $R$ a rare-earth element, have been shown to host a QBT point at the Fermi energy both via theoretical calculations \cite{WKrempa} and experimental ARPES studies \cite{kondo,PhysRevLett.117.056403}. Most members of the material class show a transition to an insulating phase with octupolar magnetic order at temperatures around 100 K \cite{OngNature}. However, the critical temperature is reduced for Nd-227, and no finite-temperature transition has been observed in Pr-227. Furthermore, Pr-227 may be close to a quantum critical point as a function of ionic radius of R, implying that its high temperature phase lies in the corresponding critical fan and thus shows nontrivial scaling of observables as a function of temperature.

A recent THz spectroscopy study \cite{2017NatCo...8.2097C} by the Armitage group on the optical response of Pr-227 in the normal phase revealed a large additive anomalous contribution to the dielectric function compared to the Drude formula, which can be traced theoretically to originate from interband transitions between the upper and lower bands of the QBT point by Broerman's formula \cite{PhysRevB.5.397}. The determination of the scattering rate shows a $\tau^{-1}\propto T^2$ temperature dependence, however, with an unusually large prefactor indicating that the system may be strongly coupled in the normal phase. The presence of a finite Fermi energy $E_{\rm F}>0$ (measured from the QBT point) in the experiment sets a limit on the intermediate frequency and temperature ranges where nontrivial scaling such as Abrikosov's NFL behavior could be observed. Measuring at larger frequencies or higher temperatures (both compared to $E_{\rm F}$), or minimizing $E_{\rm F}$ directly, will allow to experimentally test whether the NFL phase is achieved in the normal phase of Pr-227, and thus shed light onto other QBT systems where long-range interactions are important. This clearly calls for a fresh and extended view on the frequency and temperature dependence of the optical conductivity in Luttinger semimetals. Note that the existence of plasmon excitations in the normal state has recently been addressed in Ref. \cite{2018arXiv181006574M}.

In half-Heusler superconductors the presence of a QBT point close to the Fermi energy is supported by extensive density functional calculations of the band structure \cite{ChadovNatMater}. (A small linear admixture to the QBT is generally expected due to the noncentrosymmetric crystal structure \cite{PhysRevLett.116.177001}, but its effect on the low-energy physics can be estimated to be subleading for realistic $E_{\rm F}$ \cite{PhysRevLett.120.057002}.) Importantly, several compounds have an inverted band structure and become superconducting at temperatures around 1 K \cite{PhysRevB.84.220504,PhysRevB.86.064515,BayLowT,SmidmanReview}. Given the low-density in these materials, reflected by a small value of $E_{\rm F}$, such critical temperatures need to be considered high and seem to require a more complex mechanisms than phonon mediated attraction \cite{PhysRevLett.116.137001}.

The case for unconventional superconductivity in the half-Heuslers was strengthened enormously by a recent measurement of the London penetration depth in YPtBi \cite{Kimeaao4513} by the Paglione group, which shows an almost linear temperature dependence of the observable at low temperatures $T/T_{\rm c} \sim 0.1$, and thereby indicates the presence of line nodes in the gap. Whereas this eliminates the possibility for a pure s-wave gap, the spin-3/2 nature of the fermions at the QBT point allows to construct many other pairing channels (with or without even-odd-parity mixing) that feature line nodes. Since the associated orders are typically tensorial in nature, an angular resolved measurement of the optical properties appears to be a first step towards eliminating certain candidate orders. More generally, a solid understanding of how distinct superconducting orders contribute to the frequency and directional dependence of the optical conductivity in Luttinger semimetals could be central to discerning which pattern is realized in a given material in future experiments.

The scope of this work is therefore to set up a framework for studying the optical response of Luttinger semimetals in the normal and superconducting phase that allows to address the challenges described above and support future experimental explorations of QBT systems. We use a purely field theoretic approach starting from the path integral to arrive at the optical conductivity in the linear response regime within the random phase approximation (RPA). In particular, we formulate the theory such as to allow for the complex and unconventional superconducting orders that are possible in the system. We recover the expressions for the longitudinal response in normal state of Ref. \cite{PhysRevB.5.397} and extend these works by addressing  anisotropic corrections, gauge invariance, transverse  response, and momentum dependence of response functions. We derive general formulas for the response functions in superconductors with a QBT point and apply them to the s-wave singlet superconductor as a proof of principle. Since the experiments for superconducting YPtBi are in the clean limit \cite{PhysRevB.84.220504}, we do not consider the effects of disorder in the present work.

\begin{figure}[t]
\centering
\includegraphics[width=8.5cm]{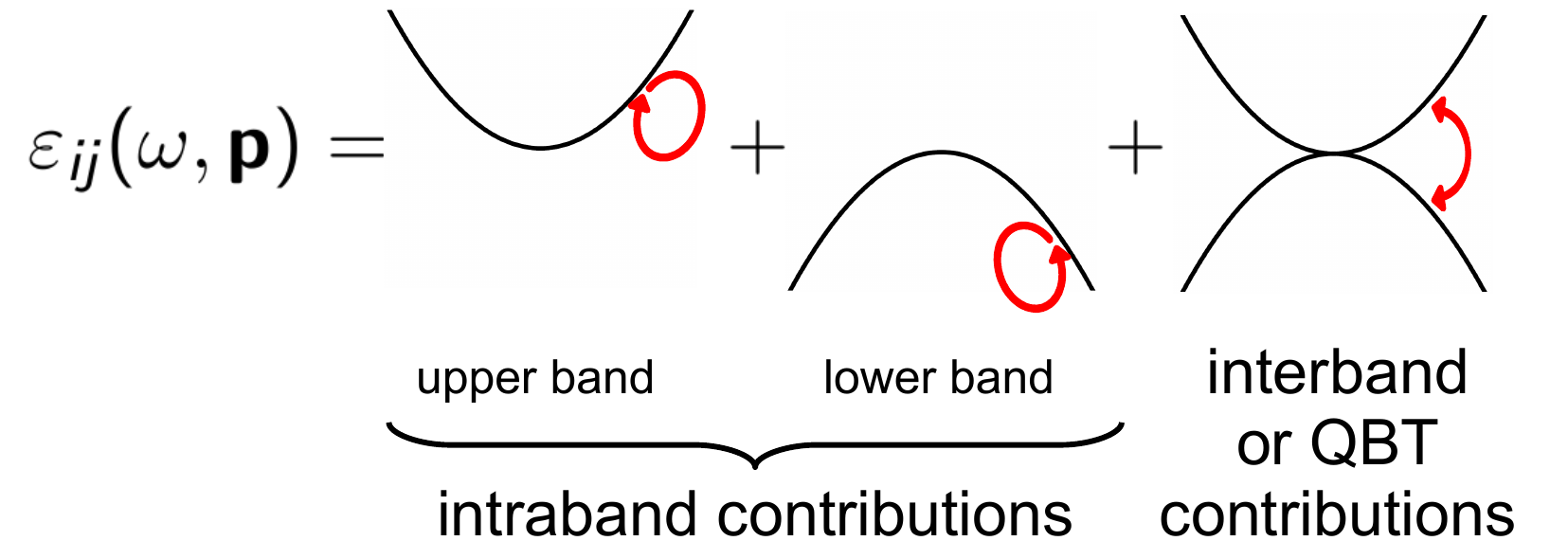}
\caption{The contributions to the dielectric tensor $\vare_{ij}(\omega,\textbf{p})$ can be split into three parts. The first two arise from intraband transitions within, respectively, the upper or lower band, and as such can be computed without knowledge of the other bands. In contrast, interband transitions or genuine QBT contributions are not captured by a single-band model. They encode, however, many important physical features of Luttinger semimetals. For instance, in the normal state they lead to a divergent contribution at low frequency as $E_{\rm F}\to 0$, or they contain the response from Bogoliubov Fermi surfaces in certain time-reversal symmetry breaking superconducting states---a feature entirely absent in single band systems.}
\label{FigBands}
\end{figure}

The picture that appears on the RPA level, and which underlies the interpretation of the experiments in Ref. \cite{2017NatCo...8.2097C}, is illustrated in Fig. \ref{FigBands}. The optical response functions, given by the dielectric tensor $\vare_{ij}(\omega,\textbf{p})$ or conductivity tensor $\sigma_{ij}(\omega,\textbf{p})$, decompose into a sum of intraband and interband transitions. The intraband contribution can be obtained from knowledge of the optical response of a single parabolic band, for instance by the usual Drude or Lindhard formulas in the normal state. The interband contribution, on the other hand, is a genuine contribution due to the QBT that cannot be captured by the theory for a single band. (We therefore also refer to it as ``QBT contribution''.) It also constitutes the anomalous contribution observed in Ref. \cite{2017NatCo...8.2097C}. We write
\begin{align}
 &\vare(\omega,\textbf{p}) = 1+\vare^{(\rm intra)}(\omega,\textbf{p}) + \vare^{(\rm QBT)}(\omega,\textbf{p}),\\
 &\vare^{(\rm intra)}(\omega,\textbf{p}) = \vare^{(\rm upper)}(\omega,\textbf{p})+ \vare^{(\rm lower)}(\omega,\textbf{p}).
\end{align}

For nonzero $E_{\rm F}$, one may expect only the band that is pierced by the chemical potential to contribute significantly to the response, whereas all other filled or empty bands are irrelevant. In Luttinger semimetals the QBT contribution quantifies how inaccurate this picture can be. On a more technical level, the interband contribution is conveniently incorporated by keeping the full $4\times 4$ structure of the underlying Luttinger Hamiltonian \cite{luttinger} instead of projecting it onto the two-dimensional basis spaces for the upper and lower band. This conveniently incorporates interband transitions. It also accounts for presence of Bogoliubov Fermi surfaces in certain time-reversal symmetry breaking superconducting states in QBT systems \cite{PhysRevLett.118.127001,PhysRevB.96.094526,2018arXiv180603773B}.

This work consists of two major parts. In the first or main part, after a review of the Luttinger Hamiltonian and optical response functions, we present the relevant formulas for the dielectric function and optical conductivity in the normal and s-wave superconducting phase and discuss their features. This presentation is intentionally left concise and does not illuminate any details how the results were obtained. The formulas are either given in fully analytic form or as one-dimensional integrals. In order to facilitate the comparison with experiment, results are presented in SI units, displaying the effective band mass $m^*$ explicitly in all formula. (We employ $\hbar=k_{\rm B}=1$ throughout the manuscript though.) 

In the second part or appendix, we give a self-contained derivation of the optical response of QBT Hamiltonians starting from the path integral, and then present the detailed calculation of the response functions presented in the main part. This extensive discussion of the setup also allows us to fix our notation and conventions, and set the stage for future works.  The results for the normal state are derived in App. \ref{AppNorm} and the results for the superconducting state in App. \ref{AppSC}. We show that the QBT contribution satisfies gauge invariance in the normal state in App. \ref{SecCont} and derive the transverse  current response in App. \ref{SecTrapp}. Algebraic conventions and matrix representations are specified in App. \ref{AppAlg}. In the appendix we work with Gauss units and set $2m^*=1$.

\section{Luttinger semimetals}
We assume the band structure of the QBT point to be described by the Luttinger model. The corresponding $4\times 4$ electronic single-particle Hamiltonian \cite{luttinger} reads
\begin{align}
 \nonumber \hat{H} ={}& \Bigl(\alpha_1+\frac{5}{2}\alpha_2\Bigr)\hat{p}^2 \mathbb{1}_4-2\alpha_3(\hat{\textbf{p}}\cdot\vec{J})^2\\
\label{lutt1} &+2(\alpha_3-\alpha_2)\sum_{i=1}^3 \hat{p}_i^2J_i^2.
\end{align}
Here $\hat{\textbf{p}}=-\rmi \nabla$ is the momentum operator and $\vec{J}=(J_1,J_2,J_3)^{\rm T}$ encompasses the spin-3/2 matrices. The Luttinger parameters $\alpha_1,\alpha_2,\alpha_3$ characterize the specific details of the QBT in a given material and may be determined experimentally or from first principle electronic band structure calculations. The number of such independent parameters is dictated by the symmetries that govern the low-energy excitations. Equation (\ref{lutt1}) captures the most general QBT Hamiltonian in the presence of time-reversal, inversion, and cubic point group symmetry. The number of independent parameters decreases upon imposing further symmetry constraints.

In order to elucidate the interplay between symmetry and band structure in the Luttinger model, we define the effective band mass $m^*$ by
\begin{align}
 \label{lutt2} \frac{1}{2m^*} &=|\alpha_2+\alpha_3|,
\end{align}
the particle-hole asymmetry parameter by
\begin{align}
 \label{lutt3} x&=\frac{\alpha_1}{|\alpha_2+\alpha_3|},
\end{align}
and the spatial anisotropy parameter by
\begin{align}
 \label{lutt4} \delta &= \frac{\alpha_3-\alpha_2}{\alpha_2+\alpha_3}\in[-1,1].
\end{align}
The single-particle energies that follow from the Luttinger Hamiltonian then take the form
\begin{align}
 \label{lutt5} E_{\pm}(\textbf{p}) &= \alpha_1 p^2 \pm \Bigl[4\alpha_2^2p^4+12(\alpha_3^2-\alpha_2^2)\sum_{i<j}p_i^2p_j^2\Bigr]^{1/2}\\
 \nonumber &= \frac{1}{2m^*}\Bigl( x p^2 \pm \Bigl[(1-\delta)^2p^4+12\delta\sum_{i<j}p_i^2p_j^2\Bigr]^{1/2}\Bigr).
\end{align}
Each eigenvalue is doubly degenerate due to time-reversal and inversion symmetry. We consider here the band inverted case which corresponds to
\begin{align}
 \label{lutt6} |x|<1.
\end{align}
The band structure then features an upper band with positive energies $E_+$ and a lower band with negative energies $E_-$ for nonzero momenta. Furthermore, for $x=0$ the spectrum of excitations becomes particle-hole symmetric, whereas $\delta=0$ implies a spatially isotropic band structure with 
\begin{align}
 \label{lutt7} E_\pm(\textbf{p}) = \frac{(x\pm 1)}{2m^*}p^2,
\end{align}
corresponding to an effective upper and lower band mass of
\begin{align}
 \label{lutt8} m_{\rm up}^* &= \frac{m^*}{1+x},\ m_{\rm low}^* = \frac{m^*}{1-x},
\end{align}
respectively. Although in a given material at hand these symmetries may not be realized exactly, it is a useful simplification to neglect $x$ and $\delta$ in calculations as long as these parameters are small compared to unity. Therefore, unless stated otherwise we set $x=\delta=0$ in this work, but discuss the influence of nonvanishing $x$ and $\delta$ on the homogeneous response functions in the normal state at the end of Sec. \ref{SecHom}.

A particularly important role for the faithful description of experimental data by means of the Luttinger model is played by the chemical potential $\mu$. For our investigation we allow $\mu$ to have either sign, and define the Fermi energy and Fermi momentum from its modulus according to
\begin{align}
 \label{lutt9} E_{\rm F} := \frac{p_{\rm F}^2}{2m^*} := |\mu|.
\end{align}
The condition that the low-energy physics are captured by the QBT in the band dispersion then implies that $E_{\rm F}\ll E_\kappa$, where $E_\kappa=\kappa^2/(2m^*)$ is an ``ultraviolet'' energy scale where either the electronic band structure deviates significantly from the quadratic dispersion for $q>\kappa$, or where other low-energy degrees of freedom such as phonons become relevant. On the other hand, the parabolic band structure may be screened by a linear band structure at low momenta that results, for instance, from adding $\hat{H}_{\rm lin}=\beta_1 (\hat{\textbf{p}}\cdot\vec{J})+\beta_2 \sum_i p_iJ_i^3$ to the Hamiltonian in Eq. (\ref{lutt1}). Such contributions arise in non-centrosymmetric materials due to asymmetric spin-orbit coupling, and their presence implies a typical ``infrared'' energy scale $E_{\rm lin}\sim |\beta_{1,2}|p_{\rm F}$. Consequently, the linear terms can be neglected if the chemical potential is sufficiently large so that $E_{\rm lin}\ll E_{\rm F}$ and, therefore, the relevant excitations at the Fermi level are dominated by the quadratic terms. Consequently, in the following the limit $\mu\to 0$ needs to be understood within the Luttinger model, meaning that the Fermi level is close enough to the QBT point so that $\mu\approx 0$ is a good approximation, but the chemical potential is still large enough so that linearly dispersing terms at even lower energies (if present) are irrelevant.

\section{Optical response functions}

The electrodynamic properties of solids in the linear response regime are encoded in the dielectric tensor $\vare_{ij}$ relating electric displacement field $\vec{D}$ and electric field $\vec{E}$ according to \cite{BookKeldysh,BookDressel,BookAltland}
\begin{align}
\label{opt1} D_i(\omega,\textbf{p}) = \eps_0 \vare_{ij}(\omega,\textbf{p}) E_j(\omega,\textbf{p}).
\end{align}
Here $\eps_0$ is the vacuum permittivity, $\omega$ and $\textbf{p}$ constitute (angular) frequency and momentum of the incident electromagnetic field, and we have defined $\vare_{ij}$ to be a dimensionless quantity. Throughout this work we use the Einstein sum convention that we sum over repeated indices. In the following we consider nonmagnetic materials with permeability equal to 1. The linear response is then equivalently expressed in terms of the conductivity $\sigma_{ij}$ given by
\begin{align}
 \label{opt2} \sigma_{ij}(\omega,\textbf{p}) = \rmi \omega \eps_0 \Bigl[\delta_{ij}-\vare_{ij}(\omega,\textbf{p})\Bigr],
\end{align}
which relates the internal current density $\vec{j}_{\rm int}$ and electric field according to
\begin{align}
 \label{opt2b} j_{{\rm int},i}(\omega,\textbf{p}) = \sigma_{ij}(\omega,\textbf{p}) E_j(\omega,\textbf{p}).
\end{align}

In a spatially isotropic medium, the tensorial response functions for nonzero $\textbf{p}$ can be decomposed into longitudinal (L) and transverse  (T) components according to
\begin{align}
 \label{opt3} \sigma_{ij}(\omega,\textbf{p}) &= \sigma_{\rm L}(\omega,p) \frac{p_ip_j}{p^2} + \sigma_{\rm T}(\omega,p) \Bigl(\delta_{ij}-\frac{p_ip_j}{p^2}\Bigr).
\end{align}
Crucially, a longitudinal (transverse) electromagnetic probe field can only induce a longitudinal (transverse) response, i.e.
\begin{align}
 \label{opt4} \vec{j}_{\rm int,L}(\omega,\textbf{p}) &=  \sigma_{\rm L}(\omega,p) \vec{E}_{\rm L}(\omega,\textbf{p}),\\
 \label{opt5} \vec{j}_{\rm int,T}(\omega,\textbf{p}) &= \sigma_{\rm T}(\omega,p) \vec{E}_{\rm T}(\omega,\textbf{p}),
\end{align}
with the usual definition of the longitudinal and transverse  parts of the vector fields. Equation (\ref{opt2}) implies
\begin{align}
 \label{opt5b} \sigma_{\rm L}(\omega,p) &= \rmi \omega \eps_0\Bigl[1-\vare_{\rm L}(\omega,p)\Bigr],\\
 \label{opt5c} \sigma_{\rm T}(\omega,p) &= \rmi \omega \eps_0\Bigl[1-\vare_{\rm T}(\omega,p)\Bigr].
\end{align}
The advantage of studying $\sigma_{\rm L,T}(\omega,p)$ over $\sigma_{ij}(\omega,\textbf{p})$ lies in the fact that the L and T components are scalar functions of $p=|\textbf{p}|$, and so the limit $p\to 0$ is defined unambiguously.

The experiments we attempt to quantify with our analysis are such that the spatial inhomogeneity of the external probe fields is irrelevant so that setting $\textbf{p}=0$ is a valid approximation. In this limit, the distinction between L and T components is meaningless and Eq. (\ref{opt2b}) provides a definition of $\sigma_{ij}(\omega,\textbf{0})$ that does not require referencing to an external momentum.  The tensorial character of this quantity is necessarily trivial and so
\begin{align}
 \label{opt6} \sigma_{ij}(\omega,\textbf{0}) = \sigma(\omega) \delta_{ij},
\end{align}
which defines the homogeneous conductivity $\sigma(\omega)$. This quantity also coincides with the $\textbf{p}\to \textbf{0}$ limit of the L and T contributions when the limit is taken for $\omega>0$, as generally the limits $p\to 0$ and $\omega\to 0$ do not commute. In fact, although any spatial dependence of the electric field is unimportant, in practice it will not be strictly zero. We can then perform the limit $\textbf{p}\to 0$ in Eq. (\ref{opt3}) explicitly by assuming (without loss of generality) that the strongest spatial inhomogeneity of $\textbf{p}$ is in the z-direction, hence $\textbf{p}\approx (0,0,p)^{\rm T}$. Then, by computing the individual components $\sigma_{ij}(\omega,\textbf{p})$ in the limit $p\to 0$ and comparing to Eq. (\ref{opt6}) we deduce that
\begin{align}
 \label{opt7a} \vare(\omega) &= \vare_{\rm L}(\omega,0) = \vare_{\rm T}(\omega,0),\\
 \label{opt7} \sigma(\omega) &= \sigma_{\rm L}(\omega,0) = \sigma_{\rm T}(\omega,0).
\end{align}
Equations (\ref{opt7a}) and (\ref{opt7}) allow us to conveniently discuss the optical response of materials in terms of a single frequency-dependent function.

In order to facilitate the comparison with experiment we employ SI units here with $\eps_0=8.854 \times 10^{-12} \text{F}\ \text{m}^{-1}$ and electric charge $e=1.602 \times 10^{-19} \text{C}$. For computing the response functions from the underlying microscopic model, as it is presented in the appendices, we conveniently use Gauss units. The corresponding electric charge in Gauss units will be denoted by an overbar, and is given by $\bar{e}=1.519 \times 10^{-14} \text{m}^{3/2} \text{kg}^{1/2}\text{s}^{-1}$. Both quantities are related by 
\begin{align}
\label{opt8b} \bar{e}^2 = \frac{e^2}{4\pi \eps_0}.
\end{align}
Further, the dielectric function and conductivity in Gauss units, denoted as $\bar{\vare}$ and $\bar{\sigma}$ with an overbar, are defined from $\vec{D}(\omega,\textbf{p})=\bar{\vare}(\omega,\textbf{p})\vec{E}(\omega,\textbf{p})$ and $\vec{j}_{\rm int}(\omega,\textbf{p})=\bar{\sigma}(\omega,\textbf{p})\vec{E}(\omega,\textbf{p})$. They are mutually related by $\bar{\sigma}(\omega,\textbf{p})=\frac{\rmi \omega}{4\pi}[1-\bar{\vare}(\omega,\textbf{p})]$, and are obtained from the response function in SI units by means of
\begin{align}
 \label{opt8c} \bar{\vare}(\omega,\textbf{p}) &= \vare(\omega,\textbf{p}),\\
 \label{opt8d} \bar{\sigma}(\omega,\textbf{p}) &= \frac{1}{4\pi \eps_0} \sigma(\omega,\textbf{p}),
\end{align}
with the charge translated according to Eq. (\ref{opt8b}).

Our approach to computing the optical response lies in a field theoretic determination of the density-density response function $-\chi(\omega,p)$ and current-current response function $-K_{ij}(\omega,\textbf{p})$ within RPA. We refer to the appendices for their definition, and limit ourselves here to a brief discussion of their key properties. We first note that gauge invariance implies
\begin{align}
 \label{opt9} \omega^2 \chi(\omega,p) = -p^2 K_{\rm L}(\omega,p).
\end{align}
The L component of the dielectric function is given by 
\begin{align}
 \label{opt10} \vare_{\rm L}(\omega,p) = 1+ 4\pi \frac{\chi(\omega,p)}{p^2},
\end{align}
and the conductivity reads
\begin{align}
 \label{opt11} \sigma_{ij}(\omega,\textbf{p}) = -\frac{4\pi \eps_0}{\rmi \omega}K_{ij}(\omega,\textbf{p}).
\end{align}
Equation (\ref{opt9}) guarantees that the L components satisfy $\sigma_{\rm L}=\rmi \omega\eps_0(1-\vare_{\rm L})$. Furthermore, it implies that  $\chi(\omega,0)=0$ for $\omega>0$. For small momenta we may then expand the density response in power of $p$ and obtain
\begin{align}
 \label{opt12} \chi(\omega,p) = Z(\omega) p^2 +\mathcal{O}(p^4).
\end{align}
Consequently, in the limit $\textbf{p}=0$ the dielectric function is given by
\begin{align}
 \label{opt13} \vare(\omega) = 1 + 4\pi Z(\omega),
\end{align}
and we have $\sigma(\omega) = -\rmi \omega\cdot 4\pi \eps_0\cdot Z(\omega)$ for the conductivity.

The function $K_{ij}(\omega,\textbf{p})$ is useful for studying several important conceptual aspects of the optical response of media \cite{BookTinkham}. First note that gauge invariance through Eq. (\ref{opt9}) implies $K_{\rm L}(0,p)=0$. Hence the static response (meaning $\omega=0$) is purely transverse. On a technical level, the absence of the static L component requires a perfect cancellation between the diamagnetic (''d'') and paramagnetic (''p'') contributions to the current-current response. Referring to the appendices for details of their definition, we note here that the response function is naturally split into the diamagnetic and paramagnetic contributions according to
\begin{align}
 \label{opt14} K_{ij}(\omega,\textbf{p}) = K_{ij}^{(\rm d)}(\omega,\textbf{p}) + K_{ij}^{(\rm p)}(\omega,\textbf{p}).
\end{align}
Whereas the perfect cancellation is also valid for the static T component in the normal state, this situation is fundamentally altered in the superconducting state. Intuitively, the diamagnetic contribution comes from all electrons of the system, whereas only electrons on the Fermi surface contribute to the paramagnetic term. Since electron excitations at the Fermi surface are gapped (hence only thermally populated) in a superconductor, the diamagnetic term then dominates over the paramagnetic one. In this context, the superfluid density $n_{\rm s}$ is defined according to
\begin{align}
 \label{opt15} \lim_{p\to 0} K_{\rm T}(0,p) = \frac{e^2n_{\rm s}}{4\pi \eps_0 m^*}.
\end{align}
Clearly we have $n_{\rm s}=0$ in the normal state. For a clean single-band superconductor in the mean-field approximation, we find that the paramagnetic contribution vanishes completely at zero temperature, and the transverse  response is entirely given by the diamagnetic term $K_{\rm T}^{(\rm d)}(\omega,p) = \frac{e^2n}{4\pi \eps_0 m^*}$, and so the superfluid density agrees with the electron density: $n_{\rm s}=n$. In a more realistic setup, considering interaction and impurity effects, we generally have $n_{\rm s}<n$ even at zero temperature.

\section{Normal state response}
We begin our analysis of optical response in Luttinger semimetals by considering systems in the normal state. Unless explicitly stated we consider the particle-hole and rotationally symmetric case with $x=\delta=0$, which encompasses the key qualitative features of the optical response within the Luttinger model as long as these parameters are small compared to unity. The formulas presented here are derived in App. \ref{AppNorm}.

\subsection{Scales and limits}
The optical response in the normal state is determined by the frequency and momentum of the probe field, $\omega$ and $p$, and the thermodynamic parameters $\mu$ and $T$. The density of charge carriers within RPA reads
\begin{align}
 \label{norm5} n = 2 \int_{\textbf{q}}\Bigl[ n_{\rm F}\Bigl(\frac{q^2}{2m^*}-\mu\Bigr) + n_{\rm F}\Bigl(\frac{q^2}{2m^*}+\mu\Bigr)\Bigr],
\end{align}
where we denote $\int_{\textbf{q}} = \int \frac{d^3q}{(2\pi)^3}$ and $n_{\rm F}(E)= (e^{E/T}+1)^{-1}$. At zero temperature we obtain
\begin{align}
 \label{scal1} n_0 := \frac{p_{\rm F}^3}{3\pi^2} = \frac{(2m^*|\mu|)^{3/2}}{3\pi^2}.
\end{align}
This coincides with the density of carriers of a single parabolic band at zero temperature since fluctuation effects between electrons in distinct bands are suppressed in our mean field approximation.

\begin{figure}[t]
\centering
\includegraphics[width=8.5cm]{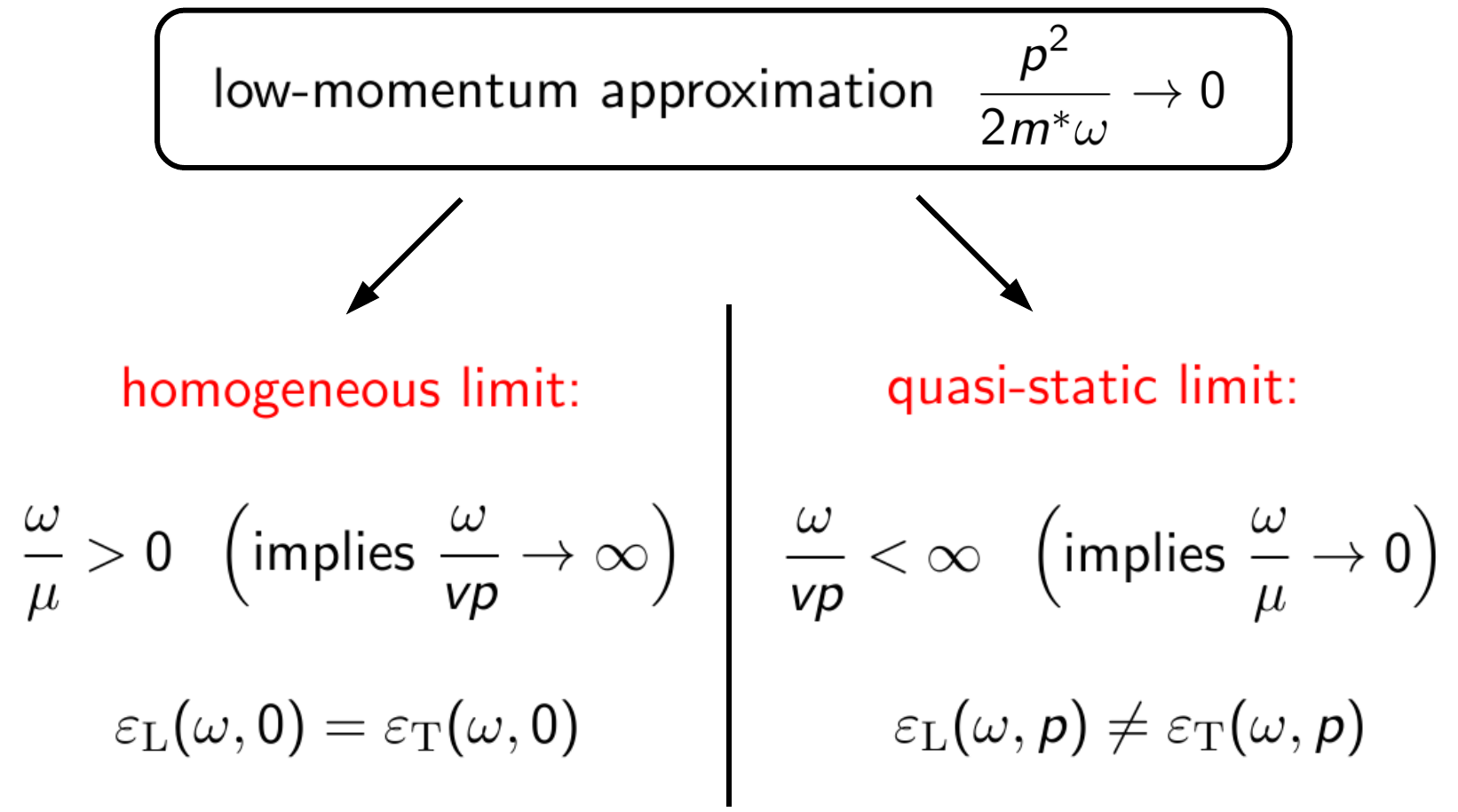}
\caption{The low-momentum regime with $p^2/(2m^*)\ll \omega$ naturally decomposes into two sectors depending on whether the product $vp$ with Fermi velocity $v\propto \sqrt{E_{\rm F}}$ is dominating or irrelevant compared to the remaining energy scales such as $\omega$ or $T$. For $vp\ll \omega$, which amounts to setting $p=0$ in practice, we obtain the homogeneous limit, where L and T response coincide. For $vp \gtrsim \omega$, on the other hand, frequencies are necessarily small compared to $\mu$ and hence this regime is labelled the quasi-static limit. The inherent momentum dependence of the response then implies that L and T contributions differ.}
\label{FigLowMom}
\end{figure}

In the following we consider two ways of taking the low-momentum limit $p^2/(2m^*\omega)\to 0$, which is typically well-satisfied for spectroscopic experiments. The first approach, which we refer to as the \emph{homogeneous limit}, corresponds to taking the limit for a fixed ratio of $\omega/\mu$. This basically corresponds to setting $p=0$ in the response functions. Importantly, in the homogeneous limit, longitudinal and transverse  response coincide. The second way to perform the limit, which we refer to as \emph{quasi-static limit}, corresponds to keeping the ratio $\omega/vp$ fixed, where
\begin{align}
\label{scal2} v := \frac{p_{\rm F}}{m^*} = \sqrt{\frac{2|\mu|}{m^*}}
\end{align}
is the Fermi velocity. Clearly, $\frac{p^2}{2m^*\omega}\to 0$ while $\frac{\omega}{vp}<\infty$ implies that $\omega\ll \mu$. The dominance of the chemical potential over all other scales, on the other hand, is a common scenario in solid state systems and thus clearly deserves consideration here. If in addition $\omega/vp\ll 1$ we are in a regime such that
\begin{align}
 \label{scal3} \frac{p^2}{2m^*} \ll \omega \ll vp.
\end{align}
These inequalities are often taken as the definition of the quasi-static limit \cite{BookDressel}, so our definition is slightly more generous. We summarize the setup in Fig. \ref{FigLowMom}.

\subsection{Homogeneous limit}\label{SecHom}

The intraband contribution from the upper and lower bands in the clean limit takes the usual form
\begin{align}
 \label{norm0} \vare^{(\rm intra)}(\omega) &= -\frac{\omega_{\rm p}^2}{\omega(\omega+\rmi 0)},\\
 \label{norm1} \sigma^{(\rm intra)}(\omega) &= -\frac{\eps_0\omega_{\rm p}^2}{\rmi (\omega+\rmi 0)},
\end{align}
with the Plasma frequency $\omega_{\rm p}$ defined from the carrier density $n$ according to
\begin{align}
 \label{norm2} \omega_{\rm p}^2 = \frac{n e^2}{\eps_0m^*}.
\end{align}
The individual contributions from the upper and lower bands to the conductivity are given by
\begin{align}
 \label{norm3} \vare^{(\rm upper)}(\omega) &= -\frac{2e^2}{\eps_0 m^*} 
\frac{1}{\omega^2}\int_{\textbf{q}} n_{\rm F}\Bigl(\frac{q^2}{2m^*}-\mu\Bigr),\\
 \label{norm4} \vare^{(\rm lower)}(\omega) &= -\frac{2e^2}{\eps_0 m^*} \frac{1}{\omega^2} \int_{\textbf{q}} n_{\rm F}\Bigl(\frac{q^2}{2m^*}+\mu\Bigr).
\end{align}
The effect of nonmagnetic impurities can be included in Eqs. (\ref{norm0}) and  (\ref{norm1}) by a shift $\omega\to \omega+\rmi /\tau$ with scattering time $\tau$, or scattering rate $\Gamma= \tau^{-1}$. Assuming for simplicity that the scattering rates for the upper and lower band are equal we obtain
\begin{align}
 \label{norm5b} \vare^{(\rm intra)}(\omega) &= -\frac{\omega_{\rm p}^2}{\omega(\omega+\rmi/\tau)},\\
 \label{norm6} \sigma^{(\rm intra)}(\omega) &= \frac{\eps_0\omega_{\rm p}^2\tau}{1-\rmi \omega\tau}.
\end{align} 
For large scattering rate, the conductivity is approximately real and frequency independent. For small scattering rate $\tau^{-1}\to 0$, on the other hand, Eq. (\ref{norm1}) implies
\begin{align}
 \label{norm6b} \sigma_1^{(\rm intra)}(\omega) &= \frac{\pi}{2} \frac{ne^2}{m^*} \delta(\omega),\\
 \label{norm6c} \sigma_2^{(\rm intra)}(\omega) &= \frac{ne^2}{m^*\omega}
\end{align}
for the real and imaginary parts. The $\delta$-function in $\sigma_1(\omega)$ is restricted to non-negative frequencies, hence the normalization with $\pi/2$.

The interband or QBT contribution to the dielectric function in the clean limit is given by \cite{PhysRevB.5.397}
\begin{align}
  \label{norm8} \vare^{(\rm QBT)}(\omega) ={}& \frac{e^2}{4\pi \eps_0}\sqrt{\frac{m^*}{\omega}}(1+\rmi) \\
 \nonumber &-\frac{2e^2}{\eps_0m^*} \int_{\textbf{q}}\frac{n_{\rm F}(\frac{q^2}{2m^*}-\mu)+n_{\rm F}(\frac{q^2}{2m^*}+\mu)}{-(\omega+\rmi 0)^2+q^4/(m^*)^2}.
\end{align}
Here the first contribution is of particular significance. Its peculiar form originates from the appearance of the square root of $\rmi \omega$ after analytic continuation from Matsubara frequencies $p_0$, $\rmi p_0\to \omega+\rmi 0$, according to
\begin{align}
 \label{norm9} \frac{1}{\sqrt{p_0}} \to \frac{1}{\sqrt{-\rmi \omega}} = \frac{1}{\sqrt{2\omega}}(1+\rmi).
\end{align}
In the limit $\mu,T\to 0$, only the first line of Eq. (\ref{norm8}) contributes to the response, and we obtain a $1/\sqrt{\omega}$-divergent low-energy response according to
\begin{align}
 \label{norm9b} \lim_{\mu,T\to 0} \vare^{(\rm QBT)}(\omega) = \frac{e^2}{4\pi \eps_0}\sqrt{\frac{m^*}{\omega}}(1+\rmi).
\end{align}
Since the intraband contribution from the upper and lower bands vanishes in this limit, the optical response is then entirely dominated by the interband transitions, and thus genuinely different from a single band system.

\begin{figure}[t!]
\centering
\begin{minipage}{0.48\textwidth}
\includegraphics[width=8cm]{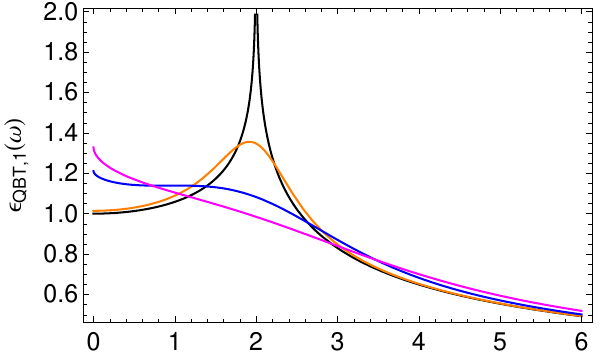}
\includegraphics[width=8cm]{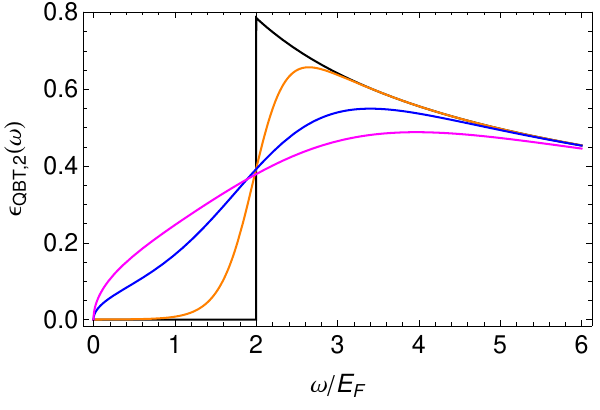}
\caption{QBT contribution to the homogeneous dielectric function $\vare(\omega)$. We show the real and imaginary part in the upper and lower plot, respectively, as a function of $\omega/E_{\rm F}$. Here we normalize the expressions by the zero temperature limit $\vare(0)=\frac{e^2}{2\pi^2\eps_0}\sqrt{2m^*/E_{\rm F}}$. The distinct curves (from bottom to top along the zero frequency axis) correspond to $T/E_{\rm F}$-values of 0 (black), 0.1 (orange), 0.3 (blue), 0.5 (magenta). At zero temperature we observe singular behavior at $\omega=2E_{\rm F}$, which extends to an anomalously large, $1/\sqrt{\omega}$-divergent contribution to both the real and imaginary parts of the optical response as $E_{\rm F}\to 0$. At nonzero temperature the functions remain regular.}
\label{FigHom}
\end{minipage}
\end{figure}

For general $\mu$ and $T$, the imaginary part of Eq. (\ref{norm8}) can be computed analytically and reads
\begin{align}
 \label{norm10} \vare_2^{(\rm QBT)}(\omega) = \frac{e^2}{4\pi\eps_0}\sqrt{\frac{m^*}{\omega}} \Bigl[1-n_{\rm F}\Bigl(\frac{\omega}{2}-\mu\Bigr)-n_{\rm F}\Bigl(\frac{\omega}{2}+\mu\Bigr)\Bigr].
\end{align}
In particular, at zero temperature we arrive at
\begin{align}
 \label{norm11} \vare_2^{(\rm QBT)}(\omega) = \frac{e^2}{4\pi \eps_0} \sqrt{\frac{m^*}{\omega}}\theta(\omega-2E_{\rm F}).
\end{align}
In order to compute the real part of Eq. (\ref{norm8}) for nonzero temperatures, the integral can be evaluated for a small finite value of $\rmi 0$ or in terms of the principal value. At zero temperature we have
\begin{align}
 \nonumber \vare_1^{(\rm QBT)}(\omega) ={}&\frac{e^2}{4\pi \eps_0} \sqrt{\frac{m^*}{\omega}} \Biggl[ 1 -\frac{2}{\pi}\arctan\Bigl(\sqrt{\frac{2E_{\rm F}}{\omega}}\Bigr)\\
 \label{norm12} &-\frac{1}{\pi} \ln\Bigl(\frac{|1-\sqrt{\omega/(2E_{\rm F})}|}{1+\sqrt{\omega/(2E_{\rm F})}}\Bigr)\Biggr].
\end{align}
In the limit $\omega\to 0$ we are left with a real response given by
\begin{align}
 \label{norm13} \vare^{(\rm QBT)}(0) = \frac{e^2}{2\pi^2\eps_0}\sqrt{\frac{2m^*}{E_{\rm F}}}.
\end{align}
We observe that a nonzero Fermi energy regularizes the $1/\sqrt{\omega}$-divergence of both the real and imaginary parts of the low-frequency response. We display the temperature dependence of the QBT contribution in Fig. \ref{FigHom}.

\begin{figure}[t]
\centering
\includegraphics[width=8cm]{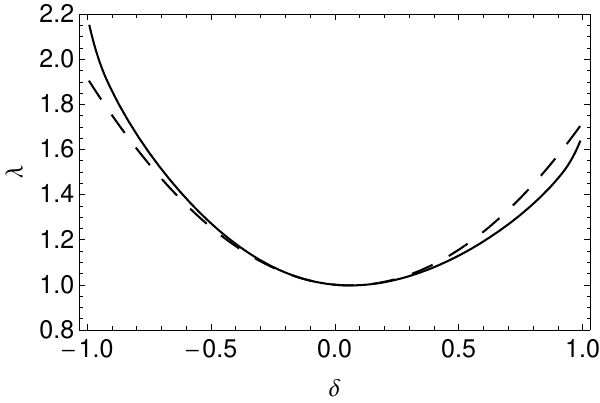}
\caption{The homogeneous optical response for nonvanishing spatial anisotropy $\delta$ gets renormalized by a prefactor $\lambda(\delta)/\sqrt{1-\delta^2}$ that diverges for strong anisotropy. This statement is true for both the intraband and interband contributions, at both zero and nonzero temperature, for $x=0$. For $\delta=0$ we have, of course, $\lambda(0)=1$. The solid line shows the function $\lambda(\delta)$ computed from the two-dimensional integral in Eq. (\ref{Aniso37}), whereas the dashed line corresponds to the expansion around $\delta=0$ to quadratic order from Eq. (\ref{norm16}). The latter should be sufficient for all practical purposes.}
\label{FigLambda}
\end{figure}

In the spatially anisotropic case with $\delta\neq 0$ (while still keeping particle-hole symmetry so that $x=0$), the intraband and interband contributions to the response functions factorize into the isotropic formula and a $\delta$-dependent prefactor. In particular, this prefactor is identical for the individual terms, and so we have an overall factorization according to
\begin{align}
 \label{norm15} \sigma(\omega) = \frac{\lambda(\delta)}{\sqrt{1-\delta^2}}\times  \sigma(\omega)|_{\delta=0}.
\end{align}
The factorization also holds for nonzero temperatures. Here $\lambda(\delta)$ is a regular function for all values of $\delta$ and can be computed numerically to arbitrary precision in terms of a two-dimensional angular integral given in Eq. (\ref{Aniso37}). For all practical purposes the quadratic approximation
\begin{align}
 \label{norm16} \lambda(\delta) = 1-\frac{1}{10}\delta +\frac{229}{280}\delta^2+\mathcal{O}(\delta^3)
\end{align}
should be sufficient, which captures the exact function with 10\% accuracy. Equation (\ref{norm15}) then implies a divergent response in the strongly anisotropic limits $\delta\to \pm 1$, resulting in an increase of conductivity. We display $\lambda(\delta)$ together with the quadratic approximation in Fig. \ref{FigLambda}.

In the particle-hole asymmetric case with $x\neq 0$ (while maintaining spatial isotropy $\delta=0$ for simplicity), the intraband contributions are obtained by replacing the mass $m^*$ with the effective band masses from Eq. (\ref{lutt8}) and thus read
\begin{align}
 \label{norm17} \vare^{(\rm upper)}(\omega) &= -\frac{2e^2}{\eps_0 m^*_{\rm up}} 
\frac{1}{\omega^2}\int_{\textbf{q}} n_{\rm F}\Bigl(\frac{q^2}{2m^*_{\rm up}}-\mu\Bigr),\\
 \label{norm18} \vare^{(\rm lower)}(\omega) &= -\frac{2e^2}{\eps_0 m^*_{\rm low}} \frac{1}{\omega^2} \int_{\textbf{q}} n_{\rm F}\Bigl(\frac{q^2}{2m^*_{\rm low}}+\mu\Bigr).
\end{align}
The corresponding QBT contribution in the absence of particle-hole symmetry is given by
\begin{align}
  \label{norm19} \vare^{(\rm QBT)}(\omega) ={}& \frac{e^2}{4\pi \eps_0}\sqrt{\frac{m^*}{\omega}}(1+\rmi) \\
 \nonumber &-\frac{2e^2}{\eps_0m^*} \int_{\textbf{q}}\frac{n_{\rm F}(\frac{q^2}{2m^*_{\rm up}}-\mu)+n_{\rm F}(\frac{q^2}{2m^*_{\rm low}}+\mu)}{-(\omega+\rmi 0)^2+q^4/(m^*)^2},
\end{align}
see our discussion at the end of App. \ref{AppNormIso}. Therein we also describe how $x\neq 0$ can be implemented easily when needed, which is necessary for studying the optical response of materials with sizeable $x$, while still keeping $|x|<1$ in order to have an inverted band structure. For the half-Heusler material YPtBi, however, $x\simeq 0.17$ is estimated to be small \cite{Kimeaao4513,PhysRevLett.120.057002}. Furthermore, $x$ is an irrelevant parameter in the sense of the renormalization group so that $x\to 0$ for $\mu=0$ and very low frequencies \cite{PhysRevB.93.205138,PhysRevB.95.075149}. Hence for the rest of the paper we assume $x=0$, which additionally implies an appealingly symmetric structure of the results.

\subsection{Quasi-static limit}
We now discuss the intraband and interband contributions in the quasi-static limit, where longitudinal and transverse  components differ. We begin with the zero temperature case as it allows to give analytical expressions for the response functions. We assume $x=\delta=0$. The intraband contributions to the longitudinal and transverse  response functions in the limit $p^2/(2m^*\omega)\to0$ with $\omega/vp$ held fixed read
\begin{align}
 \label{quasi1} \vare_{\rm L}^{(\rm intra)}(\omega,p) ={}& \frac{n_0 e^2}{\eps_0m^*} \frac{3}{v^2p^2}\Bigl[1 -\frac{\omega}{2vp}\ \ln\Bigl(\frac{\omega+vp+\rmi 0}{\omega-vp+\rmi 0}\Bigr)\Bigr],\\
 \nonumber \vare_{\rm T}^{(\rm intra)}(\omega,p) ={}& -\frac{n_0 e^2}{\eps_0m^*}\frac{3}{2v^2p^2}  \Biggl[ 1+\frac{vp}{2\omega}\Bigl[1-\Bigl(\frac{\omega}{vp}\Bigr)^2\Bigr]\\
 \label{quasi2} &\times\ln\Bigl(\frac{\omega+vp+\rmi0}{\omega-vp+\rmi0}\Bigr)\Biggr].
\end{align}
Here the logarithm for nonzero $0\neq r\in\mathbb{R}$ is defined as
\begin{align}
\label{quasi3} \ln(r\pm \rmi 0) = \begin{cases} \ln r & (r>0)\\ \ln(-r) \pm \rmi \pi & (r<0)\end{cases}.
\end{align}
Note that the longitudinal contribution is logarithmically divergent for $\omega=vp$, whereas the transverse  contributions remains finite for this frequency. We plot the functions, together with the finite temperature results presented below, in Fig. \ref{FigQuasiIntra}.

It is instructive to expand the response as a function of $\omega/vp$ in the asymptotic regimes. For $\omega\ll vp$ we obtain
\begin{align}
 \label{quasi4} \vare_{\rm L}^{(\rm intra)}(\omega,p) &= \frac{n_0 e^2}{\eps_0m^*} \frac{3}{p^2v^2}\Bigl[1+\frac{\pi}{2}\frac{\rmi \omega}{vp}-\Bigl(\frac{\omega}{vp}\Bigr)^2\Bigr],\\
 \label{quasi5} \vare_{\rm T}^{(\rm intra)}(\omega,p) &= \frac{n_0 e^2}{\eps_0m^*}\frac{3\pi}{4\omega^2} \frac{\rmi \omega}{vp} \Bigl[1+\frac{4}{\pi}\frac{\rmi \omega}{vp}-\Bigl(\frac{\omega}{vp}\Bigr)^2+\dots\Bigr].
\end{align}
We observe that the leading L contribution is real, whereas the T contribution is predominantly imaginary. Furthermore, the L component is subleading compared to the T component, as it is suppressed by an additional power of $\omega/vp$. The response functions in the quasi-static limit can also be expanded for $vp/\omega\ll 1$, which yields
\begin{align}
 \label{quasi6} \vare_{\rm L}^{(\rm intra)}(\omega,p) &= -\frac{n_0 e^2}{\eps_0m^*} \frac{1}{\omega^2}\Bigl[ 1 +\frac{3}{5}\Bigl(\frac{vp}{\omega}\Bigr)^2+\dots\Bigr],\\
 \label{quasi7} \vare_{\rm T}^{(\rm intra)}(\omega,p) &= -\frac{n_0 e^2}{\eps_0m^*} \frac{1}{\omega^2}\Bigl[ 1 +\frac{1}{5}\Bigl(\frac{vp}{\omega}\Bigr)^2+\dots\Bigr].
\end{align}
We observe to recover the homogeneous result in the limit $vp/\omega\to0$.

\begin{figure}[t!]
\centering
\begin{minipage}{0.48\textwidth}
\includegraphics[width=8cm]{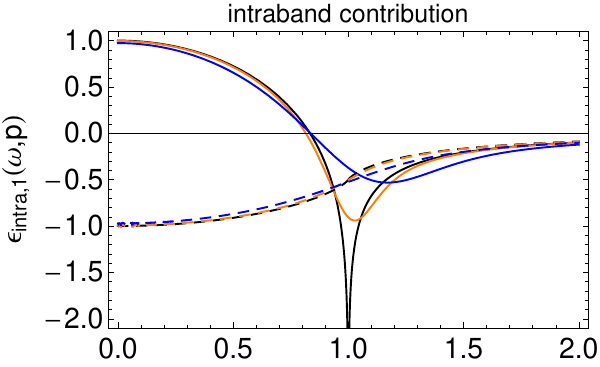}
\includegraphics[width=8cm]{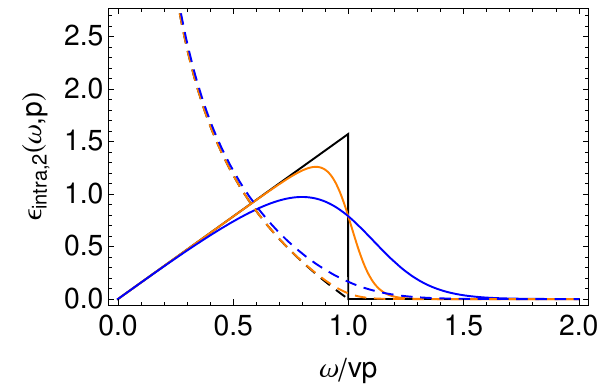}
\caption{Intraband contributions to the dielectric tensor in the quasi-static limit as a function of $\omega/vp$. Results are plotted in units of $\frac{n_0e^2}{\eps_0m^*}\frac{3}{p^2v^2}$, the solid lines constitute the longitudinal response, the dashed lines the transverse  response. The zero temperature results, shown in black, display singular behavior at $\omega=vp$. In particular, the real longitudinal component diverges logarithmically at this point. At nonzero temperature the functions are regular, shown here for $T/E_{\rm F}=0.1$ (orange) and $T/E_{\rm F}=0.3$ (blue). We observe the leading contribution at small frequencies to be imaginary and transverse. For large $\omega/vp\gg1$, longitudinal and transverse  response converge to the homogeneous limit.}
\label{FigQuasiIntra}
\end{minipage}
\end{figure}

\begin{figure}[t!]
\centering
\begin{minipage}{0.48\textwidth}
\includegraphics[width=8cm]{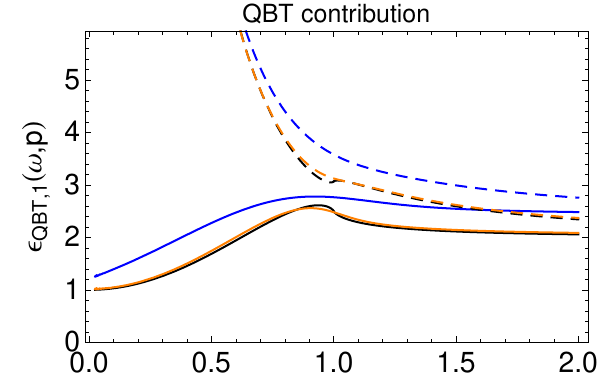}
\includegraphics[width=8cm]{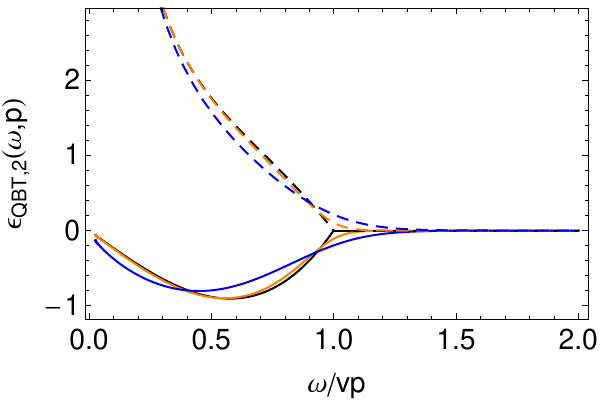}
\caption{QBT contribution to the dielectric tensor in the quasi-static limit as a function of $\omega/vp$. Curves are normalized by $\frac{e^2}{4\pi^2 \eps_0}\sqrt{2m^*/E_{\rm F}}$, and longitudinal (solid lines) and transverse  (dashed lines) contributions are shown for $T/E_{\rm F}=0$ (black), $T/E_{\rm F}=0.1$ (orange), $T/E_{\rm F}=0.3$ (blue). The interband contributions remain finite at $\omega=vp$, although showing nonanalytic behavior at zero temperature. For large $\omega/vp$ we recover the large additive contribution to the real part of $\vare(\omega)$. For small $\omega/vp$, the longitudinal contribution settles at a real value which is half the homogeneous limit. The transverse  component diverges in both the real and imaginary parts with the real part being most dominant. As a result, the limit $\omega/vp\to 0$ of $\vare_{\rm L,T}(\omega,p)$ is fully dominated by the QBT contribution, see the discussion in the main text.}
\label{FigQuasiQBT}
\end{minipage}
\end{figure}

The interband or QBT contributions at zero temperature in the quasi-static limit read
\begin{align}
 \nonumber  \vare_{\rm L}^{(\rm QBT)}(\omega,p) &=\frac{e^2}{4\pi^2\eps_0} \sqrt{\frac{2m^*}{E_{\rm F}}}\Biggl(1+\frac{3}{2}\Bigl(\frac{\omega}{vp}\Bigr)^2\\
 \label{quasi8} &+\frac{3\omega}{4vp}\Bigl[1-\Bigl(\frac{\omega}{vp}\Bigr)^2\Bigr]\ln\Bigl(\frac{\omega+vp+\rmi 0}{\omega-vp+\rmi 0}\Bigr)\Biggr),\\
 \nonumber \vare_{\rm T}^{(\rm QBT)}(\omega,p) &=\frac{5e^2}{8\pi^2\vare_0} \sqrt{\frac{2m^*}{E_{\rm F}}} \Bigl(1+\frac{3}{4}\Bigl(\frac{\omega}{vp}\Bigr)^{-2}\Bigr)\\
 \nonumber &\times \Biggl(1-\frac{3}{10}\Bigl(\frac{\omega}{vp}\Bigr)^2\\
 \label{quasi9} &-\frac{3\omega}{20vp}\Bigl[1-\Bigl(\frac{\omega}{vp}\Bigr)^2\Bigr]\ln\Bigl(\frac{\omega+vp+\rmi 0}{\omega-vp+\rmi 0}\Bigr)\Biggr).
\end{align}
The corresponding real and imaginary parts are shown in Fig. \ref{FigQuasiQBT}, together with the finite temperature results. Both longitudinal and transverse  response, although nonanalytic at $\omega=vp$, remain finite at this frequency. Expanding the QBT contribution in powers of $\omega/vp$ we obtain
\begin{align}
 \nonumber \vare_{\rm L}^{(\rm QBT)}(\omega,p)  ={}& \frac{e^2}{4\pi^2\eps_0}  \sqrt{\frac{2m^*}{E_{\rm F}}}\\
 \label{quasi10} &\ \times \Bigl[1-\frac{3\pi}{4} \frac{\rmi \omega}{vp} +3\Bigl(\frac{\omega}{vp}\Bigr)^2+\dots\Bigr],\\
 \nonumber \vare_{\rm T}^{(\rm QBT)}(\omega,p)  ={}& \frac{15e^2}{32\pi^2\eps_0}  \sqrt{\frac{2m^*}{E_{\rm F}}} \Bigl(\frac{\omega}{vp}\Bigr)^{-2}\\
 \label{quasi11} &\ \times\Bigl[1+\frac{3\pi \rmi}{20}\frac{\omega}{vp}+\frac{11}{15}\Bigl(\frac{\omega}{vp}\Bigr)^2+\dots\Bigr].
\end{align}
In contrast to the intraband response, both leading contributions are real. Furthermore, as $\omega/vp\to 0$ we observe that the longitudinal response becomes frequency-independent and settles at half the homogeneous value for $\omega\ll \mu$ given by $\vare^{(\rm QBT)}(0)=\frac{e^2}{2\pi^2\eps_0}\sqrt{\frac{2m^*}{E_{\rm F}}}$. In contrast, the transverse  contribution is divergent as $\omega/vp\to 0$. The quasi-static limit expressions for $vp/\omega\ll 1$ read
\begin{align}
 \label{quasi12} \vare_{\rm L}^{(\rm QBT)}(\omega,p)  &=\frac{e^2}{2\pi^2\eps_0} \sqrt{\frac{2m^*}{E_{\rm F}}}\Bigl[1+\frac{1}{10}\Bigl(\frac{vp}{\omega}\Bigr)^2+\dots\Bigr],\\
 \label{quasi13} \vare_{\rm T}^{(\rm QBT)}(\omega,p) &=\frac{e^2}{2\pi^2\eps_0}\sqrt{\frac{2m^*}{E_{\rm F}}}\Bigl[1+\frac{7}{10}\Bigl(\frac{vp}{\omega}\Bigr)^2+\dots\Bigr].
\end{align}
In particular, for $vp/\omega\to 0$ we obtain the homogeneous result for $\omega\ll \mu$, whereas the non-trivial frequency dependence of the homogeneous QBT contribution is lost in the quasi-static limit at zero temperature.

The very distinct behavior of the intraband and interband contributions as a function of $\omega/vp$ is striking. For large $\omega/vp$, and so in the homogeneous limit, the QBT contribution is frequency independent and amounts to the constant anomalous contribution adding to the real part of $\vare(\omega)$. For small $\omega/vp$, on the other hand, the intraband contributions are suppressed by powers of $\omega/vp$ or $(\omega/vp)^2$. The QBT contributions, on the other hand, are real and remain constant (longitudinal component) or diverge like $(\omega/vp)^{-2}$ (transverse  component). Hence the quasi-static limit is entirely dominated by the interband transitions and so genuinely different from systems with a single parabolic band.

At nonzero temperature the intraband contributions to the longitudinal and transverse  response in the quasi-static limit are given by
\begin{align}
 \label{quasi14} \vare_{\rm L}^{(\rm intra)}(\omega,p) &= \frac{2e^2}{\eps_0m^*} \int_{\textbf{q}} \frac{n_{\rm F}(\frac{q^2}{2m^*}-\mu)+n_{\rm F}(\frac{q^2}{2m^*}+\mu)}{-(\omega+\rmi 0)^2+q^2p^2/(m^*)^2},\\
 \nonumber \vare_{\rm T}^{(\rm intra)}(\omega,p) &= -\frac{e^2}{\eps_0\omega}\int_{\textbf{q}}\frac{n_{\rm F}(\frac{q^2}{2m^*}-\mu)+n_{\rm F}(\frac{q^2}{2m^*}+\mu)}{qp}\\
 \label{quasi15} &\ \times \ln\Bigl(\frac{\omega+qp^*+\rmi 0}{\omega-qp^*+\rmi 0}\Bigr),
\end{align}
with $p^*=p/m^*$. We observe that a finite temperature regularizes the logarithmic divergence of the longitudinal contribution at $\omega=vp$. The temperature dependence of the transverse  response is weak. The QBT contributions at finite temperature read
\begin{align}
 \nonumber \vare_{\rm L}^{(\rm QBT)}(\omega,p) &= \vare^{(\rm QBT)}(\omega)\\
 \nonumber & +\frac{e^2m^*}{\eps_0} \int_{\textbf{q}} \frac{n_{\rm F}(\frac{q^2}{2m^*}-\mu)+n_{\rm F}(\frac{q^2}{2m^*}+\mu)}{q^4} \\
 \nonumber &\times \Biggl[ 1-6\Bigl(\frac{\omega }{qp^*}\Bigr)^2\\
 \label{quasi16} &-\frac{3\omega}{2qp^*}\Bigl[1-2\Bigl(\frac{\omega }{qp^*}\Bigr)^2\Bigr]\ln\Bigl(\frac{\omega+qp^*+\rmi0}{\omega-qp^*+\rmi 0}\Bigr)\Biggr]
\end{align}
and
\begin{align}
 \nonumber \vare_{\rm T}^{(\rm QBT)}(\omega,p) &=  \vare^{(\rm QBT)}(\omega) \\
 \nonumber &+\frac{15e^2m^*}{8\vare_0}  \int_{\textbf{q}}  \frac{n_F(\frac{q^2}{2m^*}-\mu)+n_F(\frac{q^2}{2m^*}+\mu)}{q^4}\\
 \nonumber &\times\Biggl[ \Bigl(\frac{\omega}{qp^*}\Bigr)^{-2}+\frac{1}{3}+\frac{8}{5}\Bigl(\frac{\omega}{qp^*}\Bigr)^2\\
 \label{quasi17} &+\frac{\omega}{10qp^*}\Bigl[1-8\Bigl(\frac{\omega}{qp^*}\Bigr)^2\Bigr]\log\Bigl(\frac{\omega+qp^*+\rmi 0}{\omega-qp^*+\rmi 0}\Bigr)\Biggr].
\end{align}
Here $\vare^{(\rm QBT)}(\omega)$ is the homogeneous contribution from Eq. (\ref{norm8}). For nonzero temperature this term can have a residual (non-universal) dependence on $\omega/E_{\rm F}$. For this note that  for a generic value of $\omega/vp \sim 1$, we have $\omega/E_{\rm F} \sim p^2/(2m^*\omega)$. Hence, although $\omega/E_{\rm F}\to 0$ in the strict quasi-static limit, a finite value of $p$ implies a nonzero value of $\omega/E_{\rm F}$. This small value of $\omega/E_{\rm F}$ does not affect the zero temperature value of $\vare^{(\rm QBT)}(0)$ in Eq. (\ref{norm10}). In fact, although the integrand has a singularity at $q^2=m^*\omega$, this singularity is not resolved at $T=0$ due to the infrared cutoff provided from the Fermi--Dirac distribution, which limits the integration to $q>p_{\rm F}$. In striking contrast, for $T/E_{\rm F}>0$ the whole range of momenta is supported due to the Fermi--Dirac distribution, and so every small $\omega/E_{\rm F}\neq 0$ contributes to the integral. In the curves shown in Fig. \ref{FigQuasiQBT} we suppress this non-universal contribution by assuming $p^2/(2m^*\omega)$ to be small enough so that $\omega/E_{\rm F}\approx 0$, and so
\begin{align}
 \nonumber \vare^{(\rm QBT)}(\omega) &\approx \vare^{(\rm QBT)}(0)\\
 \label{quasi18} &=\frac{2e^2}{\eps_0m^*} \int_{\textbf{q}}\frac{1-n_{\rm F}(\frac{q^2}{2m^*}-\mu)-n_{\rm F}(\frac{q^2}{2m^*}+\mu)}{q^4/(m^*)^2},
\end{align}
which is a universal function of $T/E_{\rm F}$.

\section{Superconducting state response}

In this section, after reviewing some general facts about superconductivity in Luttinger semimetals, we compute the intraband and interband contributions to the homogeneous optical response in the s-wave superconducting state. In particular, we derive explicit expressions for the QBT contribution to the Drude weight factor and superfluid density within RPA for both finite and zero chemical potential, which comprises weak and strong coupling superconductors. The result presented here are derived in App. \ref{AppSC}.

\subsection{Superconductivity in Luttinger semimetals}

The complexity of the quadratic band touching point in Luttinger semimetals allows for a rich variety of possible superconducting ordered states. The corresponding Bogoliubov--de Gennes (BdG) Hamiltonian is given by
\begin{align}
\label{sup1} H_{\rm BdG}(\textbf{p}) = \begin{pmatrix} \hat{H}(\textbf{p})-\mu & \hat{\Delta}(\textbf{p}) \\ \hat{\Delta}(\textbf{p})^\dagger & - \hat{H}(\textbf{p})^{T}+\mu \end{pmatrix},
\end{align}
with $\hat{H}(\textbf{p})$ the Luttinger Hamiltonian from Eq. (\ref{lutt1}) and $\hat{\Delta}(\textbf{p})$ a $4\times 4$ gap matrix, so that the order parameter is given by $\langle \hat{\Delta}(\textbf{p})\rangle$. In the simplest yet far from trivial case, the ordering is local and the gap matrix momentum independent. It can then be written as a sum of two parts according to
\begin{align}
 \label{sup2} \langle \hat{\Delta}\rangle = \Bigl( \Delta \mathbb{1}_4 +\phi_{ij}J_iJ_j \Bigr) \mathcal{T},
\end{align}
where $\mathcal{T}$ is the unitary part of the time-reversal operator (see App. \ref{AppAlg} for an explicit definition). The first term in Eq. (\ref{sup2}) describes s-wave singlet superconducting order with order parameter $\Delta$, whereas $\phi_{ij}$ is a symmetric and traceless complex tensor order parameter which represents Cooper pairs having spin 2 \cite{PhysRevLett.120.057002,PhysRevB.97.064504}. The onset of complex tensor order leads to very nontrivial momentum structures of the gap, having either line nodes or inflated Bogoliubov Fermi surfaces, that should manifest in nontrivial signatures in the optical conductivity. We do not explore this highly promising direction in this work, but refer to the next section for an outlook on aspects that should be addressed in the future.

For the present work we focus on the s-wave singlet superconducting order and assume without loss of generality that the order parameter is real, $\Delta\in\mathbb{R}$. The presence of a nonzero expectation value $\Delta\neq 0$ then leads to a full gap in the excitation spectrum. For $\mu=0$, the opening of this gap requires sufficiently strong short-range interactions in the s-wave channel. At the critical coupling, the system features a quantum critical point at zero temperature, with non-Fermi liquid scaling of correlation functions and several other unusual scaling properties \cite{PhysRevB.93.205138}. For $\mu\neq0$, an infinitesimally small attraction in the s-wave channel is sufficient for ordering below an (exponentially small) critical temperature due to the Cooper instability. We therefore refer to the superconducting states that arise for $\mu=0$ and $\mu\neq0$ as strong coupling and weak coupling superconductors, respectively. In both cases the transition is of second order and the gap $\Delta(T)$ vanishes continuously at the critical temperature. The temperature dependence of the order parameter $\Delta(T)$ follows from the solution to an appropriate gap equation, which, however, requires knowledge of the coupling constant of the material. Since this quantity is generally not known in practice, we present our results as a function of independent parameters $\Delta$ and $T$, which comprises the same information and seems more accessible.

The RPA is known to yield an insufficient description of the optical response of superconductors in the single band case as it leads to expressions that violate gauge invariance. In particular, Eq. (\ref{opt9}) for the longitudinal response is not satisfied by the RPA expressions and thus leads to the question on how to interpret the outcome of the approximate calculation. It turns out that the RPA expression for the transverse  response can be used to define the optical conductivity, whereas gauge invariance of the longitudinal components is restored by including vertex corrections (see e.g. Ref. \cite{PhysRevB.96.144507} for a comprehensive discussion). We adopt this strategy for our analysis here as well and define the conductivity in the homogeneous case by
\begin{align}
 \label{sup3} \sigma(\omega) := -\frac{4\pi \vare_0}{\rmi(\omega+\rmi 0)} K_{\rm T}(\omega,0).
\end{align}
For small frequencies the conductivity behaves like \cite{BookTinkham,DresselReview}
\begin{align}
 \label{sup4}  \sigma_1(\omega) &= \frac{\pi}{2}\delta(\omega) \frac{n'e^2}{m^*},\\
 \label{sup5}  \sigma_2(\omega) &= \frac{n'e^2}{m^*\omega}
\end{align}
with Drude weight factor
\begin{align}
 \label{sup6} n' := \frac{4\pi \vare_0 m^*}{e^2} \lim_{\omega\to 0}K_{\rm T}(\omega,0).
\end{align}
Note that just like in Eq. (\ref{norm6b}) we define the $\delta$-function to be restricted to $\omega\geq 0$, which explains the prefactor of $\frac{\pi}{2}$ when going from Eq. (\ref{sup3}) to (\ref{sup4}). A quantity closely related to $n'$ is the superfluid density defined by
\begin{align}
 \label{sup7} n_{\rm s} := \frac{4\pi \vare_0 m^*}{e^2}  \lim_{p\to 0} K_{\rm T}(0,p).
\end{align}
The superfluid density allows for computing the London penetration depth.

\subsection{s-wave singlet superconductor}
Let us first discuss the superconductor with $\mu\neq 0$ and typically $\mu\gg \omega,T,\Delta$ for weak coupling, although we do not impose the latter restriction on our formulas. The intraband contribution to the conductivity is of the form of Eqs. (\ref{sup4}) and (\ref{sup5}) for all frequencies with Drude weight factor 
\begin{align}
 \label{sup8} n'^{(\rm intra)}&= \int_{\textbf{q}} \Bigl(2 -\frac{\vare_q}{E_q}[1-2n_{\rm F}(E_q)]+\frac{f_q}{F_q}[1-2n_{\rm F}(F_q)]\Bigr),
\end{align}
with upper and lower band quasiparticle dispersions
\begin{align}
 \label{sup9} \vare_q &= \frac{q^2}{2m^*}-\mu,\   E_q  = \sqrt{\vare_q^2 +\Delta^2},\\
 \label{sup10} f_q &= -\frac{q^2}{2m^*}-\mu,\    F_q  = \sqrt{f_q^2 +\Delta^2}.
\end{align}
Note that the paramagnetic term $K_{\rm T}^{(\rm p,intra)}(\omega,0)$ vanishes within RPA, and so only the diamagnetic term contributes to Eq. (\ref{sup6}). Furthermore, for $\Delta\neq0$ the cancellation between diamagnetic and paramagnetic contribution to $\lim_{p\to 0}K_{\rm T}^{(\rm intra)}(0,p)$ is not perfect, and we obtain a finite contribution to the superfluid density given by
\begin{align}
 \nonumber n_{\rm s}^{(\rm intra)} ={}& n'^{(\rm intra)}\\
 \label{sup11} &+\frac{4}{3}\int_{\textbf{q}} \frac{q^2}{2m^*} \Bigl[ \frac{\partial}{\partial E_q}n_{\rm F}(E_q)+\frac{\partial}{\partial F_q}n_{\rm F}(F_q)\Bigr].
\end{align}
Notice that the term in the second line is negative and so we have $n'^{(\rm intra)}\geq n_{\rm s}^{(\rm intra)}$, with equality at zero temperature. For vanishing gap, $\Delta\to0$, the intraband contribution to the Drude weight reproduces $n$ from Eq. (\ref{norm5}) and the superfluid density vanishes.

\begin{figure}[t]
\centering
\includegraphics[width=8cm]{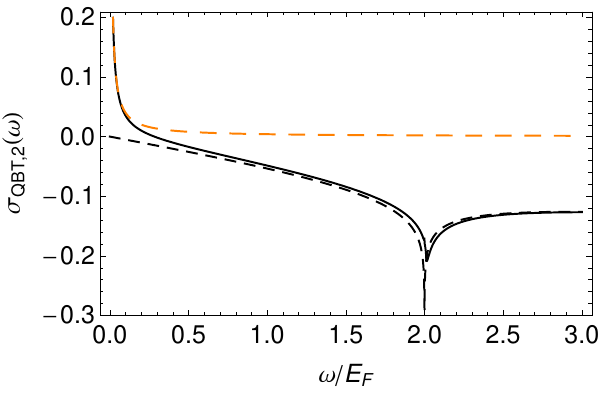}
\caption{Crossover from normal to superfluid behavior in the QBT contribution to the optical conductivity at $T=0$. The black solid line shows the result in the s-wave superconducting case with gap $\Delta/E_{\rm F}=0.1$, whereas the black dashed line shows the corresponding normal state result. The orange long-dashed line corresponds to the low-frequency behavior $n'^{(\rm QBT)}e^2/(m^*\omega)$ with QBT contribution to the Drude weight from Eq. (\ref{sup13b}). We observe that $\sigma_2(\omega)$ changes sign and so connects the negative normal state limit for $\omega\gg \Delta$ to the positive Drude like scaling at low frequencies $\omega\sim \Delta$. The same behavior is found in the strong coupling case with $\mu=0$, see Fig. \ref{FigStrong}.}
\label{FigSuper}
\end{figure}

The QBT contribution to the optical conductivity is given by
\begin{widetext}
\begin{align}
 \nonumber \sigma^{(\rm QBT)}(\omega) ={}&-\frac{e^2/m^*}{\rmi(\omega+\rmi 0)} \int_{\textbf{q}} \frac{1}{(\omega+\rmi 0)^4-4(\omega+\rmi0)^2[\frac{q^4}{(2m^*)^2}+\mu^2+\Delta^2]+16\mu^2\frac{q^4}{(2m^*)^2}}\\
 \nonumber &\times \Biggl(\Bigl[\omega^4\vare_q-4\omega^2\vare_q(\mu^2+\Delta^2)+16\mu\Delta^2 \frac{q^4}{(2m^*)^2}\Bigr]\frac{1}{E_q}[1-2n_{\rm F}(E_q)]\\
 \label{sup12} &-\Bigl[\omega^4f_q-4\omega^2f_q(\mu^2+\Delta^2)+16\mu \Delta^2\frac{q^4}{(2m^*)^2}\Bigr]\frac{1}{F_q}[1-2n_{\rm F}(F_q)]\Biggr).
\end{align}
\end{widetext}
For $\omega\gg \Delta$ the response function resembles the features of the normal state response, whereas for smaller $\omega\sim \Delta$ the conductivity has the form of Eqs. (\ref{sup4}) and (\ref{sup5}) with
\begin{align}
 \label{sup13b} n'^{(\rm QBT)} &= \frac{\Delta^2}{\mu} \int_{\textbf{q}} \Bigl(  \frac{1}{E_q}[1-2n_{\rm F}(E_q)]-\frac{1}{F_q}[1-2n_{\rm F}(F_q)]\Bigr).
\end{align}
This expression is positive for either sign of $\mu$. Remarkably, the QBT contributions to $n'$ and $n_{\rm s}$ coincide for all temperatures,
\begin{align}
 \label{sup13} n'^{(\rm QBT)} &= n_{\rm s}^{(\rm QBT)}\ \text{for }\mu\neq 0,
\end{align}
due to
\begin{align}
 \label{sup14} \lim_{p\to 0}K_{\rm T}^{(\rm QBT)}(0,p) = \lim_{\omega\to 0}K_{\rm T}^{(\rm QBT)}(\omega,0)
\end{align}
for $\mu \neq0$. This also holds in the normal phase, where $n'^{(\rm QBT)}=n_{\rm s}^{(\rm QBT)}=0$. Indeed, the normal state QBT contribution is finite for $\omega=0$ and $\mu\neq 0$, and the singular part of the optical response purely stems from the intraband terms. Note that both the intraband and QBT contributions to the Drude weight and superfluid density satisfy $n'\geq n_{\rm s}$. (This is also true in the case of $\mu=0$ discussed in the next section.) Consequently, there is no violation of the necessary requirement that the superfluid density must not exceed the density of charge carriers. In Fig. \ref{FigSuper} we show the crossover of the conductivity from the normal state behavior for $\omega\gg \Delta$ to the superfluid behavior for small $\omega\sim \Delta$.

Equation (\ref{sup13}) implies the usual exponentially weak temperature dependence $\sim e^{-\Delta/T}$ of the superfluid density and penetration depth for small temperatures that is characteristic for s-wave superconductors. In particular, for small temperatures $T\ll T_{\rm c}$ such as in the experiments of Ref. \cite{Kimeaao4513}, the temperature dependence of the gap $\Delta_0(T)$ that solves the corresponding gap equation is weak for an s-wave superconductor and so we can assume $\Delta_0(T)\approx \Delta_0(0)$ to be constant at low temperatures.

\subsection{Strong coupling superconductor}
A conceptually interesting limit of the formulas from the previous section consists in considering the case of $\mu=0$. Such a superconductor with $\Delta\neq 0$ can obviously not be caused by the Cooper instability and requires very strong coupling between fermions, but as a theoretic limit it is still worthwhile to study. The gap $\Delta$ then constitutes the only energy scale of the system at zero temperature, and thus is the only quantity that alters the universal limit $\vare(\omega)=\frac{e^2}{4\pi \eps_0}\sqrt{\frac{m^*}{\omega}}(1+\rmi)$ in Eq. (\ref{norm9b}). Note that the strong coupling required here to form the superconductor is reminiscent of the critical coupling for the existence of a bound state or dimer of two-component fermions in vacuum (i.e. for $\mu=0$) \cite{PhysRevA.73.033615,PhysRevA.75.033608}, which leads to the phenomenology of the BCS-BEC crossover for $\mu>0$ and is realized with Feshbach resonances in ultracold Fermi gases \cite{Zwerger,Boettcher:2012cm,PhysRevA.89.053630}.

\begin{figure}[t!]
\centering
\begin{minipage}{0.48\textwidth}
\includegraphics[width=8cm]{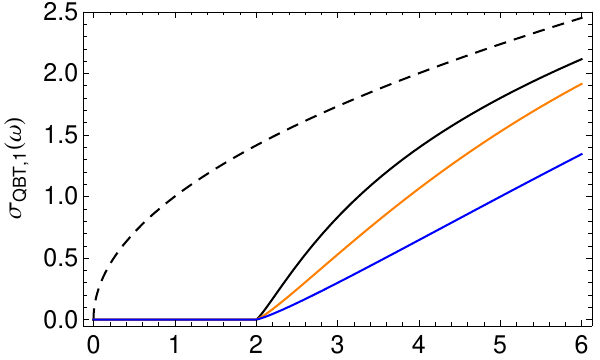}
\includegraphics[width=8cm]{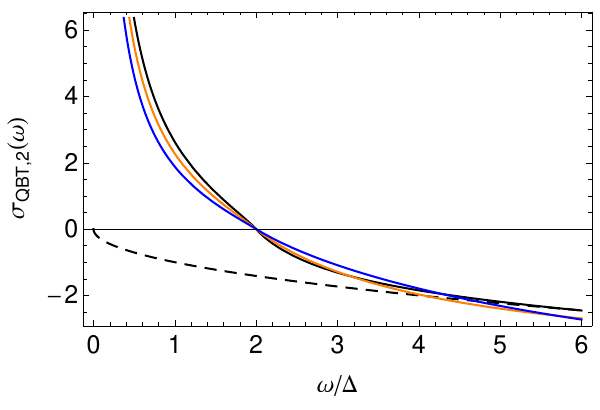}
\caption{QBT contribution to the optical conductivity of a strong coupling superconductor with $\mu=0$. The solid lines in the upper and lower panel show the real and imaginary part, respectively, for $T=0$ (black), $T/\Delta=1$ (orange), $T/\Delta=2$ (blue). We only plot the regular part of $\sigma_1(\omega)$, see Eq. (\ref{sup16}). The dashed lines show the corresponding normal state limit $\sigma(\omega)= \frac{e^2}{4\pi}\sqrt{m^*\omega}(1-\rmi)$ for $\mu=T=0$. The real part displays threshold behavior at $\omega=2\Delta$, whereas the imaginary part changes sign at this frequency. As a result, the imaginary part is negative for large frequencies---in agreement with the negative normal state limit---and it is positive with Drude-like behavior as in Eq. (\ref{sup5}) for small frequencies.}
\label{FigStrong}
\end{minipage}
\end{figure}

The transverse response function for $\mu=0$ is given by
\begin{align}
 \label{sup15} K_{\rm T}^{(\rm QBT)}(\omega,0) & = \frac{e^2(4\Delta^2-\omega^2)}{2\pi \eps_0m^*}\int_{\textbf{q}} \frac{\frac{q^2}{2m^*}[1-2n_{\rm F}(E_q)]}{E_q[-(\omega+\rmi0)^2+4E_q^2]}
\end{align}
with $E_q=\sqrt{q^4/(2m^*)^2+\Delta^2}$. We define $\sigma(\omega)$ through $K_{\rm T}^{(\rm QBT)}(\omega,0)$ by Eq. (\ref{sup3}). The corresponding optical conductivity is plotted in Fig. \ref{FigStrong} for a representative set of temperatures. The real part is given by
\begin{align}
 \nonumber &\sigma_1^{(\rm QBT)}(\omega) = \frac{\pi}{2} \delta(\omega) \frac{n'^{(\rm QBT)}e^2}{m^*} \\
 \label{sup16} &\ + \frac{e^2}{4\pi}\theta(\omega-2\Delta)\sqrt{m^*\omega}\Bigl(1-\frac{4\Delta^2}{\omega^2}\Bigr)^{5/4} [1-2n_{\rm F}(\omega/2)]
\end{align}
with Drude weight factor
\begin{align}
 \label{sup17} n'^{(\rm QBT)} &=  2\Delta^2 \int_{\textbf{q}} \frac{q^2}{2m^*}\frac{1}{E_q^3}[1-2n_{\rm F}(E_q)] .
\end{align}
Similarly, the imaginary part for small $\omega$ follows Eq. (\ref{sup5}) with $n'^{(\rm QBT)}$. Importantly, the conductivity is finite at $\omega=2\Delta$. The contribution to the superfluid density is given by
\begin{align}
  \nonumber n_{\rm s}^{(\rm QBT)} ={}&  2\Delta^2 \int_{\textbf{q}} \frac{q^2}{2m^*}\Bigl( \frac{1}{E_q^3}[1-2n_{\rm F}(E_q)] \\
  \label{sup18} &+\frac{2}{E_q^2} \frac{\partial}{\partial E_q}n_{\rm F}(E_q)\Bigr),
\end{align}
which is the $\mu\to 0$ limit of Eq. (\ref{sup13b}). We conclude that $n'^{(\rm QBT)}> n_{\rm s}^{(\rm QBT)}$ for the superconductor with $\mu=0$ at finite temperature. At zero temperature we find the explicit expression
\begin{align}
 \label{sup19} n'^{(\rm QBT)} = n_{\rm s}^{(\rm QBT)} = \frac{2\Gamma(\frac{5}{4})^2}{\pi^{5/2}} (2m^*\Delta)^{3/2}
\end{align}
with Euler's $\Gamma$-function $\Gamma(z)$.

The case of $\mu=0$ allows us to make the short-comings of the RPA with respect to gauge invariance particularly visible. In fact, Eq. (\ref{opt9}) implies that gauge invariance requires
\begin{align}
 \label{sup20} K_{\rm L}^{(\rm QBT)}(\omega,0) \stackrel{!}{=} -\omega^2 Z_{\rm QBT}(\omega).
\end{align}
However, the RPA equations for $\mu=0$ result in 
\begin{align}
 \label{sup21} K_{\rm L}^{(\rm QBT,RPA)}(\omega,0) =(4\Delta^2-\omega^2) Z_{\rm QBT}^{(\rm RPA)}(\omega),
\end{align}
which also holds at finite temperature, see Eq. (\ref{sc47b}). We added the superscript RPA to emphasize that these quantities deviate from the physical or measurable observable which satisfy gauge invariance. If we use $Z_{\rm QBT}^{(\rm RPA)}(\omega)$ to define a conductivity by means of $\tilde{\sigma}^{(\rm QBT)}(\omega) := - 4\pi \eps_0\rmi \omega Z_{\rm QBT}^{(\rm RPA)}(\omega)$, then
\begin{align}
 \label{sup22} \tilde{\sigma}^{(\rm QBT)}(\omega) = \frac{\omega^2}{\omega^2-4\Delta^2} \sigma^{(\rm QBT)}(\omega).
\end{align}
This quantity differs from $\sigma^{(\rm QBT)}(\omega)$ in two crucial aspects: First, the imaginary part $\tilde{\sigma}_2^{(\rm QBT)}(\omega)$ has a divergence at $\omega=2\Delta$. Second, for $\omega\to 0$ we have $\tilde{\sigma}_2^{(\rm QBT)}(\omega) \sim-\frac{n'^{(\rm QBT)} e^2}{4m^*\Delta^2}\omega\to 0$, and so there is no Drude-like behavior at small frequencies. We leave it for future work to study how gauge invariance can be restored by including corrections that go beyond the RPA.

\section{Summary and outlook}

In this work we have explored the optical conductivity of Luttinger semimetals in the normal and superconducting states. The motivation for this investigation is, on the one hand, recent experiments on the optical properties of Pyrochlore Iridates and half-Heusler superconductors, and, on the other hand, the recent theoretical discovery of a plethora of possible novel unconventional superconducting orders in QBT materials. Thus, although the optical properties of QBT systems in the normal state have been studied before in the context of $\alpha$-Sn \cite{PhysRevLett.20.651,PhysRevLett.25.1658,PhysRevB.5.397}, these current experimental and theoretical developments call for a more refined understanding of the electromagnetic properties of Luttinger semimetals, especially when interactions are strong or the material is in the superconducting state.

Our analysis has been built on the RPA, which constitutes the natural first step towards understanding the optical response functions. Crucially, in our analysis we have kept the full internal $4\times 4$-structure of the Luttinger Hamiltonian, which results in considerably unwieldy computations, but allows to identify both intraband and interband contributions in an unbiased way. In the normal state, the genuine QBT contribution from interband transitions is large at low-frequencies in the homogeneous limit, and it dominates the quasi-static limit. Furthermore, in the superconducting state the contribution from interband transitions is important to capture effects that are absent for single band systems. In particular, this includes Bogoliubov Fermi surfaces of certain superconducting orders in Luttinger semimetals. In the present work we have derived the general expression for the optical response in the superconducting state and applied it to the s-wave singlet superconducting case, where we find a genuine QBT contribution to the superfluid density and Drude weight.

The results that are shown in the main text of this work are either analytically evaluated or in terms of simple one-dimensional integrals. To achieve this simplicity we have restricted the presentation to the homogeneous and quasi-static limits, which are by far the most practically relevant ones. However, the full frequency and momentum dependence for the normal state response can be inferred from Eq. (\ref{trapp10}) for $K_{\rm T}^{(\rm QBT)}$ and Eq. (\ref{quapp35}) for $\chi_{\rm QBT}$ in the appendix. In particular, in App. \ref{SecCont} we show that the longitudinal QBT component satisfies the gauge invariance condition (\ref{opt9}) for all values of $\omega$ and $p$, and so $K_{\rm L}^{(\rm QBT)}$ can be deduced from $\chi_{\rm QBT}$. This leaves us with a consistent picture in the normal state, where the L component of $\vare_{ij}^{(\rm QBT)}$ can be computed from either the density or current response functions.

The consistent picture of the normal state response is absent at the RPA level in the superconducting state, where $\chi^{(\rm RPA)}$ and $K_{\rm L}^{(\rm RPA)}$ do not satisfy the gauge invariance condition (\ref{opt9}). Consequently an ambiguity arises when defining, for instance, the homogeneous conductivity $\sigma(\omega)$ from either of the two functions. This is a well-known feature for the single parabolic band, and a way around consists in either including vertex corrections to restore gauge invariance, or to use the transverse  component of the current response function to define $\sigma(\omega)$. We applied the second strategy here to infer the QBT contribution in the superconducting state, which gives the conveniently short expression for the conductivity in Eq. (\ref{sup12}), but since we have not considered the effect of vertex corrections it is too early to conclude whether this approach is correct. For the superconductor with $\mu=0$ we discussed in Eqs. (\ref{sup20})-(\ref{sup22}) how the conductivity in the homogeneous limit differs qualitatively when defined from either $K_{\rm T}$ or $\chi$.

The present work can be extended in several directions, out of which we name a few in the following. One application in the normal and superconducting state is to quantify the anomalous skin effect in Luttinger semimetals, both in the normal and superconducting phase. In fact, the quasi-static limit $q^2/(2m^*)\ll \omega\ll vp$ considered above is typically referred to as ``extreme anomalous limit'' in superconductors. The corresponding intraband contribution from the upper band has been derived in the seminal works by Mattis, Bardeen \cite{PhysRev.111.412} and by Abrikosov, Gor'kov, Khalatnikov \cite{AbrikosovGorkov}. Since we have found the normal state response in the quasi-static limit to be dominated by the QBT contribution, the behavior of Luttinger semimetals is likely to be distinctively different from single band systems in the anomalous limit, with striking observable effects in both the normal and superconducting states.

The optical response in other than s-wave singlet superconducting states can be obtained by using the general expression for the fermion propagator in the mean-field approximation in Eq. (\ref{Prop35}) with a suitable gap matrix $\hat{\Delta}$ and repeating the steps outlined in App. \ref{AppSC}. In fact, two very interesting and important cases are covered by the local gap matrix from Eq. (\ref{sup2}) with $\phi_{ij}\neq 0$: (i) By choosing a real tensor $\phi\neq 0$, the effect of nematic superconducting order on the optical response can be probed. In particular, the nematic orders feature line nodes of the gap and a spontaneous breaking of rotation symmetry. It will be exciting to see how both effects manifest in the optical response and how they relate to the measurements on half-Heusler superconductors. (ii) Choosing a genuinely complex tensor $\phi$ such that $\mbox{tr}(\phi^2)=0$ we can study superconducting orders that spontaneously break time-reversal symmetry and lead to Bogoliubov surfaces in the gap \cite{PhysRevLett.118.127001,PhysRevB.96.094526,2018arXiv180603773B}. Again, this very intriguing finding calls to be explored within the framework of electromagnetic response functions.

In order to study the effects of strong interactions and critical fluctuations on the optical response of Luttinger semimetals, it is mandatory to go beyond the RPA. First, Coulomb interactions between the electrons are relevant and famously lead to Abrikosov's non-Fermi liquid scaling of correlation functions (at least within certain regimes). Second, in the vicinity of a quantum critical point, as may be the case for Pr-227 as discussed in the Introduction, critical fluctuations of the order parameter can modify the nature of fermionic excitations. To solve such a setup self-consistently is a very challenging task and worth exploring. In a less ambitious attempt, however, it will also be interesting to assume that the mentioned strong interactions merely result in a renormalization of the fermion propagator and then use the renormalized propagator to estimate the optical response function from the fermionic one-loop diagram. Furthermore, the infrared regime can be addressed self-consistently by a scaling or renormalization group approach to infer the scaling exponents. These theoretical studies will help to design and interpret future experiments on Luttinger semimetals.

\begin{center}
 \textbf{Acknowledgements}
\end{center}

\noindent The author thanks Igor Herbut and Steve Dodge for inspiring discussions. This work was supported by the NSERC of Canada, the DoE BES Materials and Chemical Sciences Research for Quantum Information Science program, NSF Ideas Lab on Quantum Computing, DoE ASCR Quantum Testbed Pathfinder program, ARO MURI, ARL CDQI, and NSF PFC at JQI.

\begin{appendix}

\section{Clifford algebra and Gell-Mann matrices}\label{AppAlg}
In this section we collect the algebra of $\gamma$-matrices and Gell-Mann matrices that underlies the calculations that lead to the results presented in the main text. For an in-depth discussion of the algebraic structure of the Luttinger Hamiltonian and interactions in Luttinger semimetals we refer to Ref. \cite{PhysRevB.95.075149}.

Starting from the spin-3/2 matrices $\vec{J}=(J_x,J_y,J_z)^T$ we define five $\gamma$-matrices according to
\begin{align}
 \label{alg1} \gamma_1 &= \frac{1}{\sqrt{3}}(J_x^2-J_y^2),\ \gamma_2 = J_z^2-\frac{5}{4}\mathbb{1}_4,\\
 \label{alg3} \gamma_3 &= \frac{1}{\sqrt{3}}(J_zJ_x+J_xJ_z),\ \gamma_4 = \frac{1}{\sqrt{3}}(J_yJ_z+J_zJ_y),\\
 \label{alg5} \gamma_5 &=\frac{1}{\sqrt{3}}(J_xJ_y+J_yJ_x).
\end{align}
Independently of the representation  chosen for $\vec{J}$ they satisfy the Clifford algebra relation $\{\gamma_a,\gamma_b\}=2\delta_{ab}\mathbb{1}_4$, and the Luttinger Hamiltonian in Eq. (\ref{lutt1}) can be written as
\begin{align}
 \nonumber \hat{H}&= \alpha_1 \hat{p}^2 \mathbb{1}_4-(\alpha_2+\alpha_3)\sum_{a=1}^5d_a(\hat{\textbf{p}})\gamma_a\\
 \label{alg6} &\ +(\alpha_2-\alpha_3)\sum_{a=1}^5s_ad_a(\hat{\textbf{p}})\gamma_a\\
 \label{alg7} &= \frac{1}{2m^*}\Bigl(x \hat{p}^2\mathbb{1}_4+\nu\sum_{a=1}^5(1+\delta s_a)d_a(\hat{\textbf{p}})\gamma_a\Bigr)
\end{align}
with $s_{1,2}=-1$ and $s_{3,4,5}=1$, and $\nu=-\sgn(\alpha_2+\alpha_3)$. The $d_a$-functions are given by 
\begin{align}
 d_1(\textbf{p}) &= \frac{\sqrt{3}}{2}(p_x^2-p_y^2),\ d_2(\textbf{p})  = \frac{1}{2}(2p_z^2-p_x^2-p_y^2),\\
 d_3(\textbf{p}) & = \sqrt{3}p_zp_x,\ d_4(\textbf{p})  = \sqrt{3}p_yp_z,\ d_5(\textbf{p})  = \sqrt{3}p_xp_y.
\end{align}
We can write $d_a(\textbf{p}) = \frac{\sqrt{3}}{2} \Lambda^a_{ij}p_ip_j$ with the real Gell-Mann matrices
\begin{align}
 \nonumber \Lambda^1 &= \begin{pmatrix} 1 & 0 & 0 \\ 0 & -1 & 0 \\ 0 & 0 & 0 \end{pmatrix},\ \Lambda^2 = \frac{1}{\sqrt{3}} \begin{pmatrix} -1 & 0 & 0 \\ 0 & -1 & 0 \\ 0 & 0 & 2 \end{pmatrix},\\
 \label{eq6} \Lambda^3 &= \begin{pmatrix} 0 & 0 & 1 \\ 0 & 0 & 0 \\ 1 & 0 & 0 \end{pmatrix},\ \Lambda^4 = \begin{pmatrix} 0 & 0 & 0 \\ 0 & 0 & 1 \\ 0 & 1 & 0  \end{pmatrix},\ \Lambda^5 = \begin{pmatrix} 0 & 1 & 0 \\ 1 & 0 & 0 \\ 0 & 0 & 0 \end{pmatrix}.
\end{align}
We define the symbol $J_{abc}$ by
\begin{align}
 J_{abc} = \mbox{tr}(\Lambda^a\Lambda^b\Lambda^c).
\end{align}
Using the standard representation for the matrices $\vec{J}$ we have
\begin{align}
 \label{alg8} \gamma_1 &= \begin{pmatrix}0 & 0 & 1 & 0 \\ 0 & 0 & 0 & 1 \\ 1 & 0 & 0 & 0 \\ 0 & 1 & 0 & 0 \end{pmatrix},\ \gamma_2 = \begin{pmatrix} 1 & 0 & 0 & 0 \\ 0 & -1 & 0 & 0 \\ 0 & 0 & -1 & 0 \\ 0 & 0 & 0 & 1 \end{pmatrix},\\
 \label{alg9} \gamma_3 &= \begin{pmatrix} 0 & 1 & 0 & 0 \\ 1 & 0 & 0 & 0 \\ 0 & 0 & 0 & -1 \\ 0 & 0 & -1 & 0 \end{pmatrix},\ \gamma_4 = \begin{pmatrix}0 & -\rmi & 0 & 0 \\ \rmi & 0 & 0 & 0 \\0 & 0 & 0 & \rmi \\  0 & 0 & -\rmi & 0 \end{pmatrix},\\
 \label{alg10} \gamma_5 &= \begin{pmatrix}0 & 0 & -\rmi & 0 \\ 0 & 0 & 0 & -\rmi \\ \rmi & 0 & 0 & 0 \\ 0 & \rmi & 0 & 0 \end{pmatrix}.
\end{align}
In particular, the matrices $\gamma_{1,2,3}$ are real, whereas $\gamma_{4,5}$ are imaginary. The unitary part of the time-reversal operator in Eq. (\ref{sup2}) is then uniquely given by $\mathcal{T}=\rmi \gamma_4\gamma_5$ \cite{PhysRevD.87.085002,PhysRevB.93.205138}. Here we choose $\mathcal{T}$ to be Hermitean.

\section{From path integral to optical response}

In this section we derive the general formulas for the density-density and current-current linear response function for QBT systems in thermal equilibrium within RPA, starting from the path integral in real time, with subsequent analytic continuation to imaginary time. We further derive the mean-field fermion propagator in the presence of a superconducting order parameter that enters the general formulas.

\subsection{Real-time and imaginary-time formalism}\label{AppNot}

In the real-time setup we use a Minkowksi metric with signature $(-1,1,1,1)$ and denote the coordinate vector and coordinate derivative by
\begin{align}
 \label{AppNot1}  x^\mu = \begin{pmatrix} t \\ \textbf{x} \end{pmatrix}\ \text{and}\ \partial_\mu := \frac{\partial}{\partial x^{\mu}} =\begin{pmatrix} \partial_t \\ \nabla \end{pmatrix}.
\end{align}
The corresponding scalar product is given by
\begin{align}
 \label{AppNot2} x\cdot x' = x_\mu x'{}^\mu = -tt'+ \textbf{x}\cdot\textbf{x}',
\end{align}
and we write
\begin{align}
 \label{AppNot4} \int_{x} &= \int_{-\infty}^\infty \mbox{d}t \int \mbox{d}^3x\ \\
 \label{AppNot5} \int_{\omega}  &= \int_{-\infty}^\infty \frac{\mbox{d}\omega}{2\pi},\ \int_{\textbf{p}} = \int \frac{\mbox{d}^3p}{(2\pi)^3}.
\end{align}
The 4-momentum is denoted by
\begin{align}
 \label{AppNot3} p^\mu = \begin{pmatrix} \omega \\ \textbf{p} \end{pmatrix},
\end{align}
and we choose signs such that the Fourier decomposition of a field variable $\psi(t,\textbf{x})$ is given by
\begin{align}
 \label{AppNot6} \psi(t,\textbf{x}) = \int_{\omega,\textbf{p}} e^{\rmi(\textbf{p}\cdot\textbf{x}-\omega t)} \psi(\omega,\textbf{p}) = \int_{\omega,\textbf{p}} e^{\rmi p\cdot x} \psi(\omega,\textbf{p}).
\end{align}

In the imaginary-time formalism we introduce imaginary or Euclidean time $\tau$ through $\tau = \rmi t$ and the coordinate vector becomes
\begin{align}
 \label{AppNot8} X = \begin{pmatrix} \tau \\ \textbf{x} \end{pmatrix}.
\end{align}
At nonzero temperature, $\tau$ is restricted to the interval $\tau\in[0,\beta]$ with $\beta=1/T$. We have $x\cdot x' = XX'$ with Euclidean scalar product
\begin{align}
 \label{AppNot9} XX' = \tau \tau'+\textbf{x}\cdot\textbf{x}.
\end{align}
Consequently, the signature in the imaginary-time formalism is $(1,1,1,1)$ and there is no need to distinguish between upper and lower indices. All indices $i=0,1,2,3$ will be denoted as lower indices. We introduce the Euclidean 4-momentum 
\begin{align}
 \label{AppNot10} P = \begin{pmatrix} p _0 \\ \textbf{p} \end{pmatrix}
\end{align}
and write
\begin{align}
 \label{AppNot11} \int_X  &=\int_0^\beta \mbox{d}\tau \int\mbox{d}^3x\ ,\\
\label{AppNot12}  \int_P  &= T \sum_{n} \int_{\textbf{p}}.
\end{align}
We identify $p_0=\omega_n$ such that, depending on the context, $\omega_n = 2\pi n T$ or $\omega_n=2\pi (n+1/2)T$ is a bosonic or fermionic Matsubara frequency with $n\in\mathbb{Z}$, respectively. The Euclidean delta functions $\delta(X-X')$ and $\delta(P-P')$ are defined with the appropriate prefactors such that $\int_X \delta(X-X')=1$ and $\int_P\delta (P-P')=1$. The Fourier decomposition in Eq. (\ref{AppNot6}) becomes
\begin{align}
 \label{AppNot13} \psi(X) = \int_P e^{\rmi PX} \psi(P).
\end{align}
In particular, the relation $XP = x\cdot p$ with $\tau =\rmi t$ implies $\rmi p_0 = -\omega.$ The minus sign on the right hand side results from the relative sign change between temporal and spatial parts in the scalar product when going from real-time to imaginary-time. In particular, it cannot be removed by changing the overall sign of the Minkowksi metric or the overall sign of the exponent in the Fourier decomposition. The relative minus sign in $\rmi p_0=-\omega$ is, however, irrelevant for the analytic continuation of response functions, which consists of two steps: (1) After performing all Matsubara summations, express the response function as a function of the real parameter $p_0> 0$. (2) analytically continue $\rmi p_0\to \omega+\rmi 0$. In this way, $\text{Re} (p_0) = +0$ remains to have a small positive real part.

We now discuss the coupling of fermions of charge $\bar{q}$ and $N$ internal degrees of freedom to an external electromagnetic field. First consider the case of a single particle in a parabolic band with Hamiltonian $\hat{H} = H(\hat{\textbf{p}}) = \frac{1}{2m}\hat{\textbf{p}}^2 \mathbb{1}_N$ with $\hat{\textbf{p}}=-\rmi \nabla$. The real-time action for the single particle in an external field is then given by
\begin{align}
 \nonumber S &=\int_x \psi^\dagger [\rmi \partial_t-\bar{q}\phi -H(\hat{\textbf{p}}-\bar{q}\textbf{A})]\psi\\
 \nonumber &=\int_x\Biggl(\rmi \psi^\dagger(\partial_t \psi)-\bar{q}\phi \psi^\dagger \psi \\
 \label{AppNot15}&\ - \frac{1}{2m}\Bigl[(\rmi \nabla -\bar{q}\textbf{A})\psi^\dagger\Bigr]\cdot \Bigl[(-\rmi \nabla -\bar{q}\textbf{A})\psi\Bigr]\Biggr),
\end{align}
where $\psi=(\psi_1,\dots,\psi_N)^{\rm T}$ is a Grassmann field. The action is invariant under gauge transformations given by
\begin{align}
 \label{AppNot16} \psi \to e^{\rmi \bar{q}\chi}\psi,\ \psi^\dagger \to e^{-\rmi\bar{q} \chi}\psi^\dagger, \ A^\mu \to A^\mu +\partial^\mu \chi
\end{align}
with $\chi$ some scalar function. We write
\begin{align}
\label{AppNot17} A^\mu = \begin{pmatrix} \phi \\ \textbf{A} \end{pmatrix},\ j^\mu = \begin{pmatrix} \rho \\ \textbf{j} \end{pmatrix}
\end{align}
and define the current $j^\mu$ by
\begin{align}
 \label{AppNot18} j^\mu = \frac{\delta S}{\delta A_\mu}.
\end{align}
We impose the gauge condition $\nabla\cdot \textbf{A}=0$ such that $[\hat{p}_i,A_i]=0$. This yields
\begin{align}
 \label{AppNot19} \rho &= \bar{q}\ \psi^\dagger \psi,\\
 \label{AppNot20} \textbf{j} &= -\frac{\rmi \bar{q}}{2m}\Bigl[\psi^\dagger(\nabla\psi) - (\nabla\psi^\dagger)\psi\Bigr] - \frac{\bar{q}^2}{m} \textbf{A} \psi^\dagger\psi.
\end{align}
We note that to linear order in $A^\mu$ the action can be written as
\begin{align}
 \label{AppNot28b} S &=\int_x \Bigl( \psi^\dagger [\rmi \partial_t-\hat{H}]\psi+j_\mu A^\mu\Bigr)+\mathcal{O}(A^2)\\
  \label{AppNot28c} &=\int_x \Bigl( \psi^\dagger [\rmi \partial_t-\hat{H}]\psi-\rho \phi+ \textbf{j}\cdot\textbf{A}\Bigr)+\mathcal{O}(A^2).
\end{align}

Next consider a general single particle Hamiltonian quadratic in momenta,
\begin{align}
\label{AppNot21} \hat{H} =  g_{ij} \hat{p}_i \hat{p}_j,
\end{align}
where the coefficients $g_{ij}=g_{ji}$ are Hermitean $N \times N$ matrices. The case of a single parabolic band then corresponds to $g_{ij} = \frac{1}{2m}\delta_{ij}\mathbb{1}_N$. For the isotropic QBT point with $\hat{H}=d_a(\hat{\textbf{p}})\gamma_a$ we have $N=4$ and 
\begin{align}
\label{AppNot22} g_{ij} = \frac{\sqrt{3}}{2} \Lambda^a_{ij}\gamma_a.
\end{align}
The action in this more general setup reads
\begin{align}
 \nonumber S  &=\int_x\Biggl( \rmi \psi^\dagger(\partial_t \psi)-\bar{q}\phi \psi^\dagger \psi \\
 \label{AppNot23}&\ - \Bigl[(\rmi \partial_i -\bar{q}A_i)\psi^\dagger\Bigr]g_{ij} \Bigl[(-\rmi \partial_j -\bar{q}A_j)\psi\Bigr]\Biggr).
\end{align}
Defining the current $j^\mu$ through Eq. (\ref{AppNot18}) we arrive at
\begin{align}
 \label{AppNot24} \rho &= \bar{q}\ \psi^\dagger\psi,\\
 \label{AppNot25} \textbf{j} &= \textbf{j}^{(\rm p)} + \textbf{j}^{(\rm d)}
\end{align}
with the paramagnetic and diamagnetic contributions to the current given by
\begin{align}
 \label{AppNot26} j^{(\rm p)}_i & = -\rmi \bar{q} \Bigl[ \psi^\dagger g_{ij}(\partial_j \psi) - (\partial_j \psi^\dagger)g_{ij}\psi\Bigr],\\
 \label{AppNot27} j^{(\rm d)}_i &= - 2\bar{q}^2 A_j \psi^\dagger g_{ij}\psi.
\end{align}
Due to $g_{ij}^\dagger=g_{ij}$ the paramagnetic current may also be written as
\begin{align}
 \label{AppNot28} \textbf{j}^{(\rm p)} = \frac{\bar{q}}{2} \Bigl[ \psi^\dagger \Bigl(\frac{\partial H}{\partial \textbf{p}}(-\rmi \nabla)\psi\Bigr) - \Bigl(\frac{\partial H}{\partial \textbf{p}}(-\rmi \nabla)\psi^\dagger\Bigr)\psi\Bigr].
\end{align}

The equilibrium properties of the system under consideration can be obtained from the partition function
\begin{align}
 \label{AppNot29} Z[\phi,\textbf{A}] = \int \mbox{D}\psi\mbox{D}\psi^*\ e^{-S_{\rm E}}
\end{align}
for the Euclidean field $\psi(\tau=\rmi t,\textbf{x})$ with action $S_{\rm E}$. After analytic continuation to imaginary time we have $S_{\rm E}=-\rmi S$, and so for the system described by the action in Eq. (\ref{AppNot23}) we have
\begin{align}
 \nonumber S_{\rm E} ={}& \int_X \Biggl( \psi^\dagger(\partial_\tau \psi)+\bar{q}\phi \psi^\dagger \psi\\
 \label{AppNot30} &+ \Bigl[(\rmi \partial_i -\bar{q}A_i)\psi^\dagger\Bigr]g_{ij} \Bigl[(-\rmi \partial_j -\bar{q}A_j)\psi\Bigr]\Biggr).
\end{align}
The expressions for $\rho$ and $\textbf{j}$ in terms of the field $\psi$ remain invariant under analytic continuation and so does the coupling $j_\mu A^\mu$ to the electromagnetic field. To linear order in the external fields the Euclidean action is given by
\begin{align}
 S_{\rm E}=\int_X \Bigl( \psi^\dagger [\partial_\tau+\hat{H}]\psi+\rho \phi- \textbf{j}\cdot\textbf{A}\Bigr)+\mathcal{O}(A^2).
\end{align}

\subsection{Response funtions}\label{AppLin}
We next compute the linear response functions with respect to an external electromagnetic field from the Euclidean field theory. Starting from the Euclidean path integral in Eq. (\ref{AppNot29}), where $S_{\rm E}$ is assumed to generally describe an interacting theory, the expectation values for the charge density and current follow from Eq. (\ref{AppNot30}) and read
\begin{align}
  \label{Lin1} \langle \rho(X) \rangle &=  -\frac{\delta \ln Z}{\delta \phi(X)},\\
 \label{Lin2}  \langle \textbf{j}(X) \rangle &=  \frac{\delta \ln Z}{\delta \textbf{A}(X)}.
\end{align}
To linear order in $A^\mu$ we then have
\begin{align}
 \label{Lin2b}  \langle \rho(X) \rangle &= -\int_{X'} \chi(X,X') \phi(X'),\\
  \label{Lin2c}  \langle j_i(X) \rangle &= -\int_{X'} K_{ij}(X,X') A_j(X'),
\end{align}
where we define the density and current response functions, $\chi$ and $K_{ij}$, by
\begin{align}
 \label{Lin3} \chi(X,X') &= -\frac{\delta \langle\rho(X)\rangle}{\delta \phi(X')}\Bigr|_{A=0} =  \frac{\delta^2 \ln Z}{\delta \phi(X) \delta \phi(X')}\Bigr|_{A=0},\\
 \label{Lin4} K_{ij}(X,X') &= -\frac{\delta \langle j_i(X)\rangle}{\delta A_j(X')}\Bigr|_{A=0} = - \frac{\delta^2 \ln Z}{\delta A_i(X) \delta A_j(X')}\Bigr|_{A=0},
\end{align}
respectively. In the cases of interest here the system features translation invariance and thus $\chi(X,X')$ and $K_{ij}(X,X')$ only depend on the difference $X-X'$ and we can write
\begin{align}
\label{Lin5} \chi(X,X') &=\chi(X-X',0),\\
 \label{Lin6} K_{ij}(X,X') &=K_{ij}(X-X',0).
\end{align}
Equations (\ref{Lin2b}) and (\ref{Lin2c}) then imply
\begin{align}
 \label{Lin7} \langle \rho(P) \rangle &= -\chi(P) \phi(P),\\
  \label{Lin8} \langle j_i(P) \rangle &= -K_{ij}(P) A_j(P)
\end{align}
to linear order in $A^\mu$. Here the Fourier transform in the first variable is given by
\begin{align}
 \label{Lin9} \chi(P) &= \chi(p_0,\textbf{p}) =\int_X e^{-\rmi PX}\chi(X,0),\\
 \label{Lin10} K_{ij}(P) &= K_{ij}(p_0,\textbf{p}) = \int_X e^{-\rmi PX} K_{ij}(X,0).
\end{align}
The real-time linear response is obtained from these functions by analytic continuation in $p_0$ as discussed above.

After specifying the single-particle Hamiltonian $\hat{H}$, and thereby the expression for the current in terms of the microscopic field $\psi$, we can relate $\chi$ and $K_{ij}$ to specific four-point correlation functions of the field. These, on the other hand, are determined by the interactions between the fermions and may be computed within perturbation theory. For the general quadratic Hamiltonian in Eq. (\ref{AppNot21}) we obtain
\begin{align}
 \label{Lin11} \chi(X,X') &=\langle \rho(X) \rho(X')\rangle,\\
\label{Lin12} K_{ij}(X,X') &=  K_{ij}^{(\rm d)}(X,X') +K^{(\rm p)}_{ij}(X,X'),
\end{align}
with the paramagnetic contribution to the current given by
\begin{align}
\label{Lin13c} K^{(\rm p)}_{ij}(X,X') &= -\langle j^{(\rm p)}_i(X)j^{(\rm p)}_j(X')\rangle.
\end{align}
The diamagnetic contribution reads
\begin{align}
\label{Lin13b} K_{ij}^{(\rm d)}(X,X') = 2 \bar{q}^2 \langle \psi^\dagger(X^+) g_{ij}\psi(X) \rangle \delta(X-X'),
\end{align}
where we applied a regularization $X^+=(\tau+0,\textbf{x})$. For a single (upper) parabolic band the diamagnetic contribution in momentum space reads
\begin{align}
\label{Lin14} K_{ij}^{(\rm d)}(P) = \frac{\bar{q}^2}{m}n \delta_{ij}
\end{align}
with electron density $n$.

We now approximate the optical response by means of the RPA. For this we first apply a mean-field approximation where the action $S_{\rm E}$ of the interacting electron system is replaced by an effective mean-field action quadratic in the field $\psi,\psi^\dagger$. All correlation functions obtained from this mean-field action are Gaussian and so Wick's theorem applies. The RPA then simply corresponds to the one-loop diagram contributions to the response functions $\chi(P)$ and $K_{ij}(P)$. For this let the fermion two-point functions be given by
\begin{align}
 \label{Lin15a}  \langle \psi(Q) \psi^T(Q')\rangle &= F(Q) \delta(Q+Q'),\\
 \label{Lin15b}  \langle \psi(Q) \psi^\dagger(Q')\rangle &= G(Q) \delta(Q-Q'),\\
 \label{Lin15c}  \langle \psi^*(Q) \psi^T(Q')\rangle &=\hat{G}(-Q) \delta(Q-Q'),\\
  \label{Lin15d}  \langle \psi^*(Q) \psi^\dagger(Q')\rangle &= \hat{F}(-Q)\delta(Q+Q').
\end{align}
with $\hat{G}(-Q) = -G(Q)^{T}$ and $\hat{F}(Q)=F(Q)^\dagger$. Note that these expressions are $N\times N$ matrices with $N$ the number of fermion components. For short we call $G(Q)$ and $F(Q)$ the normal and anomalous contribution to the fermion propagator. An anomalous contribution can only arise in states with broken $\text{U}(1)$ symmetry. We often write the arguments of the propagators in superscript if no confusion can arise.

For the density response function we obtain
\begin{align}
 \nonumber \chi(P) &= \int_X e^{-\rmi PX} \langle \psi_\sigma^*(X)\psi_\sigma(X) \psi_{\sigma'}^*(0)\psi_{\sigma'}(0)\rangle\\
 \label{Lin16} &= -\bar{q}^2 \int_Q \mbox{tr}\Bigl[G^{Q+P}G^Q - F^{Q+P}\hat{F}^Q\Bigr].
\end{align}
The general formula for the diamagnetic contribution is given by
\begin{align}
\label{Lin16b} K_{ij}^{(\rm d)}(P) &= - 2\bar{q}^2\int_Q e^{-\rmi q_0\eta}\ \mbox{tr}[  G^Qg_{ij} ]
\end{align}
with infinitesimal $\eta=0^+$. The paramagnetic current response in the single band case reads
\begin{align}
 \nonumber K_{ij}^{(\rm p)}(P) ={}& -\int_X e^{-\rmi PX}\lim_{X'\to 0} \langle j_i^{(\rm p)}(X)j_j^{(\rm p)}(X')\rangle\\
 \nonumber ={}& \frac{\bar{q}^2}{(2m)^2} \int_Q (2q_i+p_i)(2q_j+p_j)\\
 \label{Lin17} &\times \mbox{tr}\Bigl[G^{Q+P}G^Q + F^{Q+P}\hat{F}^Q\Bigr].
\end{align}
Note the opposite relative signs between normal and anomalous contributions in $\chi$ and $K_{ij}$. For the general quadratic Hamiltonian we have
\begin{align}
 \nonumber K_{ij}^{(\rm p)}(P) ={}& \bar{q}^2 \int_{Q}(2q_k+p_k)(2q_l+p_l)\\
 \label{Lin18} &\times \mbox{tr}\Bigl[ G^{Q+P}g_{jl}G^Q g_{ik} + F^{Q+P} (g_{jl})^T \hat{F}^Q g_{ik}\Bigr],
\end{align}
where the matrix product inside the trace is over the $N$-dimensional internal space of the fermions.

\subsection{Fermion propagator}\label{AppProp}
In this section we determine the mean-field propagator for QBT systems. For this let $\psi=(\psi_1,\dots,\psi_N)^T$ be an $N$-component Grassmann field with single-particle Hamiltonian $\hat{H}=H(\hat{\textbf{p}})$ and Euclidean action
\begin{align}
 \label{Prop1} S = \int_X \Bigl[ \psi^\dagger(\partial_\tau + H(-\rmi \nabla))\psi +\frac{1}{2}\psi^\dagger \hat{\Delta} \psi^* + \frac{1}{2} \psi^T \hat{\Delta}^\dagger \psi\Bigr].
\end{align}
The $N\times N$ ``gap matrix'' $\hat{\Delta}$ breaks global $\text{U}(1)$-invariance of the theory. Only the antisymmetric part of $\hat{\Delta}$ contributes to the action and thus we assume $\hat{\Delta}^T=-\hat{\Delta}$. In the mean-field approximation it is related to the anomalous expectation value $\Delta_{\sigma\sigma'}\propto \langle \psi_\sigma\psi_{\sigma'}\rangle$ through the gap equation, but for the following derivation it plays the role of a constant parameter of the theory. A chemical potential can be implemented by shifting $H \to H - \mu \mathbb{1}_N$. 

We introduce the $2N$-component field
\begin{align}
 \label{Prop2} \Psi(X) =(\psi_1(X),\dots,\psi_N(X),\psi_1^*(X),\dots,\psi_N^*(X))^T
\end{align}
and denote the second functional derivative of the action by
\begin{align}
 \label{Prop3} S^{(2)}_{\alpha\alpha'}(X,X')  = \frac{\stackrel{\rightarrow}{\delta}}{\delta \Psi_\alpha(X)} S \frac{\stackrel{\leftarrow}{\delta}}{\delta \Psi_{\alpha'}(X')}.
\end{align}
(The arrows indicate a left- or right-derivative with respect to the Grassmann variable.) We have
\begin{align}
 \nonumber &S^{(2)}(X,X') \\
 \label{Prop4} &= \begin{pmatrix} \hat{\Delta}^\dagger & \partial_\tau\mathbb{1} -H(\rmi \nabla)^T \\ \partial_\tau\mathbb{1} +H(-\rmi \nabla) & \hat{\Delta} \end{pmatrix}\delta(X-X'),
\end{align}
with the derivative acting on $X$. Next introduce
\begin{align}
  \nonumber \mathcal{G}(X,X') &= \langle \Psi(X)\Psi^T(X')\rangle\\ 
 \label{Prop6} &= \begin{pmatrix} \langle \psi(X) \psi^T(X')\rangle & \langle \psi(X)\psi^\dagger(X')\rangle \\ \langle \psi^*(X) \psi^T(X')\rangle & \langle \psi^*(X)\psi^\dagger(X')\rangle \end{pmatrix},
\end{align}
where the expectation value is evaluated with respect to $S$. Then $\mathcal{G}$ is the inverse of $S^{(2)}$ in the sense that
\begin{align}
 \label{Prop7} \int_Y S^{(2)}(X,Y)\mathcal{G}(Y,X') = \delta(X-X') \mathbb{1}_{2N}.
\end{align}
The inverse can conveniently be computed in momentum space by employing
\begin{align}
 \nonumber S^{(2)}(Q,P) &= \int_{X,X'} e^{\rmi QX}e^{\rmi PX'} S^{(2)}(X,X') \\
 \label{Prop8} &= \mathcal{G}^{-1}(P) \delta(Q+P)
\end{align}
with
\begin{align}
 \label{Prop9} \mathcal{G}^{-1}(P) = \begin{pmatrix} \hat{\Delta}^\dagger & \rmi p_0\mathbb{1} - H(-\textbf{p})^T \\ \rmi p_0\mathbb{1} + H(\textbf{p}) & \hat{\Delta} \end{pmatrix}.
\end{align}
Equation (\ref{Prop7}) is then solved by
\begin{align}
 \label{Prop10} \mathcal{G}(Q,P) &= \int_{X,X'} e^{\rmi QX}e^{\rmi PX'} \mathcal{G}(X,X') = \mathcal{G}(P) \delta(Q+P),
\end{align}
where $\mathcal{G}(P)$ is the matrix-inverse of $\mathcal{G}^{-1}(P)$. We denote the blocks of $\mathcal{G}(P)$ as
\begin{align}
 \label{Prop11} \mathcal{G}(P) = \begin{pmatrix} F(P) & G(P) \\ \hat{G}(P) & \hat{F}(P) \end{pmatrix},
\end{align}
and by comparing Eqs. (\ref{Prop6}) and (\ref{Prop11}) we eventually read off
\begin{align}
 \label{Prop12} \langle \psi(Q)\psi^T(P)\rangle &= F(Q)\delta(Q+P)\\
 \label{Prop13} \langle \psi(Q) \psi^\dagger(P)\rangle &= G(Q) \delta(Q-P),\\
 \label{Prop14} \langle \psi^*(Q)\psi^T(P)\rangle &= \hat{G}(-Q)\delta(Q-P),\\
 \label{Prop15} \langle \psi^*(Q)\psi^\dagger(P)\rangle &= \hat{F}(-Q)\delta(Q+P).
\end{align}
We conclude that in order to compute the normal and anomalous two-point functions of the theory with action $S$, it is sufficient to invert the block matrix $\mathcal{G}{}^{-1}(P)$. Furthermore, due to the relations
\begin{align}
\label{Prop16} \hat{G}(-Q) = -G(Q)^T,\ \hat{F}(Q) = F(Q)^\dagger
\end{align}
that follow from the definition, only the functions $G(Q)$ and $F(Q)$ are independent.

We first consider systems in the normal state, where Eq. (\ref{Prop9}) reduces to
\begin{align}
 \label{Prop17} \mathcal{G}^{-1}(Q) = \begin{pmatrix} 0 & -[G^{-1}(-Q)]^T \\ G^{-1}(Q) & 0 \end{pmatrix}
\end{align}
with 
\begin{align}
 \label{Prop18} G^{-1}(Q) = (\rmi q_0-\mu)\mathbb{1}_N + H_{\textbf{q}}.
\end{align}
We included the chemical potential $\mu$. For the single parabolic band, $N=2$ and $H_{\textbf{q}} = q^2 \mathbb{1}_N$ so that
\begin{align}
\label{Prop19} G(Q) = \frac{1}{\rmi q_0 +q^2 -\mu}\mathbb{1}_2.
\end{align}
For the fully isotropic QBT point with $N=4$ we have $H_{\textbf{q}}=d_a(\textbf{q})\gamma_a$ and $H_{\textbf{q}}^2=q^4\mathbb{1}_4$ so that
\begin{align}
 \label{Prop20} G(Q) = \frac{-(\rmi q_0-\mu)\mathbb{1}_4+H_{\textbf{q}}}{(q_0+\rmi \mu)^2+q^4}.
\end{align}
We note that the chemical potential in the normal state can be implemented by means of a shift $q_0 \to q_0 + \rmi \mu$. For the general Luttinger Hamiltonian
\begin{align}
 \label{Prop21} H_{\textbf{q}} = x q^2 \mathbb{1}_4 + \sum_a (1+\delta s_a)d_a(\textbf{q})\gamma_a
\end{align}
with parameters $x$ and $\delta$, we see that particle-hole asymmetry can be included by a shift of the chemical potential according to
\begin{align}
 \label{Prop22} \mu \to \mu - xq^2.
\end{align}
Therefore we may set $x=0$ and find the propagator for nonzero $\delta$ to be 
\begin{align}
\label{Prop23} G(Q) &= \frac{-(\rmi q_0-\mu)\mathbb{1}_4+\sum_a (1+\delta s_a)d_a(\textbf{q})\gamma_a}{(q_0+\rmi \mu)^2 +\mathcal{E}^2_{\textbf{q}}}
\end{align}
with
\begin{align}
  \label{Prop24} \mathcal{E}_{\textbf{q}} &= \Bigl(\sum_a (1+\delta s_a)^2 d_a^2(\textbf{q})\Bigr)^{1/2}.
\end{align}
The poles of the propagator with $\rmi q_0 = -E$ are located at energies
\begin{align}
 \label{Prop25} E(\textbf{q}) = \pm\mathcal{E}_{\textbf{q}} -\mu.
\end{align}
Note that $\mathcal{E}_{-\textbf{q}}=\mathcal{E}_{\textbf{q}}$. Since $\mathcal{E}_{\textbf{q}}$ is strictly positive for every $\textbf{q}\neq 0$ and continuously connected to $\mathcal{E}_{\textbf{q}}\to q^2$ for $\delta\to 0$, we can identify the band dispersions of the upper and lower band through
\begin{align}
 \label{Aniso3} E_{\rm upper}(\textbf{q}) &= \mathcal{E}_{\textbf{q}}-\mu,\\
 \label{Aniso4} E_{\rm lower}(\textbf{q}) &= -\mathcal{E}_{\textbf{q}}-\mu.
\end{align}

Next we compute the propagator in the superconducting states. For the single parabolic band with $N=2$, the gap matrix can be chosen as $\hat{\Delta} = \Delta \sigma_2$ with $\Delta\in \mathbb{C}$. We then have
\begin{align}
 \label{Prop26} \mathcal{G}^{-1}(Q) = \begin{pmatrix} \Delta^* \sigma_2 & (\rmi q_0-\vare_q)\mathbb{1}_2\\ (\rmi q_0+\vare_q)\mathbb{1}_2 & \Delta \sigma_2 \end{pmatrix}
\end{align}
with $\vare_q=q^2-\mu$ and
\begin{align}
 \label{Prop27} \mathcal{G}(Q) &= \frac{1}{q_0^2+\vare_q^2 +|\Delta|^2}\begin{pmatrix} \Delta \sigma_2 & -(\rmi q_0-\vare_q)\mathbb{1}_2 \\ -(\rmi q_0 +\vare_q)\mathbb{1}_2 & \Delta^* \sigma_2\end{pmatrix}.
\end{align}
The energy spectrum of quasiparticles is isotropic and fully gapped.

For the QBT case with $N=4$, the local gap matrix $\hat{\Delta}$ can be parametrized as
\begin{align}
 \label{Prop28} \hat{\Delta} = \tilde{\Delta}\gamma_{45}
\end{align}
with $\gamma_{45}=\rmi \gamma_4\gamma_5$ and
\begin{align}
 \label{Prop29}  \tilde{\Delta} = \Delta_0 \mathbb{1}_4 + \Delta_a \gamma_a
\end{align}
and $\Delta_0,\Delta_a\in\mathbb{C}$. Since $H_{-\textbf{q}}=H_{\textbf{q}}$ is symmetric, we drop the subscript and denote $H=H_{\textbf{q}}$. The inverse propagator is then given by
\begin{align}
 \label{Prop30} \mathcal{G}^{-1}(Q) = \begin{pmatrix} \hat{\Delta}^\dagger & (\rmi q_0+\mu)\mathbb{1}_4 - H^T \\ (\rmi q_0-\mu)\mathbb{1}_4 + H & \hat{\Delta} \end{pmatrix}.
\end{align}
We omit the unit matrix $\mathbb{1}_4$ in the following derivation. We parametrize the propagator according to Eq. (\ref{Prop11}). From the condition $\mathcal{G}(Q)\mathcal{G}^{-1}(Q)=\mathbb{1}_{8}$ we obtain
\begin{align}
 \nonumber F(Q) \hat{\Delta}^\dagger +G(Q) [(\rmi q_0-\mu)+H] &= \mathbb{1}_4,\\
 \label{Prop33} F(Q) [(\rmi q_0+\mu)-H^T]+G(Q) \hat{\Delta} &=0.
\end{align}
This linear set of equations is solved by 
\begin{align}
 \nonumber G(Q) &= \Bigl[(\rmi q_0-\mu)+H-\hat{\Delta}[(\rmi q_0+\mu)-H^T]^{-1}\hat{\Delta}^\dagger\Bigr]^{-1},\\
 \label{Prop34} F(Q) &= \Bigl[ \hat{\Delta}^\dagger -[(\rmi q_0+\mu)-H^T]\hat{\Delta}^{-1}[(\rmi q_0-\mu)+H]\Bigr]^{-1}.
\end{align}
Now employ $H^T=\gamma_{45} H \gamma_{45}$ and $\hat{\Delta}=\tilde{\Delta}\gamma_{45}$ to write this as
\begin{align}
 \nonumber G(Q) &= \Bigl[(\rmi q_0-\mu)+H-\tilde{\Delta}[(\rmi q_0+\mu)-H]^{-1}\tilde{\Delta}^\dagger\Bigr]^{-1},\\
 \label{Prop35} F(Q) &= \Bigl[ \tilde{\Delta}^\dagger -[(\rmi q_0+\mu)-H]\tilde{\Delta}^{-1}[(\rmi q_0-\mu)+H]\Bigr]^{-1}\gamma_{45}.
\end{align}
The corresponding expressions for $\hat{G}(Q)$ and $\hat{F}(Q)$ are
\begin{align}
 \nonumber \hat{G}(Q) &=\gamma_{45}\Bigl[(\rmi q_0+\mu)-H-\tilde{\Delta}^\dagger [(\rmi q_0-\mu)+H]^{-1}\tilde{\Delta}\Bigr]^{-1}\gamma_{45},\\
 \label{Prop36} \hat{F}(Q) &= \gamma_{45} \Bigl[ \tilde{\Delta}-[(\rmi q_0-\mu)+H](\tilde{\Delta}^\dagger)^{-1}[(\rmi q_0+\mu)-H\Bigr]^{-1}.
\end{align}
Note that the expressions in Eqs. (\ref{Prop35}) and (\ref{Prop36}) only depend on $H-\mu\mathbb{1}$. We can account for the momentum dependence of the order parameter by replacing $\hat{\Delta}\to \hat{\Delta}_{\textbf{p}}$. Explicit forms of the propagator for the s-wave singlet superconducting case for $\mu=0$ and $\mu\neq 0$ are given at the beginning of Secs. \ref{SecSCQBTmu0} and \ref{SecSCQBT}, respectively.

\section{Normal state response}\label{AppNorm}
\subsection{Isotropic case}\label{AppNormIso}

We first compute the normal state response in the fully symmetric case with $x=\delta=0$. By inserting the fermion propagator $G(Q)$ from Eq. (\ref{Prop20}) into $\chi(P)$ from Eq. (\ref{Lin16}) with $\bar{q}=-\bar{e}$ we obtain
\begin{align}
  \nonumber &\chi(P) \\
 \label{Napp3} & =4 \bar{e}^2 \int_Q \frac{(q_0+\rmi \mu)(q_0+p_0+\rmi \mu)-d_a(\textbf{q}+\textbf{p})d_a(\textbf{q})}{[(q_0+\rmi \mu)^2+q^4][(q_0+p_0+\rmi \mu)^2+(\textbf{q}+\textbf{p})^4]}.
\end{align}
(Note that $p_0=2\pi m T$ is a bosonic Matsubara frequency, whereas $q_0=2\pi (n+1/2)T$ inside the loop is fermionic, with $m,n\in\mathbb{Z}$.) In the numerator we use
\begin{align}
\label{Napp4} d_a(\textbf{q}+\textbf{p})d_a(\textbf{q}) = (\textbf{q}+\textbf{p})^2q^2 + \frac{3}{2}\Bigl[ (\textbf{q}\cdot\textbf{p})^2-q^2p^2\Bigr].
\end{align}
Analogously, by employing the fermion propagator for the single bands from Eq. (\ref{Prop19}), and adjusting $q^2\to -q^2$ for the lower band, we find the response functions of the upper and lower bands to be
\begin{align}
 \nonumber &\chi_{\rm upper}(P) \\
 \nonumber &= -2 \bar{e}^2\int_Q \frac{1}{(\rmi q_0+q^2-\mu)[\rmi(q_0+p_0)+(\textbf{q}+\textbf{p})^2-\mu]},\\
 \nonumber  &\chi_{\rm lower}(P) \\
 \label{Napp5}&= -2 \bar{e}^2 \int_Q \frac{1}{(\rmi q_0-q^2-\mu)[\rmi(q_0+p_0)-(\textbf{q}+\textbf{p})^2-\mu]}.
\end{align}
We then have
\begin{align}
 \nonumber &\chi_{\rm upper}(P) + \chi_{\rm lower}(P) \\
 \label{Napp6} &= 4 \bar{e}^2 \int_Q \frac{(q_0+\rmi \mu)(q_0+p_0+\rmi \mu)-q^2(\textbf{q}+\textbf{p})^2}{[(q_0+\rmi \mu)^2+q^4][(q_0+p_0+\rmi \mu)^2+(\textbf{q}+\textbf{p})^4]},
\end{align}
and so the QBT contribution to $\chi(P)$ reads
\begin{align}
 \nonumber &\chi_{\rm QBT}(P) \\
   \label{Napp8}  &= 6 \bar{e}^2 \int_Q \frac{q^2p^2-(\textbf{q}\cdot\textbf{p})^2}{[(q_0+\rmi \mu)^2+q^4][(q_0+p_0+\rmi \mu)^2+(\textbf{q}+\textbf{p})^4]}.
\end{align}
In the following we compute the three contributions to $\chi(P)$ separately.

To evaluate the upper and lower band contributions we employ the Matsubara summation formula
\begin{align}
 \label{Napp9} T \sum_n \frac{1}{(\rmi q_0+a)[\rmi(q_0+p_0)+b]} = \frac{n_{\rm F}(b)-n_{\rm F}(a)}{\rmi p_0+b-a},
\end{align}
valid for every real $a,b$ and $p_0=2\pi m T$ ($m\in\mathbb{Z}$). For the upper contribution with $\vare_{\textbf{q}}=q^2-\mu$ we obtain
\begin{align}
 \nonumber &\chi_{\rm upper}(P)  = -2 \bar{e}^2 \int_Q \frac{1}{(\rmi q_0+\vare_{\textbf{q}})[\rmi(q_0+p_0)+\vare_{\textbf{q}+\textbf{p}}]}\\
 \label{Napp10} &= -2\bar{e}^2 \int_{\textbf{q}} \frac{n_{\rm F}(\vare_{\textbf{q}})}{\rmi p_0+\vare_{\textbf{q}}-\vare_{\textbf{q}+\textbf{p}}}  + \{ p_0\to -p_0\}.
\end{align}
This is the usual Lindhard expression for a single band. Note that $\chi_{\rm upper}(P)$ is symmetric in $P$, and vanishes for $p=0$. For low external momenta we have
\begin{align}
\label{Napp11} \chi_{\rm upper}(p_0,p\to 0) =  p^2 Z_{\rm upper}(p_0)
\end{align}
with
\begin{align}
 \label{Napp12} Z_{\rm upper}(p_0) =   \frac{4\bar{e}^2}{p_0^2}\int_{\textbf{q}} n_{\rm F}(q^2-\mu).
\end{align}
The lower band contribution can be computed along the same lines. We may, however, also recognize from Eq. (\ref{Napp5}) that the lower band contribution with dispersion $f_{\textbf{q}}=-q^2-\mu$ can be obtained from the upper contribution by flipping the signs of $\mu,q_0,p_0$ simultaneously. Since the final expression is symmetric in $p_0$, only the sign change in $\mu$ remains relevant, and so we conclude
\begin{align}
 \label{Napp13} Z_{\rm lower}(p_0) =  \frac{4\bar{e}^2}{p_0^2}\int_{\textbf{q}} n_{\rm F}(q^2+\mu)
\end{align}
for the low-momentum part of $\chi_{\rm lower}(p_0,p\to 0)=p^2 Z_{\rm lower}(p_0)$. The sum of the interband transitions is then given by
\begin{align}
  \label{Napp14} Z_{\rm intra}(p_0)&= \frac{2n\bar{e}^2}{p_0^2},
\end{align}
where we identified the density of charge carriers within RPA as
\begin{align}
 \label{Napp15} n = 2 \int_{\textbf{q}} \Bigl[ n_{\rm F}(q^2-\mu)+n_{\rm F}(q^2+\mu)\Bigr].
\end{align}
At zero temperature we have $n=p_{\rm F}^3/(3\pi^2)$ with $p_{\rm F}=\sqrt{|\mu|}$.

For the QBT contribution we first note that $\chi_{\rm QBT}(P)$ vanishes for $p=0$, so that for low momenta we can expand
\begin{align}
 \label{Napp17} \chi_{\rm QBT}(p_0,p\to 0) = p^2Z_{\rm QBT}(p_0)
\end{align}
in analogy to the single band contributions. From Eq. (\ref{Napp8}) we deduce
\begin{align}
 \label{Napp18} Z_{\rm QBT}(p_0) = 4\bar{e}^2\int_Q \frac{q^2}{[(q_0+\rmi \mu)^2+q^4][(q_0+p_0+\rmi \mu)^2+q^4]}.
\end{align}
We use the Matsubara sum formula
\begin{align}
 \nonumber & T \sum_n \frac{1}{[(q_0+\rmi \mu)^2+q^4][(q_0+p_0+\rmi \mu)^2+q^4]}\\ 
\label{Napp19}&=\frac{1}{q^2(p_0^2+4q^4)}\Bigl[1-n_{\rm F}(q^2-\mu)-n_{\rm F}(q^2+\mu)\Bigr],
\end{align}
valid for $p_0=2\pi mT\neq0$ ($m\in\mathbb{Z}$), to arrive at
\begin{align}
 \label{Napp20} Z_{\rm QBT}(p_0) = \bar{e}^2\int_{\textbf{q}} \frac{4}{p_0^2+4q^4}\Bigl[1-n_{\rm F}(q^2-\mu)-n_{\rm F}(q^2+\mu)\Bigr].
\end{align}
We evaluate the contribution from the first term to the integral and arrive at
\begin{align}
 \nonumber &Z_{\rm QBT}(p_0)  \\
 \label{Napp21} &=\frac{\bar{e}^2}{4\pi \sqrt{|p_0|}}-4\bar{e}^2\int_{\textbf{q}} \frac{1}{p_0^2+4q^4}\Bigl[n_{\rm F}(q^2-\mu)+n_{\rm F}(q^2+\mu)\Bigr].
\end{align}
The remaining integral on the right is limited to momenta $q^2\lesssim |\mu|$ and converges rapidly for $T>0$, making it suitable for a numerical computation. At zero temperature we have
\begin{align}
 \label{Napp22}  Z_{\rm QBT}(p_0)&= \frac{2\bar{e}^2}{\pi^2} \int_{p_{\rm F}}^\infty \mbox{d}q\ \frac{q^2}{p_0^2+4q^4}
\end{align}
For $p_0\to 0$ we arrive at
\begin{align}
 \label{Napp24} Z_{\rm QBT}(0) = \frac{\bar{e}^2}{2\pi^2p_{\rm F}}.
\end{align}

Let us briefly comment on the particle-hole asymmetric case with $x\neq 0$. The inverse propagator is then given by
\begin{align}
 \label{ph1} G^{-1}(Q) = \Bigl(\rmi q_0-\mu+\frac{x}{2m^*} q^2\Bigr)\mathbb{1}_4 +\frac{1}{2m^*}d_a(\textbf{q})\gamma_a.
\end{align}
Consequently, a finite $x$ can be implemented by a shift of the chemical potential according to $\mu \to \mu_{\textbf{q}}=\mu -\frac{x}{2m^*}q^2$. As before we set $2m^*=1$. The total density response reads
\begin{align}
 \nonumber &\chi(P) =4 \bar{e}^2\\
 \label{ph2} &\times \int_Q \frac{(q_0+\rmi \mu_{\textbf{q}})(q_0+p_0+\rmi \mu_{\textbf{q}+\textbf{p}})-d_a(\textbf{q}+\textbf{p})d_a(\textbf{q})}{[(q_0+\rmi \mu_{\textbf{q}})^2+q^4][(q_0+p_0+\rmi \mu_{\textbf{q}+\textbf{p}})^2+(\textbf{q}+\textbf{p})^4]}.
\end{align}
The upper and lower band contributions are given by
\begin{align}
 \nonumber &\chi_{\rm upper}(P) \\
 \nonumber &= -2 \bar{e}^2\int_Q \frac{1}{(\rmi q_0+\frac{q^2}{2m^*_{\rm up}}-\mu)[\rmi(q_0+p_0)+\frac{(\textbf{q}+\textbf{p})^2}{2m^*_{\rm up}}-\mu]}\\
 \label{ph3} &=  -2 \bar{e}^2\int_Q \frac{1}{(\rmi q_0+q^2-\mu_{\textbf{q}})[\rmi(q_0+p_0)+(\textbf{q}+\textbf{p})^2-\mu_{\textbf{q}+\textbf{p}}]},\\
 \nonumber &\chi_{\rm lower}(P) \\
 \nonumber &= -2 \bar{e}^2\int_Q \frac{1}{(\rmi q_0-\frac{q^2}{2m^*_{\rm low}}-\mu)[\rmi(q_0+p_0)-\frac{(\textbf{q}+\textbf{p})^2}{2m^*_{\rm low}}-\mu]}\\
 \label{ph4}&=  -2 \bar{e}^2\int_Q \frac{1}{(\rmi q_0-q^2-\mu_{\textbf{q}})[\rmi(q_0+p_0)-(\textbf{q}+\textbf{p})^2-\mu_{\textbf{q}+\textbf{p}}]},
\end{align}
with $m^*_{\rm up}=\frac{m^*}{1+x}$ and $m^*_{\rm low}=\frac{m^*}{1-x}$, see Eq. (\ref{lutt8}). Hence the QBT contribution reads
\begin{align}
 \nonumber &\chi_{\rm QBT}(P) =6 \bar{e}^2\\
 \label{ph5} &\times \int_Q \frac{q^2p^2-(\textbf{q}\cdot\textbf{p})^2}{[(q_0+\rmi \mu_{\textbf{q}})^2+q^4][(q_0+p_0+\rmi \mu_{\textbf{q}+\textbf{p}})^2+(\textbf{q}+\textbf{p})^4]}.
\end{align}
We deduce the homogeneous response as
\begin{align}
 \nonumber &Z_{\rm QBT}(p_0) = 4 \bar{e}^2 \\
 \label{ph6} &\times \int_Q \frac{q^2}{[(q_0+\rmi \mu-\rmi x q^2)^2+q^4][(q_0+p_0+\rmi \mu-\rmi x q^2)^2+q^4]},
\end{align}
which is precisely the particle-hole symmetric result from Eq. (\ref{Napp18}) but with $\mu \to \mu - xq^2$. For the upper and lower bands we find
\begin{align}
 \nonumber Z_{\rm upper}(p_0) &= \frac{2}{m^*_{\rm up}p_0^2} \int_{\textbf{q}} n_{\rm F}\Bigl(\frac{q^2}{2m^*_{\rm up}}-\mu\Bigr)\\
  \label{ph7} &= \frac{4(1+x)}{p_0^2} \int_{\textbf{q}} n_{\rm F}\Bigl(q^2-\mu+x q^2\Bigr),\\
 \nonumber Z_{\rm lower}(p_0) &= \frac{2}{m^*_{\rm low}p_0^2} \int_{\textbf{q}} n_{\rm F}\Bigl(\frac{q^2}{2m^*_{\rm low}}+\mu\Bigr)\\
  \label{ph8} &= \frac{4(1-x)}{p_0^2} \int_{\textbf{q}} n_{\rm F}\Bigl(q^2+\mu-x q^2\Bigr).
\end{align}
We see that besides the shift of the chemical potential they also get renormalized by a nontrivial prefactor, or, put differently, by replacing $m^*$ with $m^*_{\rm up/low}$, respectively.

\subsection{Anisotropic case}\label{AppAniso}
We now compute the response function for general anisotropy parameter $\delta\in[-1,1]$. We assume $x=0$. The fermion propagator $G(Q)$ is given by Eq. (\ref{Prop23}) and the associated density-density response reads
\begin{align}
\nonumber &\chi(P) = {} 4 \bar{e}^2\int_Q\frac{1}{[(q_0+\rmi \mu)^2+\mathcal{E}^2_{\textbf{q}}][(q_0+p_0+\rmi \mu)^2+\mathcal{E}^2_{\textbf{q}+\textbf{p}}]}\\
\label{Aniso1}&\times \Bigl[(q_0+\rmi \mu)(q_0+p_0+\rmi \mu)-\sum_a(1+\delta s_a)^2 d_a(\textbf{q}+\textbf{p})d_a(\textbf{q})\Bigr]
\end{align}
with $\mathcal{E}_{\textbf{q}} $ from Eq. (\ref{Prop24}). Replacing $q^2 \to \mathcal{E}_{\textbf{q}}$ in Eq. (\ref{Napp5}), we obtain the upper and lower band contributions to the response function as
\begin{align}
 \nonumber \chi_{\rm upper}(P)&= -2 \bar{e}^2\int_Q \frac{1}{(\rmi q_0+\mathcal{E}_{\textbf{q}}-\mu)[\rmi(q_0+p_0)+\mathcal{E}_{\textbf{q}+\textbf{p}}-\mu]},\\
 \label{Aniso5} \chi_{\rm lower}(P) &= -2 \bar{e}^2 \int_Q \frac{1}{(\rmi q_0-\mathcal{E}_{\textbf{q}}-\mu)[\rmi(q_0+p_0)-\mathcal{E}_{\textbf{q}+\textbf{p}}-\mu]}.
\end{align}
Adding both contributions we arrive at
\begin{align}
 \nonumber &\chi_{\rm upper}(P) + \chi_{\rm lower}(P) \\
 \label{Aniso6} &= 4 \bar{e}^2 \int_Q \frac{(q_0+\rmi \mu)(q_0+p_0+\rmi \mu)-\mathcal{E}_{\textbf{q}}\mathcal{E}_{\textbf{q}+\textbf{p}}}{[(q_0+\rmi \mu)^2+\mathcal{E}_{\textbf{q}}^2][(q_0+p_0+\rmi \mu)^2+\mathcal{E}_{\textbf{q}+\textbf{p}}^2]},
\end{align} 
and so
\begin{align}
 \nonumber &\chi_{\rm QBT}(P) \\
 \label{Aniso8} &= 4 \bar{e}^2\int_Q \frac{\mathcal{E}_{\textbf{q}}\mathcal{E}_{\textbf{q}+\textbf{p}}-\sum_a(1+\delta s_a)^2 d_a(\textbf{q}+\textbf{p})d_a(\textbf{q})}{[(q_0+\rmi \mu)^2+\mathcal{E}_{\textbf{q}}^2][(q_0+p_0+\rmi \mu)^2+\mathcal{E}_{\textbf{q}+\textbf{p}}^2]}.
\end{align}
We proceed by evaluating the individual contributions.

For the upper band contribution we note that the manipulations in Eq. (\ref{Napp10}) remain valid upon replacing $\vare_{\textbf{q}}\to \mathcal{E}_{\textbf{q}}-\mu$ and so we have
\begin{align}
 \label{Aniso9} \chi_{\rm upper}(P)= -2\bar{e}^2 \int_{\textbf{q}} \frac{n_{\rm F}(\mathcal{E}_{\textbf{q}}-\mu)}{\rmi p_0+\mathcal{E}_{\textbf{q}}-\mathcal{E}_{\textbf{q}+\textbf{p}}}  + \{ p_0\to -p_0\}.
\end{align}
We compute the corresponding coefficient $Z_{\rm upper}(p_0)$ in the expansion (\ref{Napp11}) by means of 
\begin{align}
 \label{Aniso10} Z_{\rm upper}(p_0) = \frac{1}{2p^2} \frac{\partial^2}{\partial s^2} \chi_{\rm upper}(p_0,s p)\Bigr|_{s=0}.
\end{align}
Here $s$ is some small real number.  (The following derivation parallels Eqs. (A47)-(A56) in Ref. \cite{PhysRevB.95.075149}.) We have
\begin{align}
 \label{Aniso11a} \mathcal{E}_{\textbf{q}+s\textbf{p}}^2 &= \mathcal{E}_{\textbf{q}}^2 + D_1 s +(D_2+D_3)s^2 +\mathcal{O}(s^3),\\
 \label{Aniso11b} \mathcal{E}_{\textbf{q}+s\textbf{p}} &= \mathcal{E}_{\textbf{q}} + \frac{D_1}{2\mathcal{E}_{\textbf{q}}} s -\frac{1}{2}\Bigl(\frac{D_1^2}{4\mathcal{E}_{\textbf{q}}^3}-\frac{D_2+D_3}{\mathcal{E}_{\textbf{q}}}\Bigr) s^2 +\mathcal{O}(s^3)
\end{align}
with
\begin{align}
 \label{Aniso12} D_1 &= 2\sqrt{3} \sum_a(1+\delta s_a)^2d_a(\textbf{q})(q_i\Lambda^a_{ij}p_j),\\
 \label{Aniso13} D_2 &= 2\sum_a(1+\delta s_a)^2 d_a(\textbf{q})d_a(\textbf{p}),\\
 \label{Aniso14} D_3 &= 3 \sum_a(1+\delta s_a)^2(q_i\Lambda^a_{ij}p_j)(q_k\Lambda^a_{kl}p_l).
\end{align}
This yields
\begin{align}
 \label{Aniso15} Z_{\rm upper}(p_0) = -\frac{2\bar{e}^2}{p_0^2}\int_{\textbf{q}}n_{\rm F}(\mathcal{E}_{\textbf{q}}-\mu)\frac{1}{p^2}\Bigl(\frac{D_1^2}{4\mathcal{E}_{\textbf{q}}^3}-\frac{D_2+D_3}{\mathcal{E}_{\textbf{q}}}\Bigr).
\end{align}
Since for every function $f$ we have $\int_{\textbf{q}}f(\mathcal{E}_{\textbf{q}})D_2 \propto \int_{\textbf{q}}f(\mathcal{E}_{\textbf{q}})d_a(\textbf{q})=0$, we eventually arrive at
\begin{align}
 \label{Aniso16} Z_{\rm upper}(p_0) = \frac{4\bar{e}^2}{p_0^2}\int_{\textbf{q}}n_{\rm F}(\mathcal{E}_{\textbf{q}}-\mu)\frac{1}{p^2}\Bigl(\frac{D_3}{2\mathcal{E}_{\textbf{q}}}-\frac{D_1^2}{8\mathcal{E}_{\textbf{q}}^3}\Bigr).
\end{align}
Similarly, the lower band contribution reads
\begin{align}
 \label{Aniso17} Z_{\rm lower}(p_0) = \frac{4\bar{e}^2}{p_0^2}\int_{\textbf{q}}n_{\rm F}(\mathcal{E}_{\textbf{q}}+\mu)\frac{1}{p^2}\Bigl(\frac{D_3}{2\mathcal{E}_{\textbf{q}}}-\frac{D_1^2}{8\mathcal{E}_{\textbf{q}}^3}\Bigr).
\end{align}
Before further evaluating this expression we derive the QBT contribution, which turns out to be of the same form. For this we employ Eq. (\ref{Aniso8}) and define
\begin{align}
 \label{Aniso18} Z_{\rm QBT}(p_0) = \frac{1}{2p^2} \frac{\partial^2}{\partial s^2} \chi_{\rm QBT}(p_0,sp)\Bigr|_{s=0}.
\end{align}
We have $d_a(\textbf{q}+\textbf{p}) = d_a(\textbf{q})+\sqrt{3} p_i q_j \Lambda^a_{ij} + d_a(\textbf{p})$ and so the numerator of $\chi_{\rm QBT}$ can be simplified by using
\begin{align}
 \label{Aniso19} \sum_a(1+\delta s_a)^2 d_a(\textbf{q}+\textbf{p})d_a(\textbf{q}) = \mathcal{E}_{\textbf{q}}^2+\frac{1}{2}(D_1+D_2).
\end{align}
We then arrive at
\begin{align}
 \nonumber Z_{\rm QBT}(p_0) ={}& 4\bar{e}^2\int_Q \frac{1}{p^2}\Bigl(\frac{D_3}{2\mathcal{E}_{\textbf{q}}}-\frac{D_1^2}{8\mathcal{E}_{\textbf{q}}^3}\Bigr)\\
 \label{Aniso20} &\times  \frac{\mathcal{E}_{\textbf{q}}}{[(q_0+\rmi \mu)^2+\mathcal{E}_{\textbf{q}}^2][(q_0+p_0+\rmi \mu)^2+\mathcal{E}_{\textbf{q}}^2]}.
\end{align}
We indeed observe that the same kernel function as in the upper and lower band contributions arises.

We now show that if $f$ is some well-behaved function (in particular such that the following integral is finite) then
\begin{align}
  \label{Aniso21} \int_{\textbf{q}} f(\mathcal{E}_{\textbf{q}}) \frac{1}{p^2}\Bigl(\frac{D_3}{2\mathcal{E}_{\textbf{q}}}-\frac{D_1^2}{8\mathcal{E}_{\textbf{q}}^3}\Bigr) = \int_{\textbf{q}} f(\mathcal{E}_{\textbf{q}}) K(\hat{\textbf{q}})
\end{align}
with a certain kernel $K$ that only depends on $\hat{\textbf{q}}=\textbf{q}/q$. Focussing first on the limit $\delta\to 0$, note that due to \cite{PhysRevB.92.045117,PhysRevB.93.205138}
\begin{align}
 \label{Aniso22} \Lambda^a_{ij}\Lambda^a_{kl} = \delta_{ik}\delta_{jl}+\delta_{il}\delta_{jk}-\frac{2}{3}\delta_{ij}\delta_{lk}
\end{align}
we have
\begin{align}
 \label{Aniso23} D_1 &\to 4 q^2(\textbf{q}\cdot\textbf{p}),\\
 \label{Aniso24} D_3 &\to  3q^2p^2+(\textbf{q}\cdot\textbf{p})^2
\end{align}
for $\delta\to 0$. Hence
\begin{align}
 \nonumber \int_{\textbf{q}} f(\mathcal{E}_{\textbf{q}}) \frac{1}{p^2}\Bigl(\frac{D_3}{2\mathcal{E}_{\textbf{q}}}-\frac{D_1^2}{8\mathcal{E}_{\textbf{q}}^3}\Bigr) & \to \int_{\textbf{q}} f(q^2)\frac{3}{2}\Bigl(1-\frac{(\textbf{q}\cdot\textbf{p})^2}{q^2p^2}\Bigr) \\
 \label{Aniso25} &= \int_{\textbf{q}} f(q^2),
\end{align}
and so 
\begin{align}
 \label{Aniso26} K(\hat{\textbf{q}})\to 1.
\end{align}
To obtain an explicit expression for $K(\hat{\textbf{q}})$ for nonzero $\delta$ we generalize Eqs. (A58) and (A71) from Ref. \cite{PhysRevB.95.075149} to arrive at
\begin{align}
 \label{Aniso27} \int_{\textbf{q}} f(\mathcal{E}_{\textbf{q}}) D_3 = \frac{2}{3} p^2 \Bigl[2(1-\delta)^2+3(1+\delta)^2\Bigr]\int_{\textbf{q}} f(\mathcal{E}_{\textbf{q}})q^2
\end{align}
and
\begin{align}
 \nonumber \int_{\vec{q}} f(\mathcal{E}_{\textbf{q}}) D_1^2 ={}& \frac{16}{3}p^2 \int_{\textbf{q}}f(\mathcal{E}_{\textbf{q}})\Bigl[ (1-\delta)^4q^2(d_1^2+d_2^2)\\
 \nonumber &+[(1+\delta)^4-4\delta^2]q^2 (d_3^2+d_4^2+d_5^2)\\
 \label{Aniso28} &+\frac{36}{\sqrt{3}}\delta^2 d_3d_4d_5\Bigr].
\end{align}
We conclude that $K(\hat{\textbf{q}})$ in Eq. (\ref{Aniso21}) is given by
\begin{align}
 \nonumber K(\hat{\textbf{q}}) ={}& \frac{q^2}{3\mathcal{E}_{\textbf{q}}} \Bigl[2(1-\delta)^2+3(1+\delta)^2\Bigr] \\
 \nonumber &-\frac{2}{3\mathcal{E}_{\textbf{q}}^3}\Bigl[ (1-\delta)^4q^2(d_1^2+d_2^2)\\
 \nonumber &+[(1+\delta)^4-4\delta^2]q^2 (d_3^2+d_4^2+d_5^2)\\
 \label{Aniso29} &+\frac{36}{\sqrt{3}}\delta^2 d_3d_4d_5\Bigr].
\end{align}
Using $(1-\delta)^2(d_1^2+d_2^2)=\mathcal{E}_{\textbf{q}}^2-(1+\delta)^2(d_3^2+d_4^2+d_5^2)$ we can write
\begin{align}
 \nonumber K(\hat{\textbf{q}}) ={}& (1+\delta)^2\frac{q^2}{\mathcal{E}_{\textbf{q}}} - \frac{8}{3}\delta(1+\delta+\delta^2) \frac{q^2(d_3^2+d_4^2+d_5^2)}{\mathcal{E}_{\textbf{q}}^3}\\
 \label{Aniso30} &- \frac{24}{\sqrt{3}} \delta^2 \frac{d_3d_4d_5}{\mathcal{E}_{\textbf{q}}^3}.
\end{align}
This expression makes the isotropic limit $\delta\to 0$ particularly transparent.

Equipped with the kernel $K(\hat{\textbf{q}})$, we can employ the Matsubara sum formula in Eq. (\ref{Napp19}) [with $q^2\to \mathcal{E}_{\textbf{q}}$] to write the separate contributions to the response function as
\begin{align}
 \label{Aniso31} Z_{\rm intra}(p_0) &= \frac{4\bar{e}^2}{p_0^2}\int_{\textbf{q}}K(\hat{\textbf{q}}) \Bigl[ n_{\rm F}(\mathcal{E}_{\textbf{q}}-\mu)+n_{\rm F}(\mathcal{E}_{\textbf{q}}+\mu)\Bigr],\\
 \label{Aniso32} Z_{\rm QBT}(p_0) &= \bar{e}^2\int_{\textbf{q}} \frac{4K(\hat{\textbf{q}})}{p_0^2+4\mathcal{E}_{\textbf{q}}^2}\Bigl[1-n_{\rm F}(\mathcal{E}_{\textbf{q}}-\mu)-n_{\rm F}(\mathcal{E}_{\textbf{q}}+\mu)\Bigr].
\end{align}
These expressions can be simplified even further. For this write $\mathcal{E}_{\textbf{q}} = q^2 \hat{\mathcal{E}}(\hat{\textbf{q}})$ and note that after a change of variables $q\to q'=q \hat{\mathcal{E}}^{1/2}$ we have
\begin{align}
 \label{Aniso33} \int_{\textbf{q}} f(\mathcal{E}_{\textbf{q}}) K(\hat{\textbf{q}}) &= \int_{\textbf{q}} f(q^2) \frac{K(\hat{\textbf{q}})}{\hat{\mathcal{E}}(\hat{\textbf{q}})^{3/2}} = \bar{\lambda}(\delta) \int_{\textbf{q}} f(q^2)
\end{align} 
with
\begin{align}
 \label{Aniso34} \bar{\lambda}(\delta) = \frac{1}{4\pi} \int_{\Omega} \frac{K(\hat{\textbf{q}})}{\hat{\mathcal{E}}(\hat{\textbf{q}})^{3/2}}.
\end{align}
Here $\int_\Omega$ denotes the angular integral over the unit sphere. Equation (\ref{Aniso33}) implies that the anisotropy-dependent kernel can be implemented by simultaneously replacing $K(\hat{\textbf{q}})\to \bar{\lambda}(\delta)$ and $\mathcal{E}_{\textbf{q}}\to q^2$ in the expressions for $Z_{\rm intra}(p_0)$ and $Z_{\rm QBT}(p_0)$. However, since replacing $K(\hat{\textbf{q}})\to 1$ and $\mathcal{E}_{\textbf{q}}\to q^2$ yields the isotropic result, we uncover that the anisotropic expressions \emph{factorize} into the isotropic formulas times the factor $\bar{\lambda}(\delta)$. Since the factor is common to both the intraband and QBT contributions, we further deduce a factorization of the total response function according to
\begin{align}
 \label{Aniso35} Z(\delta,p_0) = \bar{\lambda}(\delta)\cdot Z(\delta=0,p_0),
\end{align}
valid for every value of $T$ and $\mu$.

In order to evaluate the function $\bar{\lambda}(\delta)$ further we follow the route of Ref. \cite{PhysRevB.95.075149} and isolate the singular behavior of the function from the regular part. For this purpose we write
\begin{align}
 \label{Aniso36} \bar{\lambda}(\delta) = \frac{\lambda(\delta)}{\sqrt{1-\delta^2}}
\end{align}
with a regular function $\lambda(\delta)$ that is finite for all $|\delta|\leq 1$. We empirically determined the singular behavior from a power law fit of the numerical evaluation of $\bar{\lambda}(\delta)$ for $\delta\to \pm 1$. We use the usual spherical coordinates $\hat{\textbf{q}}=(\cos\phi\sin\theta,\sin\phi\sin\theta,\cos\theta)^{\rm T}$ and write
\begin{align}
 \nonumber \lambda(\delta) &= \frac{\sqrt{1-\delta^2}}{4\pi}\int_0^{2\pi}\mbox{d}\phi \int_0^\pi\mbox{d}\theta\ \sin\theta \frac{K(\hat{\textbf{q}})}{\hat{\mathcal{E}}(\hat{\textbf{q}})^{3/2}}\\
 \nonumber &= \frac{\sqrt{1-\delta^2}}{\pi}\int_0^{\pi/2}\mbox{d}\phi \int_0^\pi\mbox{d}\theta\ \sin\theta\Biggl[(1+\delta)^2\frac{1}{\hat{\mathcal{E}}(\hat{\textbf{q}})^{5/2}} \\
 \nonumber &-8\delta(1+\delta+\delta^2) \frac{\sin^2\theta(\cos^2\theta+\cos^2\phi\sin^2\phi\sin^2\theta)}{\hat{\mathcal{E}}(\hat{\textbf{q}})^{9/2}}\\
 \label{Aniso37} &- 72\delta^2 \frac{\cos^2\theta\sin^4\theta\cos^2\phi\sin^2\phi}{\hat{\mathcal{E}}(\hat{\textbf{q}})^{9/2}}\Biggr]
\end{align}
with
\begin{align}
\label{Aniso38} \hat{\mathcal{E}}(\hat{\textbf{q}}) = \sqrt{(1-\delta)^2+12\delta\sin^2\theta(\cos^2\theta+\cos^2\phi\sin^2\phi\sin^2\theta)}.
\end{align}
The integrand can now be expanded in powers of $\delta$, followed by a subsequent integration over the remaining angular variables. This yields the Taylor series
\begin{align}
 \label{Aniso39} \lambda(\delta) = 1 -\frac{1}{10}\delta +\frac{229}{280}\delta^2-\frac{1301}{6160}\delta^3-\frac{3413}{49280}\delta^4+\mathcal{O}(\delta^5),
\end{align}
Although the convergence properties of this series are rather bad, we find the quadratic order to capture the function within 10\% accuracy.

\subsection{Gauge invariance and longitudinal QBT contribution}\label{SecCont}

In the previous two sections we have derived the homogeneous electromagnetic response from the density response function $\chi(P)$. Equivalently, we can derive it from the current response function $K_{ij}(P)$. The fact that both approaches lead to the same answer is consistent with gauge invariance, which implies a simple relation between $\chi$ and $K_{\rm L}$. It is, however, nontrivial that gauge invariance is preserved within our approximations, and we will specify in detail how the individual upper band, lower band, and QBT terms contribute to a consistent picture in the normal phase.

For the calculations of this section it is convenient to introduce the following notation: If $f(P)\equiv f(p_0,p)$ is an imaginary time response function, then we define
\begin{align}
 \label{Eqf0} f(0) :=\lim_{p\to 0} f(0,p).
\end{align}
The notation is chosen to indicate that if, as in our case, $f(p_0,p)$ is given by a one-loop diagram, the limit $\lim_{p\to 0}f(0,p)$ can typically be computed by setting $p=p_0=0$ \emph{before} performing the Matsubara integration of the loop. Although this does not yield a general rule, it is true for all situations considered in this work. We define the longitudinal and transverse  components of $K_{ij}(P)$ through
\begin{align}
 K_{\rm L}(p_0,p) &= \frac{p_ip_j}{p^2} K_{ij}(P),\\
 K_{\rm T}(p_0,p) &= \frac{1}{2}\Bigl(\delta_{ij}-\frac{p_ip_j}{p^2}\Bigr) K_{ij}(P).
\end{align}

In order to quantity to which extent gauge invariance is preserved, we investigate whether the response functions in our approximation satisfy the ``continuity equation''
\begin{align}
 \label{Cont1} p_0^2 \chi(P) - p_ip_j K_{ij}(P) = 0.
\end{align}
By dividing the current response into diamagnetic and paramagnetic contributions according to
\begin{align}
 \label{Cont2} K_{ij}(P) = K_{ij}^{(\rm d)}(P) + K_{ij}^{(\rm p)}(P),
\end{align}
the relation reads
\begin{align}
 \label{Cont3} p_0^2 \chi(p_0,p) = p^2 \Bigl[K_{\rm L}^{(\rm d)}(p_0,p) + K_{\rm L}^{(\rm p)}(p_0,p)\Bigr].
\end{align}
Only the longitudinal part of $K_{ij}(P)$ enters the equation. The diamagnetic contribution is typically constant in momentum space, $K_{\rm L}^{(\rm d)}(P)=  K_{\rm L}^{(\rm d)}(0)$. If, in addition, the total response function for $P=0$ satisfies
\begin{align}
 \label{Cont4} K_{\rm L}(0) = K_{\rm L}^{(\rm d)}(0) + K_{\rm L}^{(\rm p)}(0) \stackrel{!}{=} 0,
\end{align}
then Eq. (\ref{Cont1}) is equivalent to the ``modified continuity equation''
\begin{align}
 \label{Cont5} p_0^2 \chi(p_0,p)  = p^2\Bigl[ K_{\rm L}^{(\rm p)}(p_0,p) - K_{\rm L}^{(\rm p)}(0)\Bigr].
\end{align}
In our case it turns out that the modified continuity equation is the somewhat weaker condition and it is satisfied for all components in the normal state.

It is easy to see why the continuity equation should be valid for the true response functions: For instance, applying the continuity equation $0=\partial_\mu j^\mu = \partial_t \rho +\partial_i j_i = \rmi \partial_\tau \rho+\partial_ij_i$ inside the averages of $\chi(X,X') = \langle\rho(X)\rho(X')\rangle$ and $K_{ij}(X,X')=-\langle j_i(X)j_j(X')\rangle$, we obtain
\begin{align}
 \label{Cont6} \partial_\tau \partial_{\tau'} \chi(X,X') - \partial_i \partial_j' K_{ij}(X,X')=0.
\end{align}
Since the response functions only depend on the difference $X-X'$, Fourier transforming in the first variable then yields Eq. (\ref{Cont1}). The equivalent derivation starting from requiring gauge invariance can be found in many textbooks, for instance Ref. \cite{BookAltland}.

The validity of the continuity equation (\ref{Cont3}) has some profound consequences for the electromagnetic response of the system. For one, for $p=0$ we obtain $\chi(p_0,0)=0$ for every $p_0\neq 0$. This equation reflects the conservation of particle number within our approximation.  On the other hand, for $p_0=0$  we observe $K_{\rm L}(0,p)=0$ for every $p\neq0$. Hence gauge invariance implies purely transverse  response to a static electromagnetic field. Further, by expanding the modified continuity equation according to $\chi(p_0,p)\simeq p^2 Z(p_0)$ for small $p$ we deduce
\begin{align}
 \label{Cont7}  Z(p_0) = \frac{1}{p_0^2}  \Bigl[ K_{\rm L}^{(\rm p)}(p_0,0) - K_{\rm L}^{(\rm p)}(0)\Bigr].
\end{align}
Consequently, the function $Z(p_0)$, and therefore the homogeneous response, can be obtained from either $\chi(P)$ or $K_{\rm L}^{(\rm p)}(P)$.

We begin our analysis with the upper band with propagator $G(Q)$ from Eq. (\ref{Prop19}). The density response $\chi_{\rm upper}(P)$ is given by Eq. (\ref{Napp5}) and the diamagnetic contribution derived from Eq. (\ref{Lin13b}) is $P$-independent and given by
\begin{align}
 \nonumber K_{ij}^{(\rm d,upper)}(P) &=  - 4 \bar{e}^2 \delta_{ij} \int_Q \frac{e^{-\rmi q_0\eta}}{\rmi q_0 + q^2-\mu}\\
 \label{Cont11} &= 4 \bar{e}^2\delta_{ij} \int_{\textbf{q}} n_{\rm F}(q^2-\mu).
\end{align} 
Here we identify $2\int_{\textbf{q}} n_{\rm F}(q^2-\mu)$ as the density of electrons in the upper band within our approximation. The paramagnetic contribution obtained from Eq. (\ref{Lin17}) reads
\begin{align}
 \label{Cont12} K_{ij}^{(\rm p,upper)} (P) &=2 \bar{e}^2 \int_Q\frac{(2q_i+p_i)(2q_j+p_j)}{(\rmi q_0+\vare_{\textbf{q}})[\rmi(q_0+p_0)+\vare_{\textbf{q}+\textbf{p}}]}.
\end{align}
It is now straightforward to verify the validity of the continuity equation for the upper band contribution: We have
\begin{align}
 \nonumber &p_0^2 \chi_{\rm upper}(P) - p_i p_j K_{ij}^{(\rm p,upper)}(P) \\
 \nonumber &= - 2 \bar{e}^2 \int_Q\frac{p_0^2 +(2\textbf{q}\cdot\textbf{p}+p^2)^2}{(\rmi q_0+\vare_{\textbf{q}})[\rmi(q_0+p_0)+\vare_{\textbf{q}+\textbf{p}}]}\\
 \nonumber &= -2 \bar{e}^2 \int_{\textbf{q}} \frac{ p_0^2 + (\vare_{\textbf{q}+\textbf{p}}-\vare_q)^2}{\rmi p_0 + \vare_{\textbf{q}+\textbf{p}}-\vare_q}\Bigl( n_{\rm F}(\vare_{\textbf{q}+\textbf{p}})-n_{\rm F}(\vare_q)\Bigr)\\
 \nonumber &= -2 \bar{e}^2\int_{\textbf{q}} (-\rmi p_0 + \vare_{\textbf{q}+\textbf{p}}-\vare_q)\Bigl( n_{\rm F}(\vare_{\textbf{q}+\textbf{p}})-n_{\rm F}(\vare_q)\Bigr)\\
 \nonumber &= 4 \bar{e}^2 \int_{\textbf{q}} (\vare_{\textbf{q}+\textbf{p}}-\vare_q)n_{\rm F}(\vare_q)= 4 e^2 \int_{\textbf{q}}(2\textbf{q}\cdot\textbf{p}+p^2) n_{\rm F}(\vare_q)\\
 \label{Cont13} &= 4 \bar{e}^2 p^2 \int_{\textbf{q}}n_{\rm F}(\vare_q)=  p_i p_j K_{ij}^{(\rm d,upper)}(P).
\end{align}
Furthermore, due to
\begin{align}
 \nonumber K_{\rm L}^{(\rm p,upper)}(0) &= \lim_{p\to 0}\int_Q  \frac{2 \bar{e}^2 (2\textbf{q}\cdot\textbf{p})^2}{p^2(\rmi q_0+\vare_q)^2}= \frac{8}{3}   \int_Q \frac{\bar{e}^2q^2}{(\rmi q_0+\vare_q)^2}\\
 \label{Cont14} &= - 4 \bar{e}^2 \int_{\textbf{q}} n_{\rm F}(q^2-\mu) = - K_{\rm L}^{(\rm d,upper)}(P),
\end{align}
where we used 
\begin{align}
T \sum_n \frac{1}{(\rmi q_0+\vare_q)^2}=\frac{\partial}{\partial \vare_q}n_{F}(\vare_q)
\end{align}
and a partial integration, we have
\begin{align}
 \label{Cont15} K_{\rm L}^{(\rm upper)}(0)=0,
\end{align}
and consequently also the modified continuity equation is valid. With regard to Eq. (\ref{Cont7}) we note that for $p_0 \neq 0$ we have
\begin{align}
 \label{Cont16} K_{ij}^{(\rm p,upper)} (p_0,0) &= \int_Q  \frac{\frac{8}{3} \bar{e}^2 \delta_{ij}q^2}{(\rmi q_0+\vare_q)[\rmi(q_0+p_0)+\vare_q]}=0,
\end{align}
and so
\begin{align}
 \label{Cont17} Z_{\rm upper}(p_0) =  \frac{K_{\rm L}^{(\rm d,upper)}(P)}{p_0^2} = \frac{4\bar{e}^2}{p_0^2}\int_{\textbf{q}} n_{\rm F}(q^2-\mu)
\end{align}
is only determined by the diamagnetic contribution. This result agrees with Eq. (\ref{Napp12}) found above from expanding $\chi_{\rm upper}(P)$.

Next we consider the lower band contribution obtained by replacing $q^2\to -q^2$ in the propagator (\ref{Prop19}). We employ $\chi_{\rm lower}(P)$ from Eq. (\ref{Napp5}). The diamagnetic contribution results from Eq. (\ref{Lin13b}) with $g_{ij}=-\delta_{ij}\mathbb{1}_2$ so that
\begin{align}
 \nonumber K_{ij}^{(\rm d,lower)}(P) &=   4 \bar{e}^2 \delta_{ij} \int_Q \frac{e^{-\rmi q_0\eta}}{\rmi q_0 - q^2-\mu}\\
 \label{Cont18} &= -4 \bar{e}^2\delta_{ij} \int_{\textbf{q}} n_{\rm F}(-q^2-\mu).
\end{align} 
Clearly, since $n_{\rm F}(-q^2-\mu)\to 1$ for $q\to \infty$, this expression is divergent. This divergence reflects the fact that the lower band is populated by an infinite number of electrons. We may, however, express the diamagnetic contribution as
\begin{align}
 \label{Cont19} K_{ij}^{(\rm d,lower)}(P) &=  4 \bar{e}^2\delta_{ij} \int_{\textbf{q}} \Bigl[n_{\rm F}(q^2+\mu)-1\Bigr],
\end{align} 
where the number of electrons is replaced by the number of holes subtracted for the number of single particle states in vacuum. In both Eqs. (\ref{Cont18}) and (\ref{Cont19}) we need to regularize the momentum integration with some ultraviolet cutoff $\kappa$ such that
\begin{align}
 \label{Cont20} \int_{\textbf{q}} 1 = \frac{1}{2\pi^2} \int_0^\kappa \mbox{d}q\ q^2 = \frac{\kappa^3}{6\pi^2}.
\end{align}
We will henceforth always assume this cutoff to be present if needed, see the discussion below Eq. (\ref{lutt9}).

The paramagnetic contribution to the lower band is given by
\begin{align}
 \label{Cont21} K_{ij}^{(\rm p,lower)} (P) &=2 \bar{e}^2 \int_Q\frac{(2q_i+p_i)(2q_j+p_j)}{(\rmi q_0+f_{\textbf{q}})[\rmi(q_0+p_0)+f_{\textbf{q}+\textbf{p}}]}
\end{align}
with $f_{\textbf{q}}=-q^2-\mu$. At this point we could repeat the calculations for the upper band with the formal replacement $\vare_{\textbf{q}}\to f_{\textbf{q}}$. Although a valid approach, it requires some care when handling superficial divergences that appear during the manipulations: For instance, in the third line of Eq. (\ref{Cont13}) we need to employ $n_{\rm F}(f_{\textbf{q}+\textbf{p}})-n_{\rm F}(f_\textbf{q})= n_{\rm F}(-f_{\textbf{q}})-n_{\rm F}(-f_{\textbf{q}+\textbf{p}})$ before performing the next step to ensure convergence of the integral. A much more direct approach is to note that $\chi_{\rm lower}(P)$ and $K_{ij}^{(\rm p,lower)}(P)$ result from $\chi_{\rm upper}(P)$ and $K_{ij}^{(\rm p,upper)}(P)$ upon simultaneously changing the sign of $q_0,p_0,\mu$ under the integral. In particular, since the upper band contributions are all manifestly finite, no divergences can appear from the expressions $\chi_{\rm lower}(P)$ and $K_{ij}^{(\rm p,lower)}(P)$. We have verified that both approaches for computing the contributions from the lower band yield the same results.

For vanishing external momentum and frequency, the lower band contribution to the paramagnetic current reads
\begin{align}
 \label{Cont22} K_{\rm L}^{(\rm p,lower)}(0) &=  - 4 \bar{e}^2  \int_{\textbf{q}} n_{\rm F}(q^2+\mu).
\end{align}
Comparing to the diamagnetic contribution in Eq. (\ref{Cont19}) we conclude that
\begin{align}
 \label{Cont23} K_{\rm L}^{(\rm lower)}(0) = -4 \bar{e}^2 \int_{\textbf{q}} 1,
\end{align}
and so Eq. (\ref{Cont4}) is violated for the lower band due to a constant divergent term that results from the infinite number of electrons in the lower band. On the other hand,
\begin{align}
 \nonumber &p_0^2 \chi_{\rm lower}(P) - p^2 K_{\rm L}^{(\rm p,lower)}(P) \\
 \label{Cont24} &= 4 \bar{e}^2 p^2 \int_{\textbf{q}}n_{\rm F}(q^2+\mu) = -p^2K_{\rm L}^{(\rm p,lower)}(0),
\end{align}
so that the lower band contribution satisfies the modified continuity equation. Obviously, due to $K_{\rm L}^{(\rm lower)}(0)\neq 0$, it cannot satisfy the continuity equation at the same time. Since the optical response can be deduced from Eq. (\ref{Cont24}), we see that the divergence of the diamagnetic term has no observable consequences. We note here that $K_{ij}^{(\rm p,lower)}(p_0,0)=0$ for $p_0\neq0$ as in Eq. (\ref{Cont16}), and thus
\begin{align}
 \label{Cont25} Z_{\rm lower}(p_0) =  \frac{K_{\rm L}^{(\rm d,lower)}(P)}{p_0^2} = \frac{4\bar{e}^2}{p_0^2}\int_{\textbf{q}} n_{\rm F}(q^2+\mu),
\end{align}
which confirms Eq. (\ref{Napp13}).

In order to compute the longitudinal QBT contributions we employ the propagator from Eq. (\ref{Prop20}) and $g_{ij}=\frac{\sqrt{3}}{2}\Lambda^a_{ij}\gamma_a$ in Eqs. (\ref{Lin13b}) and (\ref{Lin18}). The total diamagnetic contribution then vanishes due to
\begin{align}
 \nonumber K_{ij}^{(\rm d)}(P) &= - \sqrt{3} \bar{e}^2 \Lambda^a_{ij}\ \mbox{tr}\int_Q e^{-\rmi q_0 \eta} \gamma_a G(Q)\\
 \label{Cont26} &= - 4\sqrt{3} \bar{e}^2 \Lambda^a_{ij}\ \mbox{tr}\int_Q \frac{d_a(\textbf{q})}{(q_0+\rmi \mu)^2+q^4}=0.
\end{align}
Consequently, the QBT contribution is given by
\begin{align}
 \label{Cont27}  K_{{\rm L}}^{(\rm d,QBT)}(P) &= 4 \bar{e}^2 \int_{\textbf{q}}\Bigl[1-n_{\rm F}(q^2-\mu)-n_{\rm F}(q^2+\mu)\Bigr].
\end{align}
To determine the paramagnetic contribution according to Eq. (\ref{Lin18}) observe that $H_{\textbf{q}}=g_{ij}q_iq_j$ implies  $H_{\textbf{q}+\textbf{p}}-H_{\textbf{q}} = g_{ij}p_i(2q_j+p_j)$. Consequently,
\begin{align}
 \nonumber &p_ip_jK_{ij}^{(\rm p)}(P) \\
 \nonumber &=   \bar{e}^2 p_i p_j   \int_{Q}(2q_k+p_k)(2q_l+p_l) \ \mbox{tr}\  G^{Q+P}g_{jl}G^Q g_{ik}\\
 \nonumber &= \bar{e}^2\int_Q \mbox{tr}\ G^{Q+P}(H_{\textbf{q}+\textbf{p}}-H_{\textbf{q}})G^Q (H_{\textbf{q}+\textbf{p}}-H_{\textbf{q}})\\
  \label{ContNew1} &= \bar{e}^2\int_Q \mbox{tr}\  G^{Q+P}G^Q \Bigl( (H_{\textbf{q}+\textbf{p}} -H_{\textbf{q}})^2 -[H_{\textbf{q}},H_{\textbf{q}+\textbf{p}}]\Bigr).
\end{align}
Here $[\ ,\ ]$ denotes the commutator and we use that $G^Q$ commutes with $H_{\textbf{q}}$. We have
\begin{align}
 \label{ContNew2} (H_{\textbf{q}+\textbf{p}}-H_{\textbf{q}})^2 = \Bigl[ (\textbf{q}+\textbf{p})^4+q^4-2 d_a(\textbf{q}+\textbf{p})d_a(\textbf{q})\Bigr]\mathbb{1}_4,
\end{align}
and 
\begin{widetext}
\begin{align}
  \label{ContNew3}\mbox{tr}\  G^{Q+P}G^Q &=  -4 \frac{(q_0+\rmi \mu)(q_0+p_0+\rmi \mu)- d_a(\textbf{q}+\textbf{p})d_a(\textbf{q})}{[(q_0+\rmi \mu)^2+q^4][(q_0+p_0+\rmi \mu)^2+(\textbf{q}+\textbf{p})^4]},\\
 \label{ContNew4}  \mbox{tr}\ G^{Q+P}G^Q [H_{\textbf{q}},H_{\textbf{q}+\textbf{p}}] &= \frac{8\Bigl( (\textbf{q}+\textbf{p})^2q^2 + d_a(\textbf{q}+\textbf{p})d_a(\textbf{q})\Bigr)\Bigl( (\textbf{q}+\textbf{p})^2q^2 - d_a(\textbf{q}+\textbf{p})d_a(\textbf{q})\Bigr)}{[(q_0+\rmi \mu)^2+q^4][(q_0+p_0+\rmi \mu)^2+(\textbf{q}+\textbf{p})^4]}.
\end{align}
The intraband contributions are given by
\begin{align}
 \label{ContNew5} p_i p_j \Bigl[K_{ij}^{(\rm p,upper)}(P)+K_{ij}^{(\rm p,lower)}(P)\Bigr]  = -4 \bar{e}^2 \int_Q (2\textbf{q}\cdot\textbf{p}+p^2)^2 \frac{(q_0+\rmi \mu)(q_0+p_0+\rmi \mu)-q^2(\textbf{q}+\textbf{p})^2}{[(q_0+\rmi \mu)^2+q^4][(q_0+p_0+\rmi \mu)^2+(\textbf{q}+\textbf{p})^4]},
\end{align}
and so we eventually arrive at
\begin{align}
 \label{ContNew6}  K_{\rm L}^{(\rm p,QBT)'}(p_0,p)  = -6\bar{e}^2 \int_Q \frac{1}{p^2}[q^2p^2-(\textbf{q}\cdot\textbf{p})^2] \frac{2(q_0+\rmi \mu)(q_0+p_0+\rmi \mu) + (\textbf{q}+\textbf{p})^4+q^4}{[(q_0+\rmi \mu)^2+q^4][(q_0+p_0+\rmi \mu)^2+(\textbf{q}+\textbf{p})^4]}.
\end{align}
This completes the derivation of the longitudinal QBT contribution. The expression obtained requires regularization for $p\neq 0$ as we indicate by a prime. However, we postpone the discussion of this divergence and how to cure it to the end of this section, as it does not critically affect the following manipulations.

The formula (\ref{ContNew6}) can be used to (re)derive the homogeneous contribution for $p=0$. We have
\begin{align}
 \label{ContNew7} K_{\rm L}^{(\rm p,QBT)}(p_0,0)&= -8\bar{e}^2 \int_Q q^2\frac{(q_0+\rmi \mu)(q_0+p_0+\rmi \mu) + q^4}{[(q_0+\rmi \mu)^2+q^4][(q_0+p_0+\rmi \mu)^2+q^4]}
\end{align}
We perform the Matsubara summation according to
\begin{align}
 \label{ContNew8} T \sum_n \frac{(q_0+\rmi \mu)(q_0+p_0+\rmi \mu) +q^4}{[(q_0+\rmi \mu)^2+q^4][(q_0+p_0+\rmi \mu)^2+q^4]}&=\frac{2q^2}{p_0^2+4q^4}\Bigl[1-n_{\rm F}(q^2-\mu)-n_{\rm F}(q^2+\mu)\Bigr],
\end{align}
valid for bosonic $p_0=2\pi m T$ with $m\in \mathbb{Z}$ including $p_0=0$, and arrive at
\begin{align}
 \label{ContNew9} K_{\rm L}^{(\rm p,QBT)}(p_0,0)&=-16 \bar{e}^2 \int_{\textbf{q}}  \frac{q^4}{p_0^2+4q^4} \Bigl[1-n_{\rm F}(q^2-\mu)-n_{\rm F}(q^2+\mu)\Bigr].
\end{align}
Furthermore, setting $p_0=0$ we deduce
\begin{align}
 \label{ContNew10} K_{\rm L}^{(\rm p,QBT)}(0) &= -4 \bar{e}^2  \int_{\textbf{q}}  \Bigl[1-n_{\rm F}(q^2-\mu)-n_{\rm F}(q^2+\mu)\Bigr] = - K_{\rm L}^{(\rm d,QBT)}(0),
\end{align}
and so the QBT contribution satisfies 
\begin{align}
K_{\rm L}^{(\rm QBT)}(0) = 0.
\end{align}
Together with the vanishing of the total diamagnetic contribution this implies, as can also be verified explicitly from the original expression of $K_{ij}^{(\rm p)}(P)$ for $p_0=0$, that 
\begin{align}
 \label{ContNew11} K_{\rm L}(0) = K_{\rm L}^{(\rm p)}(0) = - 4 \bar{e}^2 \int_{\textbf{q}}1,
\end{align}
which again is understood with an ultraviolet momentum cutoff. Interestingly, the divergence of the lower band diamagnetic contribution manifests itself in the divergence of the total paramagnetic term for $P=0$. In contrast, the upper band and QBT contributions to all formulas are manifestly finite. We summarize the various contributions to the current response for $P=0$ in Table \ref{TabK0}. Also note that Eqs. (\ref{ContNew9}) and (\ref{ContNew10}) show that $Z_{\rm QBT}(p_0)$ from Eq. (\ref{Napp20}) can be written as
\begin{align}
 \label{ContNew12} Z_{\rm QBT}(p_0)&= \frac{1}{p_0^2}\Bigl[ K_{\rm L}^{(\rm p,QBT)}(p_0,0)- K_{\rm L}^{(\rm p,QBT)}(0)\Bigr],
\end{align}
which proves the modified continuity equation for the QBT contribution in the low-momentum limit.

\begin{table*}[t]
\centering
\begin{tabular}{|c||c|c|c|c|}
\hline  & total & upper band & lower band & QBT \\ 
\hline\hline $K_{\rm L}^{(\rm d)}(0)=\lim_{p\to 0}K_{\rm L}^{(\rm d)}(0,p)$ & 0 & $4\bar{e}^2\int_{\textbf{q}}^\kappa n_{\rm F}(q^2-\mu)$ & $4\bar{e}^2\int_{\textbf{q}}^\kappa\Bigl[n_{\rm F}(q^2+\mu) -1\Bigr]$ & $4\bar{e}^2\int_{\textbf{q}}^\kappa\Bigl[1- n_{\rm F}(q^2-\mu)-n_{\rm F}(q^2+\mu) \Bigr]$ \\ 
\hline $K_{\rm L}^{(\rm p)}(0)=\lim_{p\to 0}K_{\rm L}^{(\rm p)}(0,p)$ & $\ -4\bar{e}^2\int_{\textbf{q}}^\kappa1\ $ & $-4\bar{e}^2\int_{\textbf{q}}^\kappa n_{\rm F}(q^2-\mu)$ & $-4\bar{e}^2\int_{\textbf{q}}^\kappa n_{\rm F}(q^2+\mu)$ & $-4\bar{e}^2\int_{\textbf{q}}^\kappa\Bigl[1- n_{\rm F}(q^2-\mu)-n_{\rm F}(q^2+\mu) \Bigr]$ \\ 
\hline $K_{\rm L}(0)=\lim_{p\to 0}K_{\rm L}(0,p)$ & $-4\bar{e}^2\int_{\textbf{q}}^\kappa1$ & $0$ & $-4\bar{e}^2\int_{\textbf{q}}^\kappa1$ & $0$ \\ 
\hline modified  continuity equation & satisfied  & satisfied & satisfied & satisfied \\ 
\hline continuity equation & violated & satisfied & violated & satisfied \\ 
\hline 
\end{tabular} 
\caption{In this table we summarize whether the total, upper, lower, and QBT contributions satisfy the modified continuity equation (\ref{Cont5}) given by $p_0^2 \chi(p_0,p)=p^2[K_{\rm L}^{(\rm p)}(p_0,p)-K_{\rm L}^{(\rm p)}(0)]$, or the continuity equation (\ref{Cont3}) given by $p_0^2 \chi(p_0,p)=p^2[K_{\rm L}^{(\rm p)}(p_0,p)+K_{\rm L}^{(\rm d)}(0)]$. Momentum integrals are equipped with an ultraviolet cutoff $\kappa$ such that $q\leq \kappa$. We observe that each individual contribution satisfies the modified continuity equation. In particular, this allows us to determine the homogeneous response from either $\chi$ or $K_{\rm L}^{(\rm p)}$. The continuity equation for the individual components is then also satisfied if and only if there is a perfect cancellation of diamagnetic and paramagnetic terms in $K_{\rm L}(0)=0$ for that component.  We find the upper and QBT contributions to also satisfy the continuity equation. In contrast, for the lower band contribution the cancellation is imperfect due to an infinite constant that is independent of the thermodynamic parameters and thus physically irrelevant. It could be removed by a suitable renormalization of the current response function.}
\label{TabK0}
\end{table*}

To show that the QBT contribution satisfies the continuity equation for all $P$ we employ $\chi_{\rm QBT}(P)$ from Eq. (\ref{Napp8}) and obtain
\begin{align}
 \label{ContNew13} p_0^2 \chi_{\rm QBT}(P) - p^2 K_{\rm L}^{(\rm p,QBT)'}(P) &=  6\bar{e}^2 \int_Q [q^2p^2-(\textbf{q}\cdot\textbf{p})^2]\frac{p_0^2+2(q_0+\rmi \mu)(q_0+p_0+\rmi \mu) + (\textbf{q}+\textbf{p})^4+q^4}{[(q_0+\rmi \mu)^2+q^4][(q_0+p_0+\rmi \mu)^2+(\textbf{q}+\textbf{p})^4]}
\end{align}
The right hand side of this equation is independent of $p_0$, as can be seen from writing
\begin{align}
  \nonumber T\sum_n \frac{p_0^2+2(q_0+\rmi \mu)(q_0+p_0+\rmi \mu)+a^2+b^2}{[(q_0+\rmi \mu)^2+a^2][(q_0+p_0+\rmi \mu)^2+b^2]} &= T\sum_n \frac{1}{(q_0+\rmi \mu)^2+a^2} + T \sum_n \frac{1}{(q_0+p_0+\rmi \mu)^2+b^2}\\
  \label{ContNew14} &= \frac{1}{2a}\Bigl(1-n_{\rm F}(a-\mu)-n_{\rm F}(a+\mu)\Bigr) + \frac{1}{2b}\Bigl(1-n_{\rm F}(b-\mu)-n_{\rm F}(b+\mu)\Bigr),
\end{align}
which is valid for all real $a,b,\mu$ and bosonic $p_0$ including $p_0=0$. We then eventually arrive at
\begin{align}
 \nonumber p_0^2 \chi_{\rm QBT}(P) - p^2K_{\rm L}^{(\rm p,QBT)'}(P)&= 3\bar{e}^2\int_{\textbf{q}}  [q^2p^2-(\textbf{q}\cdot\textbf{p})^2] \Biggl(\frac{1}{q^2} \Bigl[1- n_{\rm F}(q^2-\mu)-n_{\rm F}(q^2+\mu)\Bigr]\\
 \label{Cont59}&+\frac{1}{(\textbf{q}+\textbf{p})^2}\Bigl[1-n_{\rm F}((\textbf{q}+\textbf{p})^2-\mu)-n_{\rm F}((\textbf{q}+\textbf{p})^2+\mu)\Bigr]\Biggr).
\end{align}
\end{widetext}
By a suitable shift of momentum, and ignoring the fact that the integrals are divergent for now, we see that the second integral is identical to the first one. This leaves us with
\begin{align}
 \nonumber &p_0^2 \chi_{\rm QBT}(P) - p^2K_{\rm L}^{(\rm p,QBT)}(P) \\
 \nonumber &\ = 6\bar{e}^2\int_{\textbf{q}}  \frac{q^2p^2-(\textbf{q}\cdot\textbf{p})^2}{q^2}\Bigl[1- n_{\rm F}(q^2-\mu)-n_{\rm F}(q^2+\mu)\Bigr]\\
 \nonumber &\ = 4 \bar{e}^2 p^2 \int_{\textbf{q}}  \Bigl[1- n_{\rm F}(q^2-\mu)-n_{\rm F}(q^2+\mu)\Bigr]\\
 \label{Cont60} &\ = p^2K_{\rm L}^{(\rm d,QBT)}(P),
\end{align}
where we inserted the diamagnetic QBT contribution from Eq. (\ref{Cont27}). This completes our proof of the validity of the continuity equation for the QBT contribution.

We close this section by discussing the regularity properties of the longitudinal QBT contribution $K_{\rm L}^{(\rm QBT)'}(p_0,p)$. We have shown that the correct expression for the homogeneous response in Eq. (\ref{ContNew12}) is recovered for $p=0$ without the occurrence of spurious divergences. For $p\neq 0$, on the other hand, the integrand in Eq. (\ref{ContNew6}) behaves like $\bar{e}^2[-4+\frac{2}{5}\frac{p^2}{q^2}+\mathcal{O}(q^{-4})]$ for large momenta. Restricting the momentum integral to a finite domain with an ultraviolet cutoff $\kappa$ such that $q\leq \kappa$, the first term $\sim 1$ is eventually cancelled by the diamagnetic term. The second term $\sim q^{-2}$, in principle, cannot be cancelled by the diamagnetic contribution, which is independent of $p$. On the other hand, this subleading divergence can be removed as in Eq. (\ref{Cont59}) by performing a formal shift of the internal momentum in the diamagnetic term. Since the diamagnetic term is not finite without ultraviolet cutoff $\kappa$, the momentum integral is not invariant under translations in $\textbf{q}$. In fact, the total integrand on the right hand side of Eq. (\ref{Cont59}) precisely behaves as $\bar{e}^2[4-\frac{2}{5}\frac{p^2}{q^2}+\mathcal{O}(q^{-4})]$, and so cancels all spurious divergences. To make the unphysical nature of the spurious divergence particularly visible, note that performing a shift of momentum $\textbf{q}\to \textbf{q}+t\textbf{p}$ with some arbitrary parameter $t$ in Eq. (\ref{ContNew6}), we find the integrand to behave like $\bar{e}^2[-4+\frac{2}{5}[1-2t(1-t)]\frac{p^2}{q^2}+\mathcal{O}(q^{-4})]$, and so the subleading divergence depends on the parameter $t$.

We thus see how the manipulation of ultraviolet divergent integrals causes a spuriously divergent expression for $K_{\rm L}^{(\rm QBT)'}(p_0,p)$. For practical purposes this is unimportant, as we can simply use $K_{\rm L}^{(\rm QBT)}(p_0,p)=\frac{p_0^2}{p^2}\chi_{\rm QBT}(p_0,p)$ to compute the longitudinal current response from the manifestly finite density response function. Still there is a precise way to remove the divergence. Since the divergent contributions arise from the large momentum part of the integral with $q^2\gg |\mu|,T,p_0$, they can be removed by equipping the formal momentum integral expression for $K_{\rm L}^{(\rm QBT)'}(p_0,p)$ with an ultraviolet cutoff $\kappa$ such that $q\leq \kappa$, and then subtracting the corresponding expression $K_{\rm L}^{(\rm QBT)'}(0,p)$ with $\mu=T=0$, i.e.
\begin{align}
 \label{ContNew15} K_{\rm L}^{(\rm QBT)}(p_0,p) = K_{\rm L}^{(\rm QBT)'}(p_0,p)-K_{\rm L}^{(\rm QBT)'}(0,p)_{\mu=T=0}.
\end{align}
This ensures that the longitudinal contribution to the current response is manifestly finite and satisfies $K_{\rm L}^{(\rm QBT)}(0,p)=0$ due to the validity of the continuity equation.

\subsection{Transverse QBT contribution}\label{SecTrapp}

In this section we compute the transverse  QBT contribution $K_{\rm T}^{(\rm QBT)}(p_0,p)$ in the isotropic and particle-hole symmetric limit. The diamagnetic T contribution coincides with the corresponding L contribution and reads
\begin{align}
 \label{trapp1} K_{\rm T}^{(\rm d,QBT)}(P) &=4\bar{e}^2\int_{\vec{q}}\Bigl[1-n_{\rm F}(q^2-\mu)-n_{\rm F}(q^2+\mu)\Bigr].
\end{align}
To determine the total contribution note that we have 
\begin{widetext}
\begin{align}
 \nonumber K_{\rm T}^{(\rm QBT)}(P) &= \frac{1}{2}\Bigl[\delta_{ij}K_{ij}^{(\rm p,QBT)}(P)-K_{\rm L}^{(\rm p,QBT)}(p_0,p)\Bigr]+K_{\rm T}^{(\rm d,QBT)}(P)\\
 \label{trapp2} &=\frac{1}{2}\Bigl[\delta_{ij}K_{ij}^{(\rm p,QBT)}(P)-\frac{p_0^2}{p^2}\chi_{\rm QBT}(P)\Bigr]+\frac{3}{2}K_{\rm T}^{(\rm d, QBT)}(P),
\end{align}
where we use that the QBT contribution satisfies the continuity equation. Hence it is sufficient to compute the trace $\delta_{ij}K_{ij}^{(\rm p,QBT)}(P)$. Starting from Eq. (\ref{Lin18}) we obtain
\begin{align}
 \nonumber  \delta_{ij}K_{ij}^{(\rm p)'}(P) &=3\bar{e}^2 \int_Q \frac{1}{[(q_0+\rmi \mu)^2+q^4][(q_0+p_0+\rmi \mu)^2+(\textbf{q}+\textbf{p})^4]}\Bigl( -\frac{10}{3}(2\textbf{q}+\textbf{p})^2(q_0+\rmi \mu)(q_0+p_0+\rmi \mu) \\
 \label{trapp3} &-2(2\textbf{q}+\textbf{p})^2d_a(\textbf{q}+\textbf{p})d_a(\textbf{q}) +\frac{2}{\sqrt{3}}J_{abc}d_a(\textbf{q}+\textbf{p})d_b(\textbf{q})d_c(2\textbf{q}+\textbf{p})\Bigr).
\end{align}
Like the longitudinal current response function, this formula requires regularization for $p\neq0$, see the discussion at the end of the previous section. We indicate the unregularized expression by a prime and continue by imposing a finite momentum cutoff $q\leq \kappa$. We readily verify that the last term in Eq. (\ref{trapp3}) can be written as
\begin{align}
 \label{trapp4} &\frac{2}{\sqrt{3}} J_{abc}d_a(\textbf{q}+\textbf{p})d_b(\textbf{q})d_c(2\textbf{q}+\textbf{p}) =(2\textbf{q}+\textbf{p})^2 \Bigl(\frac{4}{3}q^2(\textbf{q}+\textbf{p})^2-\frac{3}{2}[q^2p^2-(\textbf{q}\cdot\textbf{p})^2]\Bigr)-\frac{1}{2}p^2[q^2p^2-(\textbf{q}\cdot\textbf{p})^2].
\end{align}
We subtract the contributions from the upper and lower bands given by
\begin{align}
 \label{trapp4b} \delta_{ij}K_{ij}^{(\rm p,upper)}+\delta_{ij}K_{ij}^{(\rm p,lower)} (P) &= -4\bar{e}^2\int_Q \frac{(2\textbf{q}+\textbf{p})^2[(q_0+\rmi\mu)(q_0+p_0+\rmi \mu)-q^2(\textbf{q}+\textbf{p})^2]}{[(q_0+\rmi \mu)^2+q^4][(q_0+p_0+\rmi \mu)^2+(\textbf{q}+\textbf{p})^4]},
\end{align}
and conclude that the QBT contribution reads
\begin{align}
 \nonumber \delta_{ij}K_{ij}^{(\rm p,QBT)'}(P) &=-\frac{p^2}{4}\chi_{\rm QBT}(p_0,p) -6\bar{e}^2 \int_Q \frac{(2\textbf{q}+\textbf{p})^2(q_0+\rmi \mu)(q_0+p_0+\rmi \mu)}{[(q_0+\rmi \mu)^2+q^4][(q_0+p_0+\rmi \mu)^2+(\textbf{q}+\textbf{p})^4]}\\
 \label{trapp5} &+3\bar{e}^2 \int_Q \frac{(2\textbf{q}+\textbf{p})^2\Bigl(-2q^2(\textbf{q}+\textbf{p})^2+\frac{3}{2}[q^2p^2-(\textbf{q}\cdot\textbf{p})^2]\Bigr)}{[(q_0+\rmi \mu)^2+q^4][(q_0+p_0+\rmi \mu)^2+(\textbf{q}+\textbf{p})^4]}.
\end{align}
The density response $\chi_{\rm QBT}(p_0,p)$ is given by Eq. (\ref{Napp8}).

After we have derived a general expression for the QBT contribution to $\delta_{ij}K_{ij}^{(\rm p)}$ in terms of elementary functions, we next evaluate the Matsubara sums by means of the formulas
\begin{align}
 \nonumber T \sum_n \frac{1}{[(q_0+\rmi \mu)^2+a^2][(q_0+p_0+\rmi \mu)^2+b^2]}&=\frac{1}{4ab}\Biggl(\frac{n_{\rm F}(a+\mu)-n_{\rm F}(b+\mu)}{\rmi p_0+a-b}+\frac{n_{\rm F}(a-\mu)-n_{\rm F}(b-\mu)}{-\rmi p_0+a-b}\\
 \label{trapp6} &+\frac{1-n_{\rm F}(a+\mu)-n_{\rm F}(b-\mu)}{\rmi p_0+a+b}+\frac{1-n_{\rm F}(a-\mu)-n_{\rm F}(b+\mu)}{-\rmi p_0+a+b} \Biggr),\\
 \nonumber T \sum_n \frac{(q_0+\rmi \mu)(q_0+p_0+\rmi \mu)}{[(q_0+\rmi \mu)^2+a^2][(q_0+p_0+\rmi \mu)^2+b^2]}  &= -\frac{1}{4}\Biggl(\frac{n_{\rm F}(a+\mu)-n_{\rm F}(b+\mu)}{\rmi p_0+a-b}+\frac{n_{\rm F}(a-\mu)-n_{\rm F}(b-\mu)}{-\rmi p_0+a-b}\\
 \label{trapp7} &- \frac{1-n_{\rm F}(a+\mu)-n_{\rm F}(b-\mu)}{\rmi p_0+a+b}-\frac{1-n_{\rm F}(a-\mu)-n_{\rm F}(b+\mu)}{-\rmi p_0+a+b}\Biggr),
\end{align}
both valid for all signs of $a,b, \mu$, and bosonic frequency $p_0=2\pi m T$, $m\in\mathbb{Z}$, including $p_0=0$. Using the symmetry properties of the integrand it is further possible to remove the external momentum from the Fermi--Dirac functions, making the expression particularly suitable for further manipulations. We have
\begin{align}
 \nonumber \delta_{ij}K_{ij}^{(\rm p,QBT)'}(P) &=-\frac{p^2}{4}\chi_{\rm QBT}(p_0,p) +\frac{3}{2}\bar{e}^2 \int_{\textbf{q}} (2\textbf{q}+\textbf{p})^2\Biggl[\frac{n_{\rm F}(q^2+\mu)}{\rmi p_0+q^2-(\textbf{q}+\textbf{p})^2}\\
 \nonumber &+\frac{n_{\rm F}(q^2-\mu)}{-\rmi p_0+q^2-(\textbf{q}+\textbf{p})^2}-\frac{1-n_{\rm F}(q^2+\mu)}{\rmi p_0+q^2+(\textbf{q}+\textbf{p})^2}+\frac{n_{\rm F}(q^2-\mu)}{-\rmi p_0+q^2+(\textbf{q}+\textbf{p})^2} +\{p_0\to-p_0\}\Biggr]\\
 \nonumber &+3\bar{e}^2 \int_{\textbf{q}} \frac{(2\textbf{q}+\textbf{p})^2\Bigl(-2q^2(\textbf{q}+\textbf{p})^2+\frac{3}{2}[q^2p^2-(\textbf{q}\cdot\textbf{p})^2]\Bigr)}{4q^2(\textbf{q}+\textbf{p})^2} \Biggl[\frac{n_{\rm F}(q^2+\mu)}{\rmi p_0+q^2-(\textbf{q}+\textbf{p})^2}\\
 \label{trapp8} &+\frac{n_{\rm F}(q^2-\mu)}{-\rmi p_0+q^2-(\textbf{q}+\textbf{p})^2}+\frac{1-n_{\rm F}(q^2+\mu)}{\rmi p_0+q^2+(\textbf{q}+\textbf{p})^2}-\frac{n_{\rm F}(q^2-\mu)}{-\rmi p_0+q^2+(\textbf{q}+\textbf{p})^2} +\{p_0\to-p_0\}\Biggr]
\end{align}
By appropriately regrouping the individual terms we then verify that
\begin{align}
 \nonumber \delta_{ij}K_{ij}^{(\rm p,QBT)'}(P) &=-\frac{p^2}{4}\chi_{\rm QBT}(p_0,p) -6\bar{e}^2 \int_{\textbf{q}}\frac{(2\textbf{q}+\textbf{p})^2[q^2+(\textbf{q}+\textbf{p})^2]}{p_0^2+[q^2+(\textbf{q}+\textbf{p})^2]^2}\Bigl(1-n_{\rm F}(q^2-\mu)-n_{\rm F}(q^2+\mu)\Bigr)\\
 \nonumber &+\frac{9}{4}\bar{e}^2 \int_{\textbf{q}}\frac{(2\textbf{q}+\textbf{p})^2[q^2+(\textbf{q}+\textbf{p})^2]}{p_0^2+[q^2+(\textbf{q}+\textbf{p})^2]^2} \frac{[q^2p^2-(\textbf{q}\cdot\textbf{p})^2]}{q^2(\textbf{q}+\textbf{p})^2}  \Bigl(1-n_{\rm F}(q^2-\mu)-n_{\rm F}(q^2+\mu)\Bigr)\\
 \label{trapp9} &-\frac{9}{4}\bar{e}^2 \int_{\textbf{q}}\frac{(2\textbf{q}+\textbf{p})^2[(\textbf{q}+\textbf{p})^2-q^2]}{p_0^2+[(\textbf{q}+\textbf{p})^2-q^2]^2} \frac{[q^2p^2-(\textbf{q}\cdot\textbf{p})^2]}{q^2(\textbf{q}+\textbf{p})^2}\Bigl(n_{\rm F}(q^2-\mu)+n_{\rm F}(q^2+\mu)\Bigr).
\end{align}
We conclude that the full transverse  QBT contribution according to Eq. (\ref{trapp2}) is given by
\begin{align}
 \nonumber K_{\rm T}^{(\rm QBT)'}(p_0,p) &=-\frac{1}{2p^2}\Bigl(p_0^2+\frac{p^4}{4}\Bigr)\chi_{\rm QBT}(p_0,p) +6\bar{e}^2\int_{\textbf{q}}\Bigl(1-n_{\rm F}(q^2-\mu)-n_{\rm F}(q^2+\mu)\Bigr)\\
 \nonumber &-3\bar{e}^2 \int_{\textbf{q}}\frac{(2\textbf{q}+\textbf{p})^2[q^2+(\textbf{q}+\textbf{p})^2]}{p_0^2+[q^2+(\textbf{q}+\textbf{p})^2]^2}\Bigl(1-n_{\rm F}(q^2-\mu)-n_{\rm F}(q^2+\mu)\Bigr)\\
 \nonumber &+\frac{9}{8}\bar{e}^2 \int_{\textbf{q}}\frac{(2\textbf{q}+\textbf{p})^2[q^2+(\textbf{q}+\textbf{p})^2]}{p_0^2+[q^2+(\textbf{q}+\textbf{p})^2]^2} \frac{[q^2p^2-(\textbf{q}\cdot\textbf{p})^2]}{q^2(\textbf{q}+\textbf{p})^2}  \Bigl(1-n_{\rm F}(q^2-\mu)-n_{\rm F}(q^2+\mu)\Bigr)\\
 \label{trapp10a} &-\frac{9}{8}\bar{e}^2 \int_{\textbf{q}}\frac{(2\textbf{q}+\textbf{p})^2[(\textbf{q}+\textbf{p})^2-q^2]}{p_0^2+[(\textbf{q}+\textbf{p})^2-q^2]^2} \frac{[q^2p^2-(\textbf{q}\cdot\textbf{p})^2]}{q^2(\textbf{q}+\textbf{p})^2}\Bigl(n_{\rm F}(q^2-\mu)+n_{\rm F}(q^2+\mu)\Bigr).
\end{align}
Let us now discuss the convergence properties of the momentum integral. For $p_0=\mu=T=0$ we have
\begin{align}
 \label{trapp10b} K_{\rm T}^{(\rm QBT)'}(0,p)|_{\mu=T=0} = -\frac{p^2}{8}\chi_{\rm QBT}(0,p)|_{\mu=T=0} +3\bar{e}^2\int_{\textbf{q}}^\kappa \Biggl( 2 -\frac{(2\textbf{q}+\textbf{p})^2}{q^2+(\textbf{q}+\textbf{p})^2}\Bigl[1-\frac{3}{8} \frac{q^2p^2-(\textbf{q}\cdot\textbf{p})^2}{q^2(\textbf{q}+\textbf{p})^2}\Bigr]\Biggr).
\end{align}
The integrand behaves like $3\bar{e}^2 \frac{p^2}{q^2}+\mathcal{O}(q^{-4})$ for large momenta and thus introduces a spurious divergence for $p\neq 0$ similar to the one of the longitudinal response. This divergence can be removed by a suitable subtraction. For this note that the non-infinite physical part of Eq. (\ref{trapp10b}) $\mu=T=0$ satisfies $K_{\rm T}^{(\rm QBT)}(0,p)=C p^3$ with some constant $C$. For the present work, the value of $C$ is not important, since we only consider the homogeneous or quasi-static limits, where $p^2$ is small compared to all the other energy scales, and so $C$ cannot be resolved. We therefore chose a minimalistic subtraction where only the divergent part is removed. The manifestly finite expression for the transverse  QBT contribution for $p\geq 0$ is then given by
\begin{align}
 \nonumber K_{\rm T}^{(\rm QBT)}(p_0,p) &=-\frac{1}{2p^2}\Bigl(p_0^2+\frac{p^4}{4}\Bigr)\chi_{\rm QBT}(p_0,p) +3\bar{e}^2\int_{\textbf{q}}\Biggl\{ \Bigl(1-n_{\rm F}(q^2-\mu)-n_{\rm F}(q^2+\mu)\Bigr)\Biggl[2-\frac{(2\textbf{q}+\textbf{p})^2[q^2+(\textbf{q}+\textbf{p})^2]}{p_0^2+[q^2+(\textbf{q}+\textbf{p})^2]^2}\\
 \nonumber &+\frac{3}{8}\frac{(2\textbf{q}+\textbf{p})^2[q^2+(\textbf{q}+\textbf{p})^2]}{p_0^2+[q^2+(\textbf{q}+\textbf{p})^2]^2} \frac{[q^2p^2-(\textbf{q}\cdot\textbf{p})^2]}{q^2(\textbf{q}+\textbf{p})^2} \Biggr]-\frac{p^2}{q^2}\Biggr\} \\
 \label{trapp10} &-\frac{9}{8}\bar{e}^2 \int_{\textbf{q}}\frac{(2\textbf{q}+\textbf{p})^2[(\textbf{q}+\textbf{p})^2-q^2]}{p_0^2+[(\textbf{q}+\textbf{p})^2-q^2]^2} \frac{[q^2p^2-(\textbf{q}\cdot\textbf{p})^2]}{q^2(\textbf{q}+\textbf{p})^2}\Bigl(n_{\rm F}(q^2-\mu)+n_{\rm F}(q^2+\mu)\Bigr).
\end{align}
This formula constitutes the main result of this section. The integral in the last line vanishes in the homogeneous limit $p\to 0$, but it dominates in the quasi-static limit, as will be shown in the next section.
\end{widetext}

We can use Eq. (\ref{trapp10}) to verify that longitudinal and transverse  response coincide in the homogeneous limit. By employing Eq. (\ref{Napp20}) for the $p\to 0$ limit of $\chi_{\rm QBT}(p_0,p)$ we obtain
\begin{align}
 \nonumber &K_{\rm T}^{(\rm QBT)}(p_0,0) \\
 \label{trapp11} &= 4\bar{e}^2 \int_{\textbf{q}} \frac{p_0^2}{p_0^2+4q^4}\Bigl[1-n_{\rm F}(q^2-\mu)-n_{\rm F}(q^2+\mu)\Bigr],
\end{align}
which coincides with $K_{\rm L}^{(\rm QBT)}(p_0,0)$ derived in Sec. \ref{SecCont}. As a result, the homogeneous response in the normal state can be computed consistently from either $\chi$, $K_{\rm L}$, or $K_{\rm T}$ within our approximation.

\subsection{Quasi-static limit}
In this section we compute the normal state response functions in the quasi-static limit
\begin{align}
 \label{quapp1} p^2 \ll \omega.
\end{align}
Note that we use units $2m^*=1$ here, so that $v=\frac{p_{\rm F}}{m^*}=2\sqrt{|\mu|}$. Importantly, for $p>0$ we have to distinguish between longitudinal and transverse  response. We first discuss the upper band contribution in some detail, then easily modify this result to obtain the lower band contribution, and eventually compute the QBT contribution. Since the explicit formulas for the response at $T=0$ provide a particularly clear example to demonstrate the procedure we apply here, we will discuss the zero temperature results in considerable detail. We limit the discussion to the fully symmetric case of $x=\delta=0$.

In order to perform the quasi-static limit in a mathematically unambiguous way we introduce the dimensionless variables
\begin{align}
 \label{quapp1b} \hat{\omega} = \frac{\omega}{vp},\ \hat{p}_0=\frac{p_0}{vp},\ s = \frac{p^2}{\omega},\ s'=\frac{p^2}{p_0},
\end{align}
assuming $p_0>0$. In a scheme with natural units $\hbar=k_{\rm B}=2m^*=1$, every observable $O$ (at zero temperature for simplicity) can be expressed as
\begin{align}
  \label{quapp1c} O = O(\omega,p,\mu) = (\text{energy})^{\alpha_1} F\Bigl(\frac{\omega}{vp},\frac{p^2}{\omega}\Bigr),
\end{align}
where the prefactor is a (typically simple) fraction of powers of $\omega,p,v$, i.e. $\omega^{\alpha_2}p^{\alpha_3}v^{\alpha_4}$, with some exponents $\alpha_{1,\dots,4}$. The scaling function $F(\hat{\omega}_0,s)$ then allows to study the quasi-static limit by taking the limit $s\to 0$ (or $s'\to 0$) for fixed $\hat{\omega}$. The analytic continuation $\rmi \hat{p}_0\to \hat{\omega}+\rmi 0$ may be performed at any point in the computation and does not influence the procedure.

We start with the upper band contribution to the density response function given by Eq. (\ref{Napp10}),
\begin{align}
 \nonumber \chi_{\rm up}(p_0,p) &= -2\bar{e}^2 \int_{\textbf{q}} \frac{n_{\rm F}(\vare_q)}{\rmi p_0+\vare_q-\vare_{\textbf{q}+\textbf{p}}}  + \{ p_0\to -p_0\}\\
 \nonumber &=\frac{\bar{e}^2}{4\pi^2} \int_0^\infty\mbox{d}q\ n_{\rm F}(q^2-\mu) \\
 \label{quapp2} &\times \frac{q}{p}\ln\Bigl(\frac{\rmi p_0-p^2-2qp}{\rmi p_0-p^2+2qp}\Bigr)+\{p_0\to-p_0\},
\end{align}
where in the second line we evaluated the angular integral by means of
\begin{align}
 \label{quapp3} \int_{-1}^1\mbox{d}x \frac{1}{a+bx} = \frac{1}{b}\ln\Bigl(\frac{a+b}{a-b}\Bigr)
\end{align}
for $\text{Im}(a)\neq 0$. At zero temperature, assuming $\mu=|\mu|>0$, we can further evaluate the integral by means of 
\begin{align}
 \nonumber &\int\mbox{d}q\ q\ \ln(aq+b) \\
 \label{quapp4} &\ = \frac{b}{2a}q-\frac{1}{4}q^2+\frac{1}{2}\Bigl(q^2-\frac{b^2}{a^2}\Bigr)\ln(aq+b),
\end{align}
and so after analytic continuation with $\rmi p_0=\omega+\rmi 0$ arrive at the fully frequency and momentum dependent expression
\begin{align}
 \label{quapp5} &\chi_{\rm up}(\omega,p) =\frac{n_0\bar{e}^2}{m^*} \frac{3}{2v^2}\\
 \nonumber &\times \Biggl[ 1+\frac{p_F}{2p}\Bigl(1-\frac{(\omega-p^2)^2}{v^2p^2}\Bigr)\ln\Bigl(\frac{-\omega+p^2+vp-\rmi 0}{-\omega+p^2-vp-\rmi 0}\Bigr)\\
 \nonumber &+\frac{p_F}{2p}\Bigl(1-\frac{(\omega+p^2)^2}{v^2p^2}\Bigr)\ln\Bigl(\frac{\omega+p^2+vp+\rmi 0}{\omega+p^2-vp+\rmi0}\Bigr)\Biggr].
\end{align}
with
\begin{align}
 \label{quapp6} n_0 = \frac{p_F^3}{3\pi^2} = \frac{|\mu|^{3/2}}{3\pi^2}.
\end{align}
The density response function allows to compute the longitudinal component of the dielectric function and conductivity.

In order to access the transverse  component of the upper band response we consider the current response $K_{\rm T}$. The diamagnetic contribution reads
\begin{align}
 \label{quapp8} K_{\rm T}^{(\rm d,up)}(p_0,p) &= 4 \bar{e}^2\int_{\textbf{q}} n_{\rm F}(q^2-\mu) = 2 n \bar{e}^2 = \frac{n\bar{e}^2}{m^*}.
\end{align}
For the paramagnetic contribution we find
\begin{align}
 \nonumber &K_{\rm T}^{(\rm p,up)}(p_0,p) = \frac{1}{2}\Bigl(\delta_{ij}-\frac{p_ip_j}{p^2}\Bigr)  K_{ij}^{(\rm p,up)}(p_0,\textbf{p}) \\
 \nonumber &\ =-4 \bar{e}^2\int_{\textbf{q}} \Bigl(q^2-\frac{(\textbf{q}\cdot\textbf{p})^2}{p^2}\Bigr) \frac{n_{\rm F}(\vare_q)}{\rmi p_0+\vare_{\textbf{p}+\textbf{p}}-\vare_q}+\{p_0\to -p_0\}\\
 \nonumber &\ =-\frac{\bar{e}^2}{2\pi^2}\int_0^\infty \mbox{d}q\  n_{\rm F}(q^2-\mu) \Bigl[ q^2+\frac{q^3}{p}\Bigl(1-\frac{(\rmi p_0+p^2)^2}{4p^2q^2}\Bigr)\\
 \label{quapp9} &\times \ln\Bigl(\frac{\rmi p_0+p^2+2pq}{\rmi p_0+p^2-2pq}\Bigr)\Bigr]+\{p_0\to -p_0\},
\end{align}
where we have used
\begin{align}
 \label{quapp10} \int_{-1}^1\mbox{d}x \frac{1-x^2}{a+bx} = \frac{1}{b}\Bigl[ \frac{2a}{b}+\Bigl(1-\frac{a^2}{b^2}\Bigr)\ln\Bigl(\frac{a+b}{a-b}\Bigr)\Bigr]
\end{align}
for the angular integration with $\text{Im}(a)\neq0$. The term proportional to $[q^2+\{p_0\to -p_0\}]$ in Eq. (\ref{quapp9}) reads $-n\bar{e}^2$, and so only partially cancels the diamagnetic contribution. At zero temperature, assuming again $\mu=|\mu|>0$, we can compute the momentum integral by means of
\begin{align}
 \nonumber &\int\mbox{d}q\ q^3\Bigl(1-\frac{b^2}{a^2q^2}\Bigr) \ln\Bigl(\frac{b+aq}{b-aq}) \\
 \label{quapp11} &\ = \frac{bq}{6a}\Bigl(q^2-\frac{3b^2}{a^2}\Bigr)+\frac{1}{4}\Bigl(q^2-\frac{b^2}{a^2}\Bigr)^2\ln\Bigl(\frac{b+aq}{b-aq}\Bigr).
\end{align}
Adding the diamagnetic term and performing the analytic continuation we then arrive at
\begin{align}
 \nonumber &K_{\rm T}^{(\rm up)}(\omega,p) = \frac{n_0\bar{e}^2}{m^*}\frac{3}{8}\Biggl[ 1+3\Bigl(\frac{\omega}{vp}\Bigr)^2+\Bigl(\frac{p^2}{vp}\Bigr)^2\\
 \nonumber &\ - \frac{p_F}{2p}\Bigl[1-\Big(\frac{\omega+p^2}{vp}\Bigr)^2\Bigr]^2\ln\Bigl(\frac{\omega+p^2+vp+\rmi0}{\omega+p^2-vp+\rmi0}\Bigr)\\
 \label{quapp12} &\ - \frac{p_F}{2p}\Bigl[1-\Big(\frac{\omega-p^2}{vp}\Bigr)^2\Bigr]^2\ln\Bigl(\frac{-\omega+p^2+vp-\rmi0}{-\omega+p^2-vp-\rmi0}\Bigr)\Biggr].
\end{align}

Let us now perform the quasi-static limit according to the procedure described in Eq. (\ref{quapp1c}) for the zero temperature results with $\mu=|\mu|>0$, and then generalize the approach to $T\geq 0$ and arbitrary $\mu$. Equation (\ref{quapp5}) can be written as
\begin{align}
 \label{quapp13} \chi_{\rm up}(\omega,p) =\frac{n_0\bar{e}^2}{m^*} \frac{3}{2v^2}F_1\Bigl(\frac{\omega}{vp},\frac{p^2}{\omega}\Bigr)
\end{align}
with scaling function
\begin{align}
 \nonumber F_1(\hat{\omega},s) &= 1+ \frac{1}{4s\hat{\omega}}\Bigl[1-\hat{\omega}^2(1- s)^2\Bigr]\ln\Bigl(\frac{-1+s+\frac{1}{\hat{\omega}}-\rmi 0}{-1 +s-\frac{1}{\hat{\omega}}-\rmi 0}\Bigr)\\
 \nonumber &\ + \frac{1}{4s\hat{\omega}}\Bigl[1-\hat{\omega}^2(1+s)^2\Bigr]\ln\Bigl(\frac{1+s+\frac{1}{\hat{\omega}}+\rmi 0}{1 +s-\frac{1}{\hat{\omega}}+\rmi 0}\Bigr)\\
 \label{quapp14} &= 2-\hat{\omega}\ln\Bigl(\frac{\hat{\omega}+1+\rmi0}{\hat{\omega}-1+\rmi0}\Bigr)+\mathcal{O}(s^2).
\end{align}
In the limit $s=\frac{p^2}{\omega}\to 0$ the response function becomes
\begin{align}
 \label{quapp15} \lim_{s\to 0} \chi_{\rm up}(\omega,p) =\frac{n_0\bar{e}^2}{m^*} \frac{3}{v^2}\Bigl[1 -\frac{\omega}{2vp}\ \ln\Bigl(\frac{\omega+vp+\rmi 0}{\omega-vp+\rmi 0}\Bigr)\Bigr].
\end{align}
Expanding this expression for $\omega\ll vp$ by using $\ln(\frac{\omega+vp+\rmi0}{\omega-vp+\rmi0})=\ln(\frac{vp+\omega}{vp-\omega})-\rmi \pi$, we obtain
\begin{align}
 \label{quapp16} \chi_{\rm up}(\omega,p) &=  \frac{n_0 \bar{e}^2}{m^*} \frac{3}{v^2}\Bigl[ 1+\frac{\pi}{2} \frac{\rmi\omega}{vp}-\Bigl(\frac{\omega}{vp}\Bigr)^2+\dots\Bigr].
\end{align}
Analogously, we can write $K_{\rm T}^{(\rm up)}(\omega,p)$ as
\begin{align}
 \label{quapp17} K_{\rm T}^{(\rm up)}(\omega,p) = \frac{n_0\bar{e}^2}{m^*}\frac{3}{8} F_2\Bigl(\frac{\omega}{vp},\frac{p^2}{\omega}\Bigr)
\end{align}
with
\begin{align}
 \nonumber F_2(\hat{\omega},s) &= 1+3\hat{\omega}^2+\hat{\omega}^2s^2\\
 \nonumber &- \frac{1}{4s\hat{\omega}}\Bigl[1-\hat{\omega}^2(1-s)^2\Bigr]^2\ln\Bigl(\frac{-1+s+\frac{1}{\hat{\omega}}-\rmi 0}{-1 +s-\frac{1}{\hat{\omega}}-\rmi 0}\Bigr) \\
 \nonumber &- \frac{1}{4s\hat{\omega}}\Bigl[1-\hat{\omega}^2(1+s)^2\Bigr]^2\ln\Bigl(\frac{1+s+\frac{1}{\hat{\omega}}+\rmi 0}{1 +s-\frac{1}{\hat{\omega}}+\rmi 0}\Bigr)\\
 \label{quapp18} &= 4\hat{\omega}^2 +2\hat{\omega}(1-\hat{\omega}^2) \ln\Bigl(\frac{\hat{\omega}+1+\rmi0}{\hat{\omega}-1+\rmi 0}\Bigr)+\mathcal{O}(s^2).
\end{align}
We then arrive at
\begin{align}
 \nonumber \lim_{s\to 0} K_{\rm T}^{(\rm up)}(\omega,p) ={}& \frac{n_0\bar{e}^2}{m^*}\frac{3}{2} \Biggl[ \Bigl(\frac{\omega}{vp}\Bigr)^2+\frac{\omega}{2vp}\Bigl[1-\Bigl(\frac{\omega}{vp}\Bigr)^2\Bigr]\\
 \label{quapp19} &\times  \ln\Bigl(\frac{\omega+vp+\rmi0}{\omega-vp+\rmi0}\Bigr)\Biggr],
\end{align}
and so for $\omega\ll vp$ we have
\begin{align}
 \label{quapp20} K_{\rm T}^{(\rm up)}(\omega,p) = -\frac{n_0\bar{e}^2}{m^*} \frac{3\pi}{4} \frac{\rmi \omega}{vp} \Bigl[1+\frac{4}{\pi}\frac{\rmi \omega}{vp} -\Bigl(\frac{\omega}{vp}\Bigr)^2+\dots\Bigr].
\end{align}

Although the results of the previous paragraph relied on using the explicit response function at zero temperature, the applied method can be generalized to other systems where the momentum integration cannot necessarily be performed analytically. To illustrate this point, we rederive the $T=0$ expressions for the upper band from a different approach. To do so, we first notice that the first line of Eq. (\ref{quapp2}) can be written as
\begin{align}
 \label{quapp21} \chi_{\rm up}(p_0,p) &= \frac{\bar{e}^2}{\pi^2} \frac{\mu^{3/2}}{p_0} F_3\Bigl(\frac{p_0}{vp},\frac{p^2}{p_0}\Bigr)
\end{align}
with scaling function ($\hat{q}=q/\sqrt{\mu}$)
\begin{align}
 \label{quapp22} F_3(\hat{p}_0,s')& =\int_0^{1} \mbox{d}\hat{q} \int_{-1}^1\mbox{d}x\ \frac{\hat{q}^2(\frac{1}{\hat{p}_0}\hat{q}x+s')}{1+(\frac{1}{\hat{p}_0}\hat{q}x+s')^2}.
\end{align}
Crucially, the integrand of $F_3$ can be expanded in powers of $s'$ yielding a simplified integral that can be evaluated explicitly. We have
\begin{align}
 \nonumber F_3(\hat{p}_0,s') &=\int_0^{1} \mbox{d}\hat{q} \int_{-1}^1\mbox{d}x\ \Bigl[\frac{\hat{p}_0\hat{q}^3x}{\hat{p}_0^2+\hat{q}^2x^2}\\
\nonumber  &\ +\frac{\hat{p}_0^2\hat{q}^2(\hat{p}_0^2-\hat{q}^2x^2)}{(\hat{p}_0^2+\hat{q}^2x^2)^2}s'+\mathcal{O}(s'{}^2)\Bigr]\\
  \label{quapp23} &= 2\hat{p}_0^2s' \Bigl[1-\frac{\rmi \hat{p}_0}{2}\ln\Bigl(\frac{\rmi \hat{p}_0+1}{\rmi \hat{p}_0-1}\Bigr)\Bigr]+\mathcal{O}(s'{}^2),
\end{align}
and so
\begin{align}
 \label{quapp24} \lim_{s'\to0}\chi_{\rm up}(p_0,p) = \frac{2\bar{e}^2\mu^{3/2}}{\pi^2} \frac{1}{v^2} \Bigl[1-\frac{\rmi p_0}{2vp}\ln\Bigl(\frac{\rmi  p_0+vp}{\rmi p_0-vp}\Bigr)\Bigr],
\end{align}
which reproduces Eq. (\ref{quapp15}) after analytic continuation. For the transverse  response we rewrite Eq. (\ref{quapp9}) as
\begin{align}
 \label{quapp25} K_{\rm T}^{(\rm p,up)}(p_0,p) &= -\frac{2\bar{e}^2}{\pi^2} \frac{\mu^{5/2}}{p_0} F_4\Bigl(\frac{p_0}{vp},\frac{p^2}{p_0}\Bigr)
\end{align}
with
\begin{align}
 \nonumber F_4(\hat{p}_0,s') &=\int_0^{1}\mbox{d}\hat{q} \int_{-1}^1\mbox{d}x\ \frac{\hat{q}^4(1-x^2)(\frac{1}{\hat{p}_0}\hat{q}x+s')}{1+(\frac{1}{\hat{p}_0}\hat{q}x+s')^2}\\
 \nonumber &=\int_0^{1}\mbox{d}\hat{q} \int_{-1}^1\mbox{d}x\ \Bigl[\frac{\hat{p}_0\hat{q}^5(1-x^2)x}{\hat{p}_0^2+\hat{q}^2x^2}\\
 \nonumber &\ +\frac{\hat{p}_0^2\hat{q}^4(1-x^2)(\hat{p}_0^2-\hat{q}^2x^2)}{(\hat{p}_0^2+\hat{q}^2x^2)^2}s'+\mathcal{O}(s'{}^2)\Bigr]\\
 \nonumber &= 4\hat{p}_0^2 s' \Bigl[\frac{1}{3}+\frac{\hat{p}_0^2}{2}-\frac{\rmi \hat{p}_0}{4}\Bigl(1+\hat{p}_0^2\Bigr)\ln\Bigl(\frac{\rmi \hat{p}_0+1}{\rmi \hat{p}_0-1}\Bigr)\Bigr]\\
 \label{quapp26} &+\mathcal{O}(s'{}^2).
\end{align}
Hence
\begin{align}
 \nonumber \lim_{s'\to 0}K_{\rm T}^{(\rm p,up)}(p_0,p) &= \frac{n_0\bar{e}^2}{m^*}\Biggl[-1-\frac{3}{2}\Bigl(\frac{p_0}{vp}\Bigr)^2\\
 \label{quapp27} &+\frac{3\rmi p_0}{4vp}\Bigl[1+\Bigl(\frac{p_0}{vp}\Bigr)^2\Bigr]\ln\Bigl(\frac{\rmi p_0+vp}{\rmi p_0-vp}\Bigr)\Biggr].
\end{align}
The first term is cancelled by the diamagnetic contribution $\frac{n_0\bar{e}^2}{m^*}$. After analytic continuation we then recover Eq. (\ref{quapp19}) for the whole response function. These two examples illustrate how defining the scaling function as an integral and then expanding the integrand in powers of $s'$ (or $s$) is an efficient way to compute the quasi-static limit. Furthermore, it enables us to extend the analysis to more complicated setups, as we expound in the following.

First we aim to extend the previous discussion to nonzero temperature. Since this introduces an additional energy scale $T$, the scaling functions will also depend on $\hat{T}=T/\mu$. This dependence, on the other hand, is unimportant for the limiting procedures involved, and we may say that temperature takes a pure spectator role in the quasi-static limit. With $\hat{q}=q/\sqrt{\mu}$ we write the Fermi--Dirac distribution in dimensionless form as
\begin{align}
 \nonumber n_{\rm F}(q^2-\mu) &= (e^{\frac{q^2-\mu}{T}}+1)^{-1}\\
 \label{quapp28} &=(e^{\frac{\hat{q}^2-1}{T/\mu}}+1)^{-1}=: \hat{n}_F(\hat{q}^2-1).
\end{align}
The scaling functions $F_3$ and $F_4$ in Eqs. (\ref{quapp21}) and (\ref{quapp25}) then readily generalize to
\begin{align}
 \nonumber F_3(\hat{p}_0,s')& =\int_0^{\infty} \mbox{d}\hat{q} \int_{-1}^1\mbox{d}x\ \hat{n}_F(\hat{q}^2-1) \frac{\hat{q}^2(\frac{1}{\hat{p}_0}\hat{q}x+s')}{1+(\frac{1}{\hat{p}_0}\hat{q}x+s')^2}\\
  \label{quapp29} &=2\hat{p}_0^2 s' \int_0^\infty \mbox{d}\hat{q}\ \hat{n}_F(\hat{q}^2-1)\frac{\hat{q}^2}{\hat{p}_0^2+\hat{q}^2}+\mathcal{O}(s'{}^2)
\end{align}
and
\begin{align}
 \nonumber F_4(\hat{p}_0,s') ={}&\int_0^{\infty}\mbox{d}\hat{q} \int_{-1}^1\mbox{d}x\ \hat{n}_F(\hat{q}^2-1)\\
 \nonumber &\times  \frac{\hat{q}^4(1-x^2)(\frac{1}{\hat{p}_0}\hat{q}x+s')}{1+(\frac{1}{\hat{p}_0}\hat{q}x+s')^2}\\
 \nonumber  &=4\hat{p}_0^2s' \int_0^\infty\mbox{d}\hat{q}\ \hat{n}_F(\hat{q}^2-1) \hat{q}^2\\
 \label{quapp30} &\times \Bigl[ 1-\frac{\rmi \hat{p}_0}{2\hat{q}}\ln\Bigl(\frac{\rmi \hat{p}_0+\hat{q}}{\rmi \hat{p}_0-\hat{q}}\Bigr)\Bigr]+\mathcal{O}(s'{}^2).
\end{align}
In these limits the response functions become
\begin{align}
\label{quapp31} \lim_{s'\to0} \chi_{\rm up} (p_0,p) &= 4\bar{e}^2p^2 \int_{\textbf{q}} \frac{n_{\rm F}(q^2-\mu)}{p_0^2+4q^2p^2},\\
\label{quapp32} \lim_{s'\to 0} K_{\rm T}^{(\rm up)}(p_0,p) &= \bar{e}^2 \int_{\textbf{q}} n_{\rm F}(q^2-\mu)\frac{\rmi p_0}{qp}\ln\Bigl(\frac{\rmi p_0+2qp}{\rmi p_0-2qp}\Bigr).
\end{align}
In addition, these formulas can be applied for any sign of the chemical potential.

The contribution from the lower band can now easily be obtained by a sign change $\mu\to -\mu$. We are left with
\begin{align}
\label{quapp33} \lim_{s'\to0} \chi_{\rm low} (p_0,p) &= 4\bar{e}^2p^2 \int_{\textbf{q}} \frac{n_{\rm F}(q^2+\mu)}{p_0^2+4q^2p^2},\\
\label{quapp34} \lim_{s'\to 0} K_T^{(\rm low)}(p_0,p) &= \bar{e}^2 \int_{\textbf{q}} n_{\rm F}(q^2+\mu)\frac{\rmi p_0}{qp}\ln\Bigl(\frac{\rmi p_0+2qp}{\rmi p_0-2qp}\Bigr).
\end{align}
In the sum of the contributions from the upper and lower bands, the chemical potential only enters through the Fermi functions by means of  $n_{\rm F}(q^2-\mu)+n_{\rm F}(q^2+\mu)$. At zero temperature this reduces to $\theta(|\mu|-q^2)$. Consequently, the zero temperature formulas derived above for $\mu>0$ apply to the whole intraband contribution after replacing $\mu\to|\mu|$ in the formulas.

Let us now turn to the longitudinal QBT contribution. Starting from Eq. (\ref{Napp8}) and performing the Matsubara sum according to Eq. (\ref{trapp6}) we arrive at
\begin{align}
 \nonumber &\chi_{\rm QBT}(p_0,p) = 6 \bar{e}^2 \int_{\textbf{q}} \frac{q^2p^2-(\textbf{q}\cdot\textbf{p})^2}{4q^2(\textbf{q}+\textbf{p})^2}\Biggl[\frac{n_{\rm F}(q^2+\mu)}{\rmi p_0+q^2-(\textbf{q}+\textbf{p})^2}\\
 \nonumber &\ +\frac{n_{\rm F}(q^2-\mu)}{-\rmi p_0+q^2-(\textbf{q}+\textbf{p})^2}+\frac{1-n_{\rm F}(q^2+\mu)}{\rmi p_0+q^2+(\textbf{q}+\textbf{p})^2}\\
 \nonumber &\ +\frac{-n_{\rm F}(q^2-\mu)}{-\rmi p_0+q^2+(\textbf{q}+\textbf{p})^2} \Biggr]+\{p_0\to-p_0\}\\
 \nonumber &\ = 12\bar{e}^2 \int_{\textbf{q}} \frac{q^2p^2-(\textbf{q}\cdot\textbf{p})^2}{4q^2(\textbf{q}+\textbf{p})^2}\Biggl[\frac{q^2+(\textbf{q}+\textbf{p})^2}{p_0^2+[q^2+(\textbf{q}+\textbf{p})^2]^2}\\
 \nonumber &\ -\Bigl(n_{\rm F}(q^2-\mu)+n_{\rm F}(q^2+\mu)\Bigr) \\
 \label{quapp35} &\ \times \Bigl(\frac{[q^2+(\textbf{q}+\textbf{p})^2]}{p_0^2+[q^2+(\textbf{q}+\textbf{p})^2]^2}+\frac{(\textbf{q}+\textbf{p})^2-q^2}{p_0^2+[(\textbf{q}+\textbf{p})^2-q^2]^2}\Bigr)\Biggr].
\end{align}
The zero temperature response reads
\begin{align}
 \nonumber \chi_{\rm QBT}(p_0,p) ={}&  \frac{3\bar{e}^2}{4\pi^2} \frac{p^2}{p_0} \int_0^\infty\mbox{d}q\int_{-1}^1\mbox{d}x\\
 \label{quapp37} &\times \Bigl[\theta(q^2-|\mu|) \hat{X} -\theta(|\mu|-q^2)\hat{Y}\Bigr]
\end{align}
with
\begin{align}
 \label{quapp38} \hat{X} &= p_0\frac{q^2(1-x^2)}{q^2+2qpx+p^2}\frac{2q^2+2qpx+p^2}{p_0^2+(2q^2+2qpx+p^2)^2}\\
 \nonumber &=\frac{\frac{1}{4\hat{p}_0^2s'}\hat{q}^2(1-x^2)}{\frac{1}{4\hat{p}_0^2s'} \hat{q}^2+\frac{1}{\hat{p}_0}\hat{q}x+s'} \frac{\frac{1}{2\hat{p}_0^2s'} \hat{q}^2+\frac{1}{\hat{p}_0}\hat{q}x+s'}{1+[\frac{1}{2\hat{p}_0^2s'}\hat{q}^2+\frac{1}{\hat{p}_0}\hat{q}x+s']^2},\\
 \label{quapp39} \hat{Y} &= p_0 \frac{q^2(1-x^2)}{q^2+2qpx+p^2}\frac{p^2+2qpx}{p_0^2+(p^2+2qpx)^2}\\
 \nonumber &= \frac{\frac{1}{4\hat{p}_0^2s'}\hat{q}^2(1-x^2)}{\frac{1}{4\hat{p}_0^2s'} \hat{q}^2+\frac{1}{\hat{p}_0}\hat{q}x+s'} \frac{s'+\frac{1}{\hat{p}_0}\hat{q}x}{1+[s'+\frac{1}{\hat{p}_0}\hat{q}x]^2}.
\end{align}
The expansion of $\hat{X}$ misses an infrared singularity of $\int_{-1}^1\mbox{d}x\ \hat{X}$ at $q=\sqrt{\omega/2}$ after analytic continuation. This effect, however, is only visible at finite temperature, and we discuss it below when computing the quasi-static result for $T>0$. Here we take the limit $s'\to0$ and have
\begin{align}
 \nonumber \hat{X} &= \frac{2\hat{p}_0^2(1-x^2)}{\hat{q}^2}s'-\frac{12\hat{p}_0^3x(1-x^2)}{\hat{q}^3}s'{}^2 +\mathcal{O}(s'{}^3),\\
 \nonumber \hat{Y}&=\frac{\hat{p}_0\hat{q}x(1-x^2)}{\hat{p}_0^2+\hat{q}^2x^2}\\
 \label{quapp40} &+\frac{\hat{p}_0^2(1-x^2)(\hat{p}_0^2-4\hat{p}_0^2x^2-\hat{q}^2x^2-4\hat{q}^2x^4)}{(\hat{p}_0^2+\hat{q}^2x^2)^2}s'+\mathcal{O}(s'{}^2).
\end{align}
We introduce the scaling function $F_7$ according to
\begin{align}
 \label{quapp43}  \chi_{\rm QBT}(p_0,p) &= \frac{3\bar{e}^2}{4\pi^2} \frac{p^2\sqrt{|\mu|}}{p_0} F_7\Bigl(\frac{p_0}{vp},\frac{p^2}{p_0}\Bigr)
\end{align}
with
\begin{align}
 \nonumber F_7(\hat{p}_0,s') &= \int_0^\infty\mbox{d}\hat{q}\int_{-1}^1\mbox{d}x\ \Bigl[\theta(\hat{q}^2-1) \hat{X} -\theta(1-\hat{q}^2)\hat{Y}\Bigr]\\
 \nonumber &\simeq \frac{4}{3}s'\Biggl(2\hat{p}_0^2 + \int_0^1\mbox{d}\hat{q}\ \frac{\hat{p}_0^2}{\hat{q}^5}\Bigl[ \hat{q}^3+6\hat{p}_0^2\hat{q}\\
 \nonumber &-\frac{3\rmi\hat{p}_0}{2}(\hat{q}^2+2\hat{p}_0^2) \ln\Bigl(\frac{\rmi \hat{p}_0+\hat{q}}{\rmi \hat{p}_0-\hat{q}}\Bigr)\Bigr]\Biggr)\\
 \label{quapp45}  &= \frac{4}{3}s'\hat{p}_0^2\Bigl[1 -\frac{3}{2}\hat{p}_0^2+\frac{3\rmi \hat{p}_0}{4}(1+\hat{p}_0^2)\ln\Bigl(\frac{\rmi \hat{p}_0+1}{\rmi \hat{p}_0-1}\Bigr)\Bigr].
\end{align}
After analytic continuation we are left with
\begin{align}
 \nonumber \lim_{s' \to 0}  \chi_{\rm QBT}(\omega,p) &=\frac{\bar{e}^2}{4\pi^2} \frac{p^2}{\sqrt{|\mu|}} \Biggl(1+\frac{3}{2}\Bigl(\frac{\omega}{vp}\Bigr)^2\\
 \label{quapp46} &+\frac{3\omega}{4vp}\Bigl[1-\Bigl(\frac{\omega}{vp}\Bigr)^2\Bigr]\ln\Bigl(\frac{\omega+vp+\rmi0}{\omega-vp+\rmi0}\Bigr)\Biggr).
\end{align}
In the limit $\omega\ll vp$ we obtain
\begin{align}
 \label{quapp47} \chi_{\rm QBT}(\omega,p) &=\frac{\bar{e}^2}{4\pi^2} \frac{p^2}{\sqrt{|\mu|}} \Bigl[1-\frac{3\pi}{4} \frac{\rmi \omega}{vp} +3\Bigl(\frac{\omega}{vp}\Bigr)^2+\dots\Bigr].
\end{align}

We now turn to the transverse  QBT response given in Eq. (\ref{trapp10}). We introduce the function $k_{\rm T}(p_0,p)$ according to
\begin{align}
 \label{quapp48} K_{\rm T}^{(\rm QBT)}(p_0,p) &=-\frac{1}{2p^2}\Bigl(p_0^2+\frac{p^4}{4}\Bigr)\chi_{\rm QBT}(P) +k_{\rm T}(p_0,p).
\end{align}
In the quasi-static limit we neglect $p^4/4$ compare to $p_0^2$ and so
\begin{align}
 \label{quapp48b} \lim_{s'\to 0} K_{\rm T}^{(\rm QBT)}(p_0,p) &=\lim_{s'\to 0}\Bigl[-\frac{p_0^2}{2p^2}\chi_{\rm QBT}(P) +k_{\rm T}(p_0,p)\Bigr].
\end{align}
Since we already know the $s'\to 0$ limit of $\chi_{\rm QBT}$, we only need to consider $k_{\rm T}$ in the following. At zero temperature we have
\begin{align}
 \nonumber k_{\rm T}(p_0,p) &=3\bar{e}^2\int_{\textbf{q}}\Biggl( \Bigl[2-\frac{(2\textbf{q}+\textbf{p})^2[q^2+(\textbf{q}+\textbf{p})^2]}{p_0^2+[q^2+(\textbf{q}+\textbf{p})^2]^2}\\
 \nonumber &+\frac{3}{8}(2\textbf{q}+\textbf{p})^2\frac{p^2}{p_0q^2}\hat{X} \Bigr]\theta(q ^2-|\mu|)-\frac{p^2}{q^2}\Biggr) \\
 \label{quapp49} &-\frac{9}{8}\bar{e}^2 \int_{\textbf{q}}\theta(|\mu|-q^2)(2\textbf{q}+\textbf{p})^2\frac{p^2}{p_0q^2}\hat{Y}
\end{align}
with $\hat{X}$ and $\hat{Y}$ from Eqs. (\ref{quapp38})-(\ref{quapp40}). We introduce the scaling function $F_8$ through
\begin{align}
 \label{quapp50} k_{\rm T}(p_0,p) = \frac{3\bar{e}^2}{4\pi^2} |\mu|^{3/2} F_8\Bigl(\frac{p_0}{vp},\frac{p^2}{p_0}\Bigr)
\end{align}
with
\begin{align}
 \nonumber &F_8(\hat{p}_0,s') =\int_0^\infty\mbox{d}\hat{q}\int_{-1}^1\mbox{d}x\ \Biggl\{\theta(\hat{q}^2-1)\\
 \nonumber &\times \Bigl[ 2\hat{q}^2 -\hat{q}^2\frac{(\frac{1}{\hat{p}_0^2s'}\hat{q}^2+\frac{2}{\hat{p}_0}\hat{q}x+s')(\frac{1}{2\hat{p}_0^2s'}\hat{q}^2+\frac{1}{\hat{p}_0}\hat{q}x+s')}{1+(\frac{1}{2\hat{p}_0^2s'}\hat{q}^2+\frac{1}{\hat{p}_0}\hat{q}x+s')^2} \\
 \nonumber &+\frac{3}{2}\Bigl( \hat{q}^2 +2\hat{p}_0\hat{q}xs'+\hat{p}_0^2s'{}^2\Bigr)s'\hat{X}\Bigr]-4s'{}^2\hat{p}_0^2\Biggr\}\\
 \label{quapp51} &- \frac{3}{2}\int_0^\infty\mbox{d}\hat{q}\int_{-1}^1\mbox{d}x\  \theta(1-\hat{q}^2) \Bigl( \hat{q}^2 +2\hat{p}_0\hat{q}xs'+\hat{p}_0^2s'{}^2\Bigr)s'\hat{Y}.
\end{align}
Using the above expansion of $\hat{X}$ and $\hat{Y}$ in powers of $s'$ we identify the leading contribution to be given by
\begin{align}
 \nonumber &F_8(\hat{p}_0,s') \\
 \nonumber &= s'{}^2\int_0^\infty\mbox{d}\hat{q}\int_{-1}^1\mbox{d}x\ \Biggl\{\theta(\hat{q}^2-1)\Bigl[5\hat{p}_0^2+\frac{8\hat{p}_0^4}{\hat{q}^2} -3\hat{p}_0^2x^2\Bigr]\\
 \nonumber &-4\hat{p}_0^2\Biggr\}- \frac{3}{2}s'{}^2\int_0^\infty\mbox{d}\hat{q}\int_{-1}^1\mbox{d}x\ \theta(1-\hat{q}^2)\\
 \label{quapp52} &\times  \frac{\hat{p}_0^2\hat{q}^2(1-x^2)[\hat{p}_0^2(1-2x^2)-\hat{q}^2x^2(1+2x^2)]}{(\hat{p}_0^2+\hat{q}^2x^2)^2}+\mathcal{O}(s'{}^3).
\end{align}
The remaining integrals can be evaluated explicitly. Importantly, the regularization introduced in Eq. (\ref{trapp10}) for the transverse  response is critical to make the first integral finite. The contribution from the first integral to $F_8$ is given by
\begin{align}
 \label{quapp52b} \int_0^\infty\mbox{d}\hat{q}\Bigl[ -8\hat{p}_0^2\theta(1-\hat{q}^2)+\frac{16\hat{p}_0^4}{\hat{q}^2}\theta(\hat{q}^2-1)\Bigr]=-\hat{p}_0^2(1-2\hat{p}_0^2).
\end{align}
The full scaling function reads
\begin{align}
 \label{quapp53} F_8(\hat{p}_0,s') &= -2s'{}^2\hat{p}_0^2\Biggl( 5-\frac{13}{2}\hat{p}_0^2\\
 \nonumber &-\frac{3\rmi \hat{p}_0}{4}(1+\hat{p}_0^2)\ln\Bigl(\frac{\rmi \hat{p}_0+1}{\rmi \hat{p}_0-1}\Bigr)\Biggr)+\mathcal{O}(s'{}^3).
\end{align}
We include the density response according to Eq. (\ref{quapp48b}) and analytically continue to arrive at
\begin{align}
 \nonumber \lim_{s \to 0}& K_{\rm T}^{(\rm QBT)}(\omega,p) =  - \frac{15\bar{e}^2}{8\pi^2}p^2\sqrt{\mu}\Bigl[1+\frac{4}{3}\Bigl(\frac{\omega}{vp}\Bigr)^2\Bigr]\\
 \nonumber &\times  \Biggl[1-\frac{3}{10}\Bigl(\frac{\omega}{vp}\Bigr)^2\\
 \label{quapp56} & -\frac{3\omega}{20vp}\Bigl(1-\Bigl(\frac{\omega}{vp}\Bigr)^2\Bigr)\ln\Bigl(\frac{\omega+vp+\rmi 0}{\omega-vp+\rmi 0}\Bigr)\Biggr].
\end{align}
In the limit $\omega\ll vp$ we find
\begin{align}
 \label{quapp57}K_{\rm T}^{(\rm QBT)}(\omega,p) ={}&-\frac{15\bar{e}^2}{8\pi^2}p^2\sqrt{\mu}\\
  \nonumber &\ \times \Bigl[ 1+\frac{3\pi \rmi}{20}\frac{\omega}{vp}+\frac{11}{15}\Bigl(\frac{\omega}{vp}\Bigr)^2+\dots\Bigr].
\end{align}

For computing the quasi-static limit of the QBT contributions at nonzero temperatures, it is crucial to first study the angular integral over ($p_0$ times)  $\hat{X}$ from Eq. (\ref{quapp38}). We define
\begin{align}
 \nonumber f_X(q) &= \int_{-1}^1\mbox{d}x\ \frac{q^2(1-x^2)}{q^2+2qpx+p^2}\\
 \label{quapp58} &\ \times \frac{2q^2+2qpx+p^2}{-(\omega+\rmi 0)^2+(2q^2+2qpx+p^2)^2}.
\end{align}
The density response after analytic continuation can then be expressed as
\begin{align}
 \nonumber &\chi_{\rm QBT}(\omega,p) \\
 \nonumber &=  \frac{3\bar{e}^2}{4\pi^2} p^2 \int_0^\infty\mbox{d}q\ [1-n_{\rm F}(q^2-\mu)-n_{\rm F}(q^2+\mu)] f_X(q)\\
 \label{quapp59} &-\frac{3\bar{e}^2}{4\pi^2} \frac{p^2}{p_0} \int_0^\infty\mbox{d}q\int_{-1}^1\mbox{d}x\ [n_{\rm F}(q^2-\mu)+n_{\rm F}(q^2+\mu)] \hat{Y}.
\end{align}
It is easy to verify that the function $f_X(q)$ becomes singular at $q=\sqrt{\frac{\omega}{2}}$ for $p\to 0$ in both its real and imaginary part. For instance, the imaginary part can be obtained by applying $ \text{Im}\frac{1}{E+\rmi 0} = -\pi\ \delta(E)$ and reads
\begin{align}
 \label{quapp61} \text{Im} f_X(q) =\theta(2qp-|\omega-2q^2-p^2|) \frac{\pi q(1-[\frac{\omega-2q^2-p^2}{2qp}]^2)}{4p(\omega-q^2)}.
\end{align}
This expression is nonzero for $\omega>p^2/2$, in which case its support is limited to the interval $q^2\in[\frac{\omega-p\sqrt{2\omega-p^2}}{2},\frac{\omega+p\sqrt{2\omega-p^2}}{2}]$. In particular, for $p\to 0$ it becomes sharply peaked at $q^2=\omega/2$ with
\begin{align}
\label{quapp62} \lim_{p\to 0} \text{Im} f_X(q) =\frac{\pi}{3\sqrt{2\omega}} \delta\Bigl(q-\sqrt{\frac{\omega}{2}}\Bigr)
\end{align}
Similarly, also the real part of $f_X(q)$ has a singularity at $q=\sqrt{\omega/2}$. This implies a nontrivial contribution to the first integral in Eq. (\ref{quapp59}) which is missed upon expanding $\hat{X}$ according to Eq. (\ref{quapp40}) before evaluating the $x$-integration. At zero temperature, the Fermi distribution limits the momentum integration to the region with $q^2>|\mu|$. This misses the singularity of $f_X(q)$ since $\frac{\omega}{2}<|\mu|$ in the quasi-static limit, and so the expansion of $\hat{X}$ is valid. At nonzero temperature, on the other hand, the Fermi distribution gives a nonzero weight to all momenta $q$, and thus the singular region of the integrand contributes to the final expression. Furthermore, it is easy to see that the analytical structure of this $p^2/\omega\to 0$ contribution is identical to the homogeneous result, although for $\omega\ll |\mu|$, and so Eq. (\ref{quapp59}) can generally be written as
\begin{align}
 \nonumber &\chi_{\rm QBT}(\omega,p) = \chi_{\rm QBT}(\omega)\\
 \label{quapp63} &-\frac{3\bar{e}^2}{4\pi^2} \frac{p^2}{p_0} \int_0^\infty\mbox{d}q\int_{-1}^1\mbox{d}x\ [n_{\rm F}(q^2-\mu)+n_{\rm F}(q^2+\mu)]\hat{Y},
\end{align}
where $\chi_{\rm QBT}(\omega)$ is the homogeneous result. The contribution from the remaining integral is regular and can be obtained by expanding $\hat{Y}$ as in Eq. (\ref{quapp40}). The final result for the longitudinal QBT contribution to $\vare_{\rm L}=\frac{4\pi \chi}{p^2}$ as a function of $\hat{\omega}=\omega/(vp)$ reads
\begin{align}
 \nonumber &\lim_{s\to 0}\vare_{\rm L}^{(\rm QBT)}(\omega,p) = \vare^{(\rm QBT)}(\omega) \\
 \nonumber &+\frac{e^2}{4\pi^2\eps_0}\sqrt{\frac{2m^*}{|\mu|}} \int_0^\infty\mbox{d}\hat{q}\  \frac{\hat{n}_F(\hat{q}^2-1)+\hat{n}_F(\hat{q}^2+1)}{\hat{q}^2}\\
 \label{quapp64} &\times \Biggl[ 1-6\Bigl(\frac{\hat{\omega}}{\hat{q}}\Bigr)^2-\frac{3\hat{\omega}}{2\hat{q}}\Bigl[1-2\Bigl(\frac{\hat{\omega} }{\hat{q}}\Bigr)^2\Bigr]\ln\Bigl(\frac{\hat{\omega}+\hat{q}+\rmi0}{\hat{\omega}-\hat{q}+\rmi 0}\Bigr)\Biggr],
\end{align}
with $\hat{n}_F(\hat{q}^2\pm 1)=n_{\rm F}(q^2\pm \mu)$ defined in Eq. (\ref{quapp28}). The homogeneous result is given by
\begin{align}
 \nonumber \vare^{(\rm QBT)}(\omega) &= \frac{e^2}{2\pi^2\eps_0}\sqrt{\frac{2m^*}{|\mu|}} \int_0^\infty\mbox{d}\hat{q}\ \hat{q}^2\\
 \label{quapp65} &\times  \frac{1-\hat{n}_F(\hat{q}^2-1)-\hat{n}_F(\hat{q}^2+1)}{-(\frac{\omega}{2\mu}+\rmi 0)^2+\hat{q}^4}.
\end{align}

The  transverse  contribution for $T>0$ can be treated in a similar fashion by isolating the homogeneous contribution from $k_{\rm T}(\omega,p)$ and expanding the remaining terms in powers of $s$. We write $k_{\rm T}(p_0,p) = k_{\rm T}(p_0,0) + \tilde{k}_{\rm T}(p_0,p)$ with
\begin{align}
  \nonumber \tilde{k}_{\rm T}(p_0,p) &= k_{\rm T}(p_0,p)-k_{\rm T}(p_0,0) \\
 \label{quapp65b}&= \frac{3\bar{e}^2}{4\pi^2} |\mu|^{3/2} F_8\Bigl(\frac{p_0}{vp},\frac{p^2}{p_0}\Bigr)-\{p_0\to0\},
\end{align}
where $F_8(\hat{p}_0,s')$ is obtained from the formal expansion of
\begin{align}
 \nonumber &F_8 =\int_0^\infty\mbox{d}\hat{q}\int_{-1}^1\mbox{d}x\ \Biggl\{[1-\hat{n}_F(\hat{q}^2-1)-\hat{n}_F(\hat{q}^2+1)]\\
 \nonumber &\times \Bigl[ 2\hat{q}^2 -\hat{q}^2\frac{(\frac{1}{\hat{p}_0^2s'}\hat{q}^2+\frac{2}{\hat{p}_0}\hat{q}x+s')(\frac{1}{2\hat{p}_0^2s'}\hat{q}^2+\frac{1}{\hat{p}_0}\hat{q}x+s')}{1+(\frac{1}{2\hat{p}_0^2s'}\hat{q}^2+\frac{1}{\hat{p}_0}\hat{q}x+s')^2}\\
 \nonumber &+\frac{3}{2}\Bigl( \hat{q}^2 +2\hat{p}_0\hat{q}xs'+\hat{p}_0^2s'{}^2\Bigr)s'\hat{X}\Bigr]-4s'{}^2\hat{p}_0^2\Biggr\}\\
 \nonumber &- \frac{3}{2}\int_0^\infty\mbox{d}\hat{q}\int_{-1}^1\mbox{d}x\ [\hat{n}_F(\hat{q}^2-1)+\hat{n}_F(\hat{q}^2+1)]\\
 \label{quapp65c} &\times \Bigl( \hat{q}^2 +2\hat{p}_0\hat{q}xs'+\hat{p}_0^2s'{}^2\Bigr)s'\hat{Y}
\end{align}
in powers of $s'$. We obtain
\begin{align}
 \nonumber F_8(\hat{p}_0,s') &=16s'{}^2\hat{p}_0^4\int_0^\infty\mbox{d}\hat{q}\ \frac{1-\hat{n}_F(\hat{q}^2-1)-\hat{n}_F(\hat{q}^2+1)}{\hat{q}^2}\\
 \nonumber &-2s'{}^2\hat{p}_0^2 \int_0^\infty\mbox{d}\hat{q}\ [\hat{n}_F(\hat{q}^2-1)+\hat{n}_F(\hat{q}^2+1)] \\
 \label{quapp65d} &\times \Bigl[5-\frac{3\hat{p}_0^2}{\hat{q}^2}+\frac{3\rmi \hat{p}_0^3}{2\hat{q}^3}\log\Bigl(\frac{\rmi \hat{p}_0+\hat{q}}{\rmi \hat{p}_0-\hat{q}}\Bigr)\Bigr]+\mathcal{O}(s'{}^3).
\end{align}
Crucially, after multiplication with $|\mu|^{3/2}$ in Eq. (\ref{quapp65b}), the first term is proportional to $p_0^2/\sqrt{|mu|}$, and so is cancelled upon subtracting the homogeneous result with $p=0$. In contrast, the second term is proportional to $\sqrt{\mu}p$, and thus not eliminated by the subtraction. We conclude that after analytic continuation we have
\begin{align}
 \nonumber \lim_{s\to 0}k_{\rm T}(\omega,p) &= k_{\rm T}(\omega) -  \frac{3\bar{e}^2}{8\pi^2}\sqrt{\mu}p^2\\
 \nonumber & \times \int_0^\infty\mbox{d}\hat{q}\ [\hat{n}_F(\hat{q}^2-1)+\hat{n}_F(\hat{q}^2+1)]\\
 \label{quapp65e} &\times \Bigl[5+\frac{3\hat{\omega}^2}{\hat{q}^2}-\frac{3\hat{\omega}^3}{2\hat{q}^3}\log\Bigl(\frac{\hat{\omega}+\hat{q}+\rmi 0}{\hat{\omega}-\hat{q}+\rmi0}\Bigr)\Bigr],
\end{align}
where $k_{\rm T}(\omega)$ is the homogeneous contribution. Adding the contribution from the density response, we obtain the transverse  contribution to $\vare_{\rm T}=-\frac{4\pi}{\omega^2}K_{\rm T}$ as
\begin{align}
 \nonumber &\lim_{s\to 0}\vare_{\rm T}^{(\rm QBT)}(\omega,p) = \vare_{\rm QBT}(\omega) \\
 \nonumber &+ \frac{15e^2}{32\pi^2\vare_0} \sqrt{\frac{2m^*}{|\mu|}} \int_0^\infty\mbox{d}\hat{q}\  \frac{\hat{n}_F(\hat{q}^2-1)+\hat{n}_F(\hat{q}^2+1)}{\hat{q}^2}\\
 \nonumber &\times \Biggl[ \Bigl(\frac{\hat{\omega}}{\hat{q}}\Bigr)^{-2}+\frac{1}{3}+\frac{8}{5}\Bigl(\frac{\hat{\omega}}{\hat{q}}\Bigr)^2\\
 \label{quapp67} &+\frac{\hat{\omega}}{10\hat{q}}\Bigl[1-8\Bigl(\frac{\hat{\omega}}{\hat{q}}\Bigr)^2\Bigr]\log\Bigl(\frac{\hat{\omega}+\hat{q}+\rmi 0}{\hat{\omega}-\hat{q}+\rmi 0}\Bigr)\Biggr].
\end{align}

\section{Superconducting state response}\label{AppSC}

In this section we apply the general formulas for the superconducting case derived in Sec. \ref{AppProp} to the case of an s-wave singlet superconductor. We confine the discussion to the homogeneous limit and assume $x=\delta=0$.

\subsection{Intraband contribution}
We first derive the response functions for the single bands. For the dispersions of the upper and lower band, respectively, we write
\begin{align}
 \label{sc1} \vare_{\textbf{q}} &= q^2-\mu,\ E_{\textbf{q}} = \sqrt{\vare_{\textbf{q}}^2+\Delta^2},\\
 \label{sc2} f_{\textbf{q}} &= -q^2-\mu,\ F_{\textbf{q}} = \sqrt{f_{\textbf{q}}^2+\Delta^2}.
\end{align}
The order parameter is assumed to be real and we denote the gap amplitude as $|\Delta|=\Delta$. For the upper band we have
\begin{align}
 \nonumber G_{\rm up}(Q)&= \frac{-\rmi q_0 +\vare_q}{q_0^2+E_q^2} \mathbb{1}_2 =\Bigl(\frac{u_q^2}{\rmi q_0+E_q}+\frac{v_q^2}{\rmi q_0-E_q}\Bigr)\mathbb{1}_2,\\
 \nonumber F_{\rm up}(Q) &= \frac{\Delta}{q_0^2+E_q^2} \sigma_2 = u_q v_q \Bigl(\frac{1}{\rmi q_0+E_q}-\frac{1}{\rmi q_0-E_q}\Bigr)\sigma_2,
\end{align}
with the usual factors
\begin{align}
 \label{sc4} u_q^2 = \frac{1}{2}\Bigl( 1+\frac{\vare_q}{E_q}\Bigr),\ v_q^2 = \frac{1}{2}\Bigl( 1-\frac{\vare_q}{E_q}\Bigr).
\end{align}
The propagator in the superconducting states is given by Eq. (\ref{Prop27}), with $\vare_q\to f_q$ for the lower band. The diamagnetic contribution resulting from Eq. (\ref{Lin16b}) for the upper band is given by
\begin{align}
 \nonumber K_{ij}^{(\rm d,upper)}(P) &= - 2\bar{e}^2\delta_{ij}\ \mbox{tr}\int_Q e^{-\rmi q_0\eta} G_{\rm up}(Q)\\
 \label{sc8} &= 4\bar{e}^2\delta_{ij}\int_{\textbf{q}} \Bigl[\frac{1}{2}\Bigl(1-\frac{\vare_q}{E_q}\Bigr) + \frac{\vare_q}{E_q}n_{\rm F}(E_q)\Bigr].
\end{align}
It is of the form $2\bar{e}^2 \delta_{ij}n_{\rm upper}$ with
\begin{align}
\label{sc9} n_{\rm upper} = 2 \int_{\textbf{q}} \Bigl[\frac{1}{2}\Bigl(1-\frac{\vare_q}{E_q}\Bigr) + \frac{\vare_q}{E_q}n_{\rm F}(E_q)\Bigr]
\end{align}
the density of quasiparticles in the upper band. The corresponding diamagnetic term for the lower band reads
\begin{align}
 \label{sc10} K_{ij}^{(\rm d,lower)}(P) &= - 4\bar{e}^2\delta_{ij}\int_{\textbf{q}} \Bigl[\frac{1}{2}\Bigl(1-\frac{f_q}{F_q}\Bigr) + \frac{f_q}{F_q}n_{\rm F}(F_q)\Bigr].
\end{align}
The opposite overall sign compared to Eq. (\ref{sc8}) is due to  $g_{ij}= - \frac{1}{2m}\delta_{ij}\mathbb{1}_2$ for the lower band. For $\Delta\to 0$ we recover the normal state results
\begin{align}
 \label{sc11} K_{ij}^{(\rm d,upper)}(P) &\to 4\bar{e}^2\delta_{ij}\int_{\textbf{q}} n_{\rm F}(q^2-\mu),\\
 \label{sc12} K_{ij}^{(\rm d,lower)}(P) &\to -4\bar{e}^2\delta_{ij}\int_{\textbf{q}} \Bigl[ 1- n_{\rm F}(q^2+\mu)\Bigr],
\end{align}
given in Eqs. (\ref{Cont11}) and (\ref{Cont19}). In both the normal and superconducting state, the lower band contribution has an infinite part $-4\bar{e}^2 \delta_{ij} \int_{\textbf{q}}1$. We already encountered this unphysical divergence in Sec. \ref{SecCont}, and discuss how to treat it at the end of this section.

The density and paramagnetic response functions for the upper band result from Eqs. (\ref{Lin16}) and (\ref{Lin17}). We have
\begin{align}
  \nonumber &\chi_{\rm upper}(P) \\
 \label{sc13} &\hspace{5mm} = -2\bar{e}^2 \int_Q \frac{(-\rmi q_0+\vare_q)[-\rmi (q_0+p_0) +\vare_{\textbf{q}+\textbf{p}}]-\Delta^2}{[(q_0+p_0)^2+E_{\textbf{q}+\textbf{p}}^2](q_0^2+E_q^2)}
\end{align}
and
\begin{align}
 \nonumber K_{ij}^{(\rm p,upper)}(P) &= 2 \bar{e}^2 \int_Q (2q_i+p_i)(2q_j+p_j) \\
 \label{sc14} &\ \times \frac{(-\rmi q_0+\vare_q)[-\rmi (q_0+p_0) +\vare_{\textbf{q}+\textbf{p}}]+\Delta^2}{[(q_0+p_0)^2+E_{\textbf{q}+\textbf{p}}^2](q_0^2+E_q^2)}.
\end{align}
The resulting response functions are real and we can drop the imaginary part in the following considerations. To see this employ the Matsubara sum formula 
\begin{align}
 \label{sc15} T \sum_n \frac{-\rmi q_0 b -\rmi (q_0+p_0)a}{[q_0^2+a^2][(q_0+p_0)^2+b^2]} &= \frac{\rmi p_0[n_{\rm F}(a)-n_{\rm F}(b)]}{p_0^2+(a-b)^2}
\end{align}
and subsequently use that
\begin{align}
 \label{sc16} \int_{\textbf{q}} (\dots) \frac{\rmi p_0}{p_0^2+(E_q-E_{\textbf{q}+\textbf{p}})^2}\Bigl(n_{\rm F}(E_q)-n_{\rm F}(E_{\textbf{q}+\textbf{p}})\Bigr)=0
\end{align}
due to the symmetry of the integrands. We conclude that
\begin{align}
  \label{sc17} \chi_{\rm upper}(P) &= -2\bar{e}^2 \int_Q \frac{-(q_0+p_0)q_0 +\vare_q\vare_{\textbf{q}+\textbf{p}}-\Delta^2}{[(q_0+p_0)^2+E_{\textbf{q}+\textbf{p}}^2](q_0^2+E_q^2)},\\
 \nonumber K_{ij}^{(\rm p,upper)}(P) &= 2 \bar{e}^2 \int_Q (2q_i+p_i)(2q_j+p_j) \\
 \label{sc18} &\ \times \frac{-(q_0+p_0)q_0 +\vare_q\vare_{\textbf{q}+\textbf{p}}+\Delta^2}{[(q_0+p_0)^2+E_{\textbf{q}+\textbf{p}}^2](q_0^2+E_q^2)}.
\end{align}
Analogously, the lower band contributions are given by
\begin{align}
  \label{sc19} \chi_{\rm lower}(P) &= -2\bar{e}^2 \int_Q \frac{-(q_0+p_0)q_0 +f_qf_{\textbf{q}+\textbf{p}}-\Delta^2}{[(q_0+p_0)^2+F_{\textbf{q}+\textbf{p}}^2](q_0^2+F_q^2)},\\
 \nonumber K_{ij}^{(\rm p,lower)}(P) &= 2 \bar{e}^2 \int_Q (2q_i+p_i)(2q_j+p_j) \\
 \label{sc20} &\ \times \frac{-(q_0+p_0)q_0 +f_qf_{\textbf{q}+\textbf{p}}+\Delta^2}{[(q_0+p_0)^2+F_{\textbf{q}+\textbf{p}}^2](q_0^2+F_q^2)}.
\end{align}
These expressions for the single bands result in the usual phenomenology of conventional superconductors.

In the homogeneous limit with nonvanishing frequency $p_0\neq0$ we have
\begin{align}
 \nonumber &K_{ij}^{(\rm p,upper)}(p_0\neq0,0) \\
 \label{sc21} &\ =  \frac{8}{3} \bar{e}^2 \delta_{ij} \int_Q q^2 \frac{-(q_0+p_0)q_0 +E_q^2}{[(q_0+p_0)^2+E_q^2](q_0^2+E_q^2)}=0,\\
\nonumber &K_{ij}^{(\rm p,lower)}(p_0\neq0,0) \\
 \label{sc21b} &\ =  \frac{8}{3} \bar{e}^2 \delta_{ij} \int_Q q^2 \frac{-(q_0+p_0)q_0 +F_q^2}{[(q_0+p_0)^2+F_q^2](q_0^2+F_q^2)}=0,
\end{align}
which is analogous to $K_{ij}^{(\rm p,intra)}(p_0\neq0,0)=0$ in the normal phase, see Eq. (\ref{Cont16}). Here we use the Matsubara sum
\begin{align}
 \label{sc22} T \sum_n \frac{-(q_0+p_0)q_0+a^2}{[(q_0+p_0)^2+a^2](q_0^2+a^2)}=0,
\end{align}
which holds for every bosonic frequency $p_0=2\pi m T\neq 0$. For $p_0=0$ we find
\begin{align}
 \nonumber K_{\rm L,T}^{(\rm p,upper)}(0) &:= \lim_{p\to 0} K_{\rm L,T}^{(\rm p,upper)}(0,p)\\
 \nonumber &= \frac{8}{3} \bar{e}^2 \int_Q q^2 \frac{-q_0^2 +E_q^2}{(q_0^2+E_q^2)^2}\\
 \label{sc23} &= \frac{8}{3} \bar{e}^2 \int_{\textbf{q}} q^2 \frac{\partial}{\partial E_q} n_{\rm F}(E_q),
\end{align}
which vanishes for $T=0$ and reduces to $-2\bar{e}^2n_{\rm upper} $ for $\Delta=0$ (after partial integration). As a result, both the total longitudinal and transverse  current response
\begin{align}
 \label{sc24} K_{\rm L,T}^{(\rm upper)}(0) = K_{\rm L,T}^{(\rm d,upper)}(0) + K_{\rm L,T}^{(\rm p,upper)}(0) 
\end{align}
do not vanish for $\Delta\neq 0$ within RPA. For the lower band contribution we similarly have
\begin{align}
 \label{sc25} K_{\rm L,T}^{(\rm p,lower)}(0)&= \frac{8}{3} \bar{e}^2 \int_{\textbf{q}} q^2 \frac{\partial}{\partial F_q} n_{\rm F}(F_q).
\end{align}

Since $K_{\rm T}^{(\rm p,intra)}(p_0,0)$ vanishes for $p_0\neq 0$ we conclude that the intraband contribution to the homogeneous response function $\sigma(\omega) = \frac{4\pi \rmi \vare_0}{\omega} K_{\rm T}(\omega,0)$ is given entirely by the diamagnetic term. As observed below Eq. (\ref{sc10}) the integral for the lower band contribution is not finite and leads to the same infinite additive contribution $K_{\rm T}^{(\rm d, intra)}(0)\sim -4\bar{e}^2\delta_{ij}\int_{\textbf{q}}1$ as in the normal state. As expounded in detail in Sec. \ref{SecCont}, this infinite constant does not contribute to the optical response function in the normal state, and is absent when calculating $\sigma(\omega)$ from the density response function. The correct interpretation of Eq. (\ref{sc10}) is thus to subtract this infinite term when computing $\sigma(\omega)$ by replacing
\begin{align}
 \label{sc26c} K_{\rm T}^{(\rm d,lower)}(P) &\to - 4\bar{e}^2\int_{\textbf{q}} \Bigl[\frac{1}{2}\Bigl(1-\frac{f_q}{F_q}\Bigr) + \frac{f_q}{F_q}n_{\rm F}(F_q)-1\Bigr],
\end{align}
and the corresponding total intraband contribution for $p_0\neq 0$ reads
\begin{align}
 \nonumber &K_{\rm T}^{(\rm intra)}(p_0,0) = K_{\rm T}^{(\rm d,intra)}(p_0,0)\\
 \label{sc26d} &\to4\bar{e}^2\int_{\textbf{q}} \Bigl(1-\frac{\vare_q}{2E_q}[1-2n_{\rm F}(E_q)]+\frac{f_q}{2F_q}[1-2n_{\rm F}(F_q)]\Bigr).
\end{align}

\subsection{QBT contribution (zero chemical potential)}\label{SecSCQBTmu0}
We analyse the QBT contributions to the optical response in the s-wave singlet superconducting case for $\mu=0$. The vanishing of the chemical potential allows us to discuss the qualitative features and problems of the RPA in a simple case. The gap matrix reads $\hat{\Delta}=\Delta\gamma_{45}$ with $\Delta$ chosen real, and the matrix in Eq. (\ref{Prop30}) is easily inverted for $\mu=0$ to yield the propagator in the parametrization (\ref{Prop11}) with
\begin{align}
 \label{sc27} G(Q) &= \frac{-\rmi q_0\mathbb{1}_4+H}{q_0^2+q^4+\Delta^2},\\
 \label{sc28} F(Q)&=  \hat{F}(Q)= \frac{\Delta\gamma_{45}}{q_0^2+q^4+\Delta^2}.
\end{align}
The density response is then found to be
\begin{align}
 \nonumber &\chi(p_0,p) = -\bar{e}^2 \int_Q \mbox{tr} \Bigl[ G^{Q+P}G^Q - F^{Q+P}\hat{F}^Q\Bigr]\\
 \label{sc29} &=-4\bar{e}^2\int_Q \frac{-(q_0+p_0)q_0+d_a(\textbf{q}+\textbf{p})d_a(\textbf{q}) -\Delta^2}{[(q_0+p_0)^2+(\textbf{q}+\textbf{p})^4+\Delta^2](q_0^2+q^4+\Delta^2)}.
\end{align}
Using Eq. (\ref{Napp4}) and the intraband contributions (they are equal for $\mu=0$) given by
\begin{align}
 \nonumber &\chi_{\rm upper}(p_0,p) = \chi_{\rm lower}(p_0,p) \\
 \label{sc30} &= -2 \bar{e}^2\int_Q \frac{-(q_0+p_0)q_0+(\textbf{q}+\textbf{p})^2q^2-\Delta^2}{[(q_0+p_0)^2+(\textbf{q}+\textbf{p})^4+\Delta^2][q_0^2+q^4+\Delta^2]},
\end{align}
we arrive at the QBT contribution
\begin{align}
 \nonumber &\chi_{\rm QBT}(p_0,p) \\
 \label{sc31} &=6\bar{e}^2\int_Q \frac{q^2p^2-(\textbf{q}\cdot\textbf{p})^2}{[(q_0+p_0)^2+(\textbf{q}+\textbf{p})^4+\Delta^2](q_0^2+q^4+\Delta^2)}.
\end{align}
Importantly we have $\chi_{\rm QBT}(p_0\neq0,0)=0$, in contrast to the intraband expressions. For small $p$ we then have $Z_{\rm QBT}(p_0)=\lim_{p\to 0}\frac{\chi_{\rm QBT}(p_0,p)}{p^2}$ with
\begin{align}
 \nonumber Z_{\rm QBT}(p_0)&=4\bar{e}^2 \int_Q \frac{q^2}{[(q_0+p_0)^2+q^4+\Delta^2](q_0^2+q^4+\Delta^2)}\\
 \label{sc32} &=8\bar{e}^2 \int_Q \frac{q^2}{[p_0^2+4(q^4+\Delta^2)](q_0^2+q^4+\Delta^2)},
\end{align}
where in the last line we assumed $p_0$ to be nonzero.

The total diamagnetic contribution vanishes according to
\begin{align}
  \label{sc35}  K_{ij}^{(\rm d)}(P) &= - \sqrt{3} \bar{e}^2 \Lambda^a_{ij}\ \mbox{tr}\int_Q e^{-\rmi q_0 \eta} \gamma_a G(Q)&= 0.
\end{align}
The single band diamagnetic contributions for $\mu=0$ read
\begin{align}
 \label{sc36} K_{\rm L,T}^{(\rm d,upper)}(P) &= 4\bar{e}^2\int_{\textbf{q}} \Bigl[\frac{1}{2}\Bigl(1-\frac{q^2}{E_q}\Bigr) + \frac{q^2}{E_q}n_{\rm F}(E_q)\Bigr],\\
 \label{sc37} K_{\rm L,T}^{(\rm d,lower)}(P) &= - 4\bar{e}^2\int_{\textbf{q}} \Bigl[\frac{1}{2}\Bigl(1+\frac{q^2}{E_q}\Bigr) - \frac{q^2}{E_q}n_{\rm F}(E_q)\Bigr]
\end{align}
with $E_q=F_q=\sqrt{q^4+\Delta^2}$, and so the intraband diamagnetic contribution is given by
\begin{align}
\nonumber K_{\rm L,T}^{(\rm d,intra)}(P) &= -4\bar{e}^2\int_{\textbf{q}}\frac{q^2}{\sqrt{q^4+\Delta^2}}[1-2n_{\rm F}(\sqrt{q^4+\Delta^2})]\\
 \label{sc38} &=-8 \bar{e}^2\int_Q \frac{q^2}{q_0^2+q^4+\Delta^2}.
\end{align}
We deduce the diamagnetic QBT contribution to be
\begin{align}
\label{sc39} K_{\rm L,T}^{(\rm d,QBT)}(P) =8 \bar{e}^2\int_Q \frac{q^2}{q_0^2+q^4+\Delta^2}.
\end{align}

In order to compute the paramagnetic QBT contribution recall that in the homogeneous limit ($p=0, p_0\neq0$) we have
\begin{align}
 \label{sc40} K_{\rm L}^{(\rm QBT)}(p_0,0) = K_{\rm T}^{(\rm QBT)}(p_0,0)
\end{align}
Using the same manipulations as in Eqs. (\ref{Cont29})-(\ref{Cont35}) we obtain
\begin{align}
 \nonumber K_{\rm L,T}^{(\rm p)}(p_0,0) &=\frac{1}{3}\delta_{ij}K_{ij}^{(\rm p)}(p_0,0)\\
 \nonumber &= \frac{1}{4} \bar{e}^2\delta_{ij} \Lambda_{jl}^a\Lambda_{ik}^b \int_Q 4q_kq_l\mbox{tr}\Bigl[ G^{Q+p_0}\gamma_aG^Q \gamma_b \\
 \nonumber &+ F^{Q+p_0} (\gamma_a)^T \hat{F}^Q \gamma_b\Bigr]\\
 \nonumber &=\bar{e}^2\Bigl(\frac{2}{3}\delta_{ab}\delta_{kl}+\frac{1}{2}J_{abc}\Lambda^c_{kl}\Bigr) \int_Q q_k q_l \\
 \label{sc42} &\times \mbox{tr}\Bigl[ G^{Q+p_0}\gamma_aG^Q \gamma_b + F^{Q+p_0} (\gamma_a)^T \hat{F}^Q \gamma_b\Bigr]
\end{align}
with
\begin{align}
 \nonumber & \mbox{tr}\Bigl[ G^{Q+p_0}\gamma_aG^Q \gamma_b + F^{Q+p_0} (\gamma_a)^T \hat{F}^Q \gamma_b\Bigr] \\
 \label{sc43} &= 4\frac{[-(q_0+p_0)q_0-q^4+\Delta^2]\delta_{ab} +2d_a(\textbf{q})d_b(\textbf{q})}{[(q_0+p_0)^2+q^4+\Delta^2](q_0^2+q^4+\Delta^2)}.
\end{align}
We used $\gamma_a^T=\gamma_{45}\gamma_a\gamma_{45}$. After angular integration we then arrive at
\begin{align}
 \nonumber &K_{\rm L,T}^{(\rm p)}(p_0,0) \\
 \label{sc44} &=4\bar{e}^2 \int_Q q^2 \frac{-\frac{10}{3}(q_0+p_0)q_0-\frac{2}{3}q^4+\frac{10}{3}\Delta^2}{[(q_0+p_0)^2+q^4+\Delta^2](q_0^2+q^4+\Delta^2)}.
\end{align}
We subtract the intraband contributions
\begin{align}
 \nonumber &K_{\rm L,T}^{(\rm p,upper)}(p_0,0) =  K_L^{(\rm p,lower)}(p_0,0)\\
 \label{sc45} &=\frac{8}{3}\bar{e}^2\int_Q q^2 \frac{-(q_0+p_0)q_0+q^4+\Delta^2}{[(q_0+p_0)^2+q^4+\Delta^2](q_0^2+q^4+\Delta^2)},
\end{align}
which again are identical, and arrive at
\begin{align}
 \nonumber &K_{\rm L,T}^{(p,\rm QBT)}(p_0,0) \\
 \label{sc46} &= 8\bar{e}^2\int_Q q^2\frac{-(q_0+p_0)q_0-q^4+\Delta^2}{[(q_0+p_0)^2+q^4+\Delta^2](q_0^2+q^4+\Delta^2)}.
\end{align}
This result holds for both $p_0\neq0$ and $p_0=0$ as we have not yet evaluated the Matsubara sum of the loop. For $p_0\neq0$ we can use Eq. (\ref{sc22}) to arrive at
\begin{align}
 \nonumber &K_{\rm L,T}^{(p,\rm QBT)}(p_0\neq0,0) \\
 \nonumber &= -16\bar{e}^2\int_Q \frac{q^6}{[(q_0+p_0)^2+q^4+\Delta^2](q_0^2+q^4+\Delta^2)}\\
 \label{sc47} &=-32\bar{e}^2\int_Q \frac{q^6}{[p_0^2+4(q^4+\Delta^2)](q_0^2+q^4+\Delta^2)},
\end{align}
and so
\begin{align}
  \nonumber K_{\rm L,T}^{(\rm QBT)}(p_0\neq0,0) &= 8\bar{e}^2\int_{\textbf{q}} \frac{p_0^2+4\Delta^2}{p_0^2+4(q^4+\Delta^2)} \frac{q^2}{q_0^2+q^4+\Delta^2}\\
 \label{sc47b} &=(p_0^2+4\Delta^2) Z_{\rm QBT}(p_0,0).
\end{align}
In the last line we compare the expression with Eq. (\ref{sc32}) obtained from the density response. Setting $p_0=0$ in Eq. (\ref{sc46}), on the other hand, we obtain
\begin{align}
 \nonumber &K_{\rm L,T}^{(\rm p,QBT)}(0) = 8\bar{e}^2\int_Q q^2\frac{-q_0^2-q^4+\Delta^2}{(q_0^2+q^4+\Delta^2)^2}\\
 \nonumber &=-8\bar{e}^2\int_Q \frac{q^2}{q_0^2+q^4+\Delta^2}+16\bar{e}^2\Delta^2\int_Q \frac{q^2}{(q_0^2+q^4+\Delta^2)^2}\\
 \label{sc48} &= -K_{\rm L,T}^{(\rm d,QBT)}(0)+16\bar{e}^2\Delta^2\int_Q \frac{q^2}{(q_0^2+q^4+\Delta^2)^2}.
\end{align}
This implies that the total QBT contribution $K_{\rm L,T}^{(\rm QBT)}(0)>0$ for $\Delta\neq0$, implying a nonzero contribution to the superfluid density. Furthermore, evaluating the Matsubara summations we find
\begin{align}
 \nonumber K_{\rm T}^{(\rm QBT)}(p_0\neq0,0) &= 4\bar{e}^2\int_{\textbf{q}} q^2 \frac{(p_0^2+4\Delta^2)[1-2n_{\rm F}(E_q)]}{E_q(p_0^2+4E_q^2)} ,\\
 \nonumber K_{\rm T}^{(\rm QBT)}(0)&= 4\bar{e}^2\Delta^2 \int_{\textbf{q}} q^2 \Bigl(\frac{1}{E_q^3}[1-2n_{\rm F}(E_q)]\\
 \label{sc48b} &{}+\frac{2}{E_q^2}\frac{\partial}{\partial E_q}n_{\rm F}(E_q)\Bigr),
\end{align}
and so
\begin{align}
 \label{sc48c} \lim_{p_0\to 0} K_{\rm T}^{(\rm QBT)}(p_0,0) \neq K_{\rm T}^{(\rm QBT)}(0)\ \text{for }\mu=0.
\end{align}
Compare this to the contrary result in Eq. (\ref{sc71}) for finite chemical potential.

\subsection{QBT contribution (finite chemical potential)}\label{SecSCQBT}
Let us now turn to the superconducting response for finite chemical potential. We restrict the analysis to the transverse  component of the current response function. The total  paramagnetic response from Eq. (\ref{Lin18}) for $p=0$ reads
\begin{align}
 \nonumber K_{\rm T}^{(\rm p)}(p_0,0) ={}& \frac{1}{3}\delta_{ij}K_{ij}^{(\rm p)}(p_0,0) = \bar{e}^2\ \Lambda^a_{li}\Lambda^b_{ik}\int_{Q}q_kq_l\\
 \label{sc49} &\times \mbox{tr} \Bigl[ G^{Q+p_0}\gamma_aG^Q \gamma_b+ F^{Q+p_0} (\gamma_a)^T \hat{F}^Q \gamma_b\Bigr].
\end{align}
We use the shorthand $Q+p_0=(q_0+p_0,\textbf{q})$. The propagator  derived from Eqs. (\ref{Prop35}) is given by
\begin{align}
 \label{sc50} G^Q &= M^Q [-(\rmi q_0+\mu)\mathbb{1}+H],\\
 \label{sc51} F^Q &= \Delta M^Q \gamma_{45},\\
 \label{sc52} \hat{F}^Q &= (F^Q)^\dagger = \Delta \gamma_{45} M^Q,
\end{align}
where we introduce
\begin{align}
 \label{sc53} M^Q &= \frac{X^Q\mathbb{1}+2\mu H}{(q_0^2+E_q^2)(q_0^2+F_q^2)},\\
 \label{sc54} X^Q &= q_0^2+q^4+\mu^2+\Delta^2.
\end{align}
We also write
\begin{align*}
 D_{\rm up}^Q &= q_0^2+E_q^2,\\
 D_{\rm low}^Q &= q_0^2+F_q^2
\end{align*}
for the denominators of the single band propagators. We only need to evaluate the real part of the diagram as the imaginary part vanishes upon integration over $Q$. We have
\begin{widetext}
\begin{align}
  \nonumber \text{Re}\ \mbox{tr}(G^{Q+p_0}\gamma_aG^Q\gamma_b) &= [-(q_0+p_0)q_0+\mu^2]\mbox{tr}(M^{Q+p_0}\gamma_aM^Q\gamma_b) -\mu\Bigl[\mbox{tr}(M^{Q+p_0}H\gamma_a M^Q\gamma_b)+\mbox{tr}(M^{Q+p_0}\gamma_a M^QH\gamma_b)\Bigr]\\
 \label{sc55} &+ \mbox{tr}(M^{Q+p_0}H\gamma_a M^QH\gamma_b),\\
 \label{sc56} \mbox{tr}(F^{Q+p_0}(\gamma_a)^T\hat{F}^Q\gamma_b) &= \Delta^2\ \mbox{tr}(M^{Q+p_0}\gamma_aM^Q\gamma_b),
\end{align} 
and these traces can be evaluated according to
\begin{align}
  \label{sc57} \mbox{tr}(M^{Q+p_0}\gamma_aM^Q\gamma_b) &= \frac{1}{D_{\rm up}^{Q+p_0}D_{\rm up}^Q D_{\rm low}^{Q+p_0}D_{\rm low}^Q}\Bigl[ X^{Q+p_0}X^Q\mbox{tr}(\gamma_a\gamma_b)+4\mu^2 \mbox{tr}(H\gamma_aH\gamma_b)\Bigr],\\
 \label{sc58} \mbox{tr}(M^{Q+p_0}H\gamma_a M^Q\gamma_b)&= \frac{1}{D_{\rm up}^{Q+p_0}D_{\rm up}^Q D_{\rm low}^{Q+p_0}D_{\rm low}^Q}2\mu \Bigl[ X^Q q^4 \mbox{tr}(\gamma_a\gamma_b)+X^{Q+p_0}\mbox{tr}(H\gamma_a H\gamma_b)\Bigr],\\
 \label{sc59} \mbox{tr}(M^{Q+p_0}\gamma_a M^QH\gamma_b) &= \frac{1}{D_{\rm up}^{Q+p_0}D_{\rm up}^Q D_{\rm low}^{Q+p_0}D_{\rm low}^Q}2\mu \Bigl[ X^{Q+p_0} q^4 \mbox{tr}(\gamma_a\gamma_b)+X^Q\mbox{tr}(H\gamma_a H\gamma_b)\Bigr],\\
 \label{sc60} \mbox{tr}(M^{Q+p_0}H\gamma_a M^QH\gamma_b) &= \frac{1}{D_{\rm up}^{Q+p_0}D_{\rm up}^Q D_{\rm low}^{Q+p_0}D_{\rm low}^Q}\Bigl[ 4\mu^2 q^8 \mbox{tr}(\gamma_a\gamma_b) + X^{Q+p_0}X^Q \mbox{tr}(H\gamma_a H \gamma_b)\Bigr],
\end{align}
and subsequent Clifford algebra for the traces on the right hand side. The remaining contractions and angular integrations can be performed by means of
\begin{align}
 \label{sc61} \Lambda^a_{li}\Lambda^b_{ik} q_k q_l \mbox{tr}(\gamma_a\gamma_b) &= \frac{40}{3} q^2,\\
 \label{sc62} \Lambda^a_{li}\Lambda^b_{ik}\int_{\textbf{q}}q_kq_l\ \mbox{tr}(H\gamma_a H \gamma_b)\ f(q^2) &=-\frac{8}{3}\int_{\textbf{q}} q^6\ f(q^2),
\end{align}
where $f(q^2)$ is some function, and so we arrive at
\begin{align}
 \nonumber K_{\rm T}^{(\rm p)}(p_0,0) &= \frac{8}{3}\bar{e}^2\int_{Q}q^2 \frac{1}{D_{\rm up}^{Q+p_0}D_{\rm up}^Q D_{\rm low}^{Q+p_0}D_{\rm low}^Q} \Biggl( \Bigl[-5(q_0+p_0)q_0+5\mu^2+5\Delta^2- q^4\Bigr]X^{Q+p_0}X^Q\\
 \label{sc64} &-8\mu^2 q^4(X^{Q+p_0}+X^Q) +4\mu^2q^4\Bigl[(q_0+p_0)q_0-\mu^2-\Delta^2+5q^4\Bigr]\Biggr).
\end{align}
Now we subtract the single band contributions given by
\begin{align}
 \label{sc65} K_{\rm T}^{(\rm p,upper)}(p_0,0) &= \frac{8}{3}\bar{e}^2 \int_Q q^2 \frac{-(q_0+p_0)q_0+E_q^2}{D_{\rm up}^{Q+p_0}D_{\rm up}^Q},\\
 \label{sc66} K_{\rm T}^{(\rm p,lower)}(p_0,0) &= \frac{8}{3}\bar{e}^2 \int_Q q^2 \frac{-(q_0+p_0)q_0+F_q^2}{D_{\rm low}^{Q+p_0}D_{\rm low}^Q},
\end{align}
which leaves us with the superconductor QBT contribution
\begin{align}
 \nonumber K_{\rm T}^{(\rm p,QBT)}(p_0,0) &= 4\bar{e}^2\int_{Q}q^2\Bigl[-(q_0+p_0)q_0-q^4+\mu^2+\Delta^2\Bigr]\Bigl(\frac{1}{D_{\rm up}^{Q+p_0}D_{\rm low}^Q}+\frac{1}{D_{\rm low}^{Q+p_0}D_{\rm up}^Q}\Bigr)\\
 \nonumber &+\frac{16}{3}\mu\bar{e}^2\int_{Q}q^4 \Bigl( \frac{1}{D_{\rm up}^{Q+p_0}D_{\rm up}^Q}-\frac{1}{D_{\rm low}^{Q+p_0}D_{\rm low}^Q}\Bigr)\\
 \label{sc67} &-\frac{64}{3}\mu^2\bar{e}^2\int_{Q}q^6 \frac{(q_0+p_0)^2+q_0^2+2(q^4+\mu^2+\Delta^2)}{D_{\rm up}^{Q+p_0}D_{\rm up}^QD_{\rm low}^{Q+p_0}D_{\rm low}^Q}.
\end{align}
Note that since we have not yet performed the Matsubara summation in this expression, we can use this formula to compute both $K_{\rm T}^{(\rm p,QBT)}(p_0\neq0,0)$ and $K_{\rm T}^{(\rm p,QBT)}(0)$. In the latter case we have to set $p_0=0$ in the integral before performing the Matsubara summation. It is then easy to verify that for \emph{both} $p_0=2\pi mT\neq0$ and $p_0=0$ we have the following two identities of sums:
\begin{align}
 \label{sc68} T\sum_n \frac{-(q_0+p_0)q_0-q^4+\mu^2+\Delta^2}{D_{\rm up}^{Q+p_0}D_{\rm low}^Q}&= T \sum_n \frac{-(q_0+p_0)q_0-q^4+\mu^2+\Delta^2}{D_{\rm low}^{Q+p_0}D_{\rm up}^Q},\\
 \label{sc69} T \sum_n \frac{(q_0+p_0)^2+q_0^2+2(q^4+\mu^2+\Delta^2)}{D_{\rm up}^{Q+p_0}D_{\rm up}^QD_{\rm low}^{Q+p_0}D_{\rm low}^Q} &= \frac{1}{4q^2\mu}\ T \sum_n\Bigl( \frac{1}{D_{\rm up}^{Q+p_0}D_{\rm up}^Q}-\frac{1}{D_{\rm low}^{Q+p_0}D_{\rm low}^Q}\Bigr).
\end{align}
Whereas the first identity implies that we can simplify the first line in Eq. (\ref{sc67}), the second identity implies that the second and third line in Eq. (\ref{sc67}) cancel. We then arrive at the conveniently short expression
\begin{align}
 \label{sc70} K_{\rm T}^{(\rm p,QBT)}(p_0,0) &= 8\bar{e}^2\int_{Q}q^2\frac{-(q_0+p_0)q_0-q^4+\mu^2+\Delta^2}{D_{\rm up}^{Q+p_0}D_{\rm low}^Q},
\end{align}
\end{widetext}
which constitutes the main result of this section.

The simple expression for $K_{\rm T}^{(\rm p,QBT)}(p_0,0)$ obtained in the previous paragraph allows us to compute the QBT contribution to the superfluid density and optical conductivity. Since the total diamagnetic contribution also vanishes for $\mu\neq 0$ (due to a similar argument as in Eq. (\ref{sc35}) for $\mu=0$) we have
\begin{align}
 \nonumber K_{\rm T}^{(\rm d,QBT)}(0) &= - K_{\rm T}^{(\rm d,intra)}(0)\\
 \nonumber &=  4\bar{e}^2\int_Q \Big(\frac{q^2-\mu}{D_{\rm up}^Q}+\frac{q^2+\mu}{D_{\rm low}^Q}\Bigr)\\
 \nonumber &=2 \bar{e}^2 \int_{\textbf{q}} \Bigl( \frac{q^2-\mu}{E_q}[1-2n_{\rm F}(E_q)]\\
 \label{sc72} &+\frac{q^2+\mu}{F_q}[1-2n_{\rm F}(F_q)]\Bigr).
\end{align}
The paramagnetic contribution to the $p\to 0$ limit, on the other hand, is given by
\begin{align}
 \nonumber K_{\rm T}^{(\rm p,QBT)}(0) &= 8\bar{e}^2\int_{Q}q^2\frac{-q_0^2-q^4+\mu^2+\Delta^2}{D_{\rm up}^{Q}D_{\rm low}^Q}\\
 \nonumber &=-2\bar{e}^2\int_{\textbf{q}} \Bigl( \frac{q^2-\mu-\frac{\Delta^2}{\mu}}{E_q}[1-2n_{\rm F}(E_q)]\\
 \label{sc73} &+\frac{q^2+\mu+\frac{\Delta^2}{\mu}}{F_q}[1-2n_{\rm F}(F_q)]\Bigr).
\end{align}
Note that, like in the normal phase, both expressions individually contain ultraviolet divergent momentum integrals but their sum is finite. Hence the expressions again need to be understood with a finite but large momentum cutoff. Equations (\ref{sc72}) and (\ref{sc73}) reveal that the diamagnetic term is nullified by parts of the paramagnetic term and the remainder is given by
\begin{align}
 \nonumber K_{\rm T}^{(\rm QBT)}(0) &= \frac{2\Delta^2\bar{e}^2}{\mu}\int_{\textbf{q}} \Bigl(  \frac{1}{E_q}[1-2n_{\rm F}(E_q)]\\
 \label{sc74} &-\frac{1}{F_q}[1-2n_{\rm F}(F_q)]\Bigr)\\
 \nonumber &=: 2 \bar{e}^2n_{\rm s}^{(\rm QBT)}
\end{align}
with QBT contribution to the superfluid density
\begin{align}
 \label{sc75} n_{\rm s}^{(\rm QBT)}=\frac{\Delta^2}{\mu} \int_{\textbf{q}} \Bigl(  \frac{1}{E_q}[1-2n_{\rm F}(E_q)]-\frac{1}{F_q}[1-2n_{\rm F}(F_q)]\Bigr).
\end{align}
This expression remains finite for vanishing chemical potential (since $E_q-F_q\to 0$ in this limit) and we have
\begin{align}
 \label{sc76} n_{\rm s}^{(\rm QBT)}\to2\Delta^2 \int_{\textbf{q}} q^2\Bigl[ \frac{1}{E_q^3}[1-2n_{\rm F}(E_q)] +\frac{2}{E_q^2} \frac{\partial}{\partial E_q}n_{\rm F}(E_q)\Bigr]
\end{align}
for $\mu\to 0$, with $E_q=\sqrt{q^4+\Delta^2}$, see Eq. (\ref{sc48b}).

To compute the conductivity we evaluate Eq. (\ref{sc70}) for arbitrary $p_0$ which yields
\begin{widetext}
\begin{align}
 \nonumber K_{\rm T}^{(\rm p,QBT)}(p_0,0) &=8 \bar{e}^2 \int_{\textbf{q}} \frac{q^2}{p_0^4+4p_0^2(q^4+\mu^2+\Delta^2)+16\mu^2q^4} \Biggl[ \Bigl(-p_0^2q^2(q^2-\mu)+4\mu q^2 (\mu^2+\Delta^2-\mu q^2)\Bigr)\frac{1}{E_q}[1-2n_{\rm F}(E_q)]\\
 \label{sc77} &+ \Bigl(-p_0^2q^2(q^2+\mu)-4\mu q^2 (\mu^2+\Delta^2+\mu q^2)\Bigr) \frac{1}{F_q}[1-2n_{\rm F}(F_q)]\Biggr].
\end{align}
Combining this with the diamagnetic contribution from Eq. (\ref{sc72}) we obtain
\begin{align}
 \nonumber K_{\rm T}^{(\rm QBT)}(p_0,0) &=2\bar{e}^2\int_{\textbf{q}} \frac{1}{p_0^4+4p_0^2(q^4+\mu^2+\Delta^2)+16\mu^2q^4}\\
 \nonumber &\times \Biggl[\Bigl(p_0^4(q^2-\mu)+4p_0^2(q^2-\mu)(\mu^2+\Delta^2)+16\mu q^4\Delta^2\Bigr)\frac{1}{E_q}[1-2n_{\rm F}(E_q)]\\
 \label{sc78} &+\Bigl(p_0^4(q^2+\mu)+4p_0^2(q^2+\mu)(\mu^2+\Delta^2)-16\mu q^4\Delta^2\Bigr)\frac{1}{F_q}[1-2n_{\rm F}(F_q)]\Biggr].
\end{align}
\end{widetext}
 In particular we verify that
\begin{align}
 \label{sc71}  \lim_{p_0\to 0} K_{\rm T}^{(\rm p,QBT)}(p_0,0)=K_{\rm T}^{(\rm p,QBT)}(0)\ \text{for }\mu\neq0.
\end{align}
After analytic continuation the conductivity is given by
\begin{align}
 \label{sc79} \sigma^{(\rm QBT)}(\omega) = -\frac{4\pi\eps_0}{\rmi (\omega+\rmi 0) } K_{\rm T}^{(\rm QBT)}(\omega,0).
\end{align}
For small $\omega$, the conductivity diverges like
\begin{align}
 \nonumber \sigma^{(\rm QBT)}(\omega\to 0 ) &=4\pi\rmi \eps_0\Bigl(\frac{1}{\omega}-\frac{\pi}{2}\rmi \delta(\omega)\Bigr)\lim_{\omega\to 0} K_{\rm T}^{(\rm QBT)}(\omega,0)\\
 \label{sc80}&=: \Bigl(\frac{\pi}{2}\delta(\omega)+\rmi \frac{1}{\omega}\Bigr)\frac{ n'^{(\rm QBT)} e^2}{m^*}
\end{align}
with 
\begin{align}
 \label{sc81} n'^{(\rm QBT)} &= \frac{4\pi \eps_0m^*}{\bar{e}^2} \lim_{\omega \to 0}K_{\rm T}^{(\rm QBT)}(\omega,0) = n_{\rm s}^{(\rm QBT)},
\end{align}
where we used Eq. (\ref{sc71}).

\end{appendix}

\bibliography{refs_opt}

\begin{thebibliography}{63}%
\makeatletter
\providecommand \@ifxundefined [1]{%
 \@ifx{#1\undefined}
}%
\providecommand \@ifnum [1]{%
 \ifnum #1\expandafter \@firstoftwo
 \else \expandafter \@secondoftwo
 \fi
}%
\providecommand \@ifx [1]{%
 \ifx #1\expandafter \@firstoftwo
 \else \expandafter \@secondoftwo
 \fi
}%
\providecommand \natexlab [1]{#1}%
\providecommand \enquote  [1]{``#1''}%
\providecommand \bibnamefont  [1]{#1}%
\providecommand \bibfnamefont [1]{#1}%
\providecommand \citenamefont [1]{#1}%
\providecommand \href@noop [0]{\@secondoftwo}%
\providecommand \href [0]{\begingroup \@sanitize@url \@href}%
\providecommand \@href[1]{\@@startlink{#1}\@@href}%
\providecommand \@@href[1]{\endgroup#1\@@endlink}%
\providecommand \@sanitize@url [0]{\catcode `\\12\catcode `\$12\catcode
  `\&12\catcode `\#12\catcode `\^12\catcode `\_12\catcode `\%12\relax}%
\providecommand \@@startlink[1]{}%
\providecommand \@@endlink[0]{}%
\providecommand \url  [0]{\begingroup\@sanitize@url \@url }%
\providecommand \@url [1]{\endgroup\@href {#1}{\urlprefix }}%
\providecommand \urlprefix  [0]{URL }%
\providecommand \Eprint [0]{\href }%
\providecommand \doibase [0]{http://dx.doi.org/}%
\providecommand \selectlanguage [0]{\@gobble}%
\providecommand \bibinfo  [0]{\@secondoftwo}%
\providecommand \bibfield  [0]{\@secondoftwo}%
\providecommand \translation [1]{[#1]}%
\providecommand \BibitemOpen [0]{}%
\providecommand \bibitemStop [0]{}%
\providecommand \bibitemNoStop [0]{.\EOS\space}%
\providecommand \EOS [0]{\spacefactor3000\relax}%
\providecommand \BibitemShut  [1]{\csname bibitem#1\endcsname}%
\let\auto@bib@innerbib\@empty
\bibitem [{\citenamefont {Qi}\ and\ \citenamefont
  {Zhang}(2011)}]{RevModPhys.83.1057}%
  \BibitemOpen
  \bibfield  {author} {\bibinfo {author} {\bibfnamefont {Xiao-Liang}\
  \bibnamefont {Qi}}\ and\ \bibinfo {author} {\bibfnamefont {Shou-Cheng}\
  \bibnamefont {Zhang}},\ }\bibfield  {title} {\enquote {\bibinfo {title}
  {Topological insulators and superconductors},}\ }\href {\doibase
  10.1103/RevModPhys.83.1057} {\bibfield  {journal} {\bibinfo  {journal} {Rev.
  Mod. Phys.}\ }\textbf {\bibinfo {volume} {83}},\ \bibinfo {pages}
  {1057--1110} (\bibinfo {year} {2011})}\BibitemShut {NoStop}%
\bibitem [{\citenamefont {Witczak-Krempa}\ \emph {et~al.}(2014)\citenamefont
  {Witczak-Krempa}, \citenamefont {Chen}, \citenamefont {Kim},\ and\
  \citenamefont {Balents}}]{WKrempa}%
  \BibitemOpen
  \bibfield  {author} {\bibinfo {author} {\bibfnamefont {William}\ \bibnamefont
  {Witczak-Krempa}}, \bibinfo {author} {\bibfnamefont {Gang}\ \bibnamefont
  {Chen}}, \bibinfo {author} {\bibfnamefont {Yong~Baek}\ \bibnamefont {Kim}}, \
  and\ \bibinfo {author} {\bibfnamefont {Leon}\ \bibnamefont {Balents}},\
  }\bibfield  {title} {\enquote {\bibinfo {title} {{Correlated Quantum
  Phenomena in the Strong Spin-Orbit Regime}},}\ }\href {\doibase
  10.1146/annurev-conmatphys-020911-125138} {\bibfield  {journal} {\bibinfo
  {journal} {Annu. Rev. Condens. Matter Phys.}\ }\textbf {\bibinfo {volume}
  {5}},\ \bibinfo {pages} {57--82} (\bibinfo {year} {2014})}\BibitemShut
  {NoStop}%
\bibitem [{\citenamefont {Smidman}\ \emph {et~al.}(2017)\citenamefont
  {Smidman}, \citenamefont {Salamon}, \citenamefont {Yuan},\ and\ \citenamefont
  {Agterberg}}]{SmidmanReview}%
  \BibitemOpen
  \bibfield  {author} {\bibinfo {author} {\bibfnamefont {M}~\bibnamefont
  {Smidman}}, \bibinfo {author} {\bibfnamefont {M~B}\ \bibnamefont {Salamon}},
  \bibinfo {author} {\bibfnamefont {H~Q}\ \bibnamefont {Yuan}}, \ and\ \bibinfo
  {author} {\bibfnamefont {D~F}\ \bibnamefont {Agterberg}},\ }\bibfield
  {title} {\enquote {\bibinfo {title} {{Superconductivity and spin–orbit
  coupling in non-centrosymmetric materials: a review}},}\ }\href
  {http://stacks.iop.org/0034-4885/80/i=3/a=036501} {\bibfield  {journal}
  {\bibinfo  {journal} {Rep. Prog. Phys.}\ }\textbf {\bibinfo {volume} {80}},\
  \bibinfo {pages} {036501} (\bibinfo {year} {2017})}\BibitemShut {NoStop}%
\bibitem [{\citenamefont {{Chadov}}\ \emph {et~al.}(2010)\citenamefont
  {{Chadov}}, \citenamefont {{Qi}}, \citenamefont {{K{\"u}bler}}, \citenamefont
  {{Fecher}}, \citenamefont {{Felser}},\ and\ \citenamefont
  {{Zhang}}}]{ChadovNatMater}%
  \BibitemOpen
  \bibfield  {author} {\bibinfo {author} {\bibfnamefont {S.}~\bibnamefont
  {{Chadov}}}, \bibinfo {author} {\bibfnamefont {X.}~\bibnamefont {{Qi}}},
  \bibinfo {author} {\bibfnamefont {J.}~\bibnamefont {{K{\"u}bler}}}, \bibinfo
  {author} {\bibfnamefont {G.~H.}\ \bibnamefont {{Fecher}}}, \bibinfo {author}
  {\bibfnamefont {C.}~\bibnamefont {{Felser}}}, \ and\ \bibinfo {author}
  {\bibfnamefont {S.~C.}\ \bibnamefont {{Zhang}}},\ }\bibfield  {title}
  {\enquote {\bibinfo {title} {{Tunable multifunctional topological insulators
  in ternary Heusler compounds}},}\ }\href {\doibase 10.1038/nmat2770}
  {\bibfield  {journal} {\bibinfo  {journal} {Nat. Mater.}\ }\textbf {\bibinfo
  {volume} {9}},\ \bibinfo {pages} {541--545} (\bibinfo {year}
  {2010})}\BibitemShut {NoStop}%
\bibitem [{\citenamefont {{Cheng}}\ \emph {et~al.}(2017)\citenamefont
  {{Cheng}}, \citenamefont {{Ohtsuki}}, \citenamefont {{Chaudhuri}},
  \citenamefont {{Nakatsuji}}, \citenamefont {{Lippmaa}},\ and\ \citenamefont
  {{Armitage}}}]{2017NatCo...8.2097C}%
  \BibitemOpen
  \bibfield  {author} {\bibinfo {author} {\bibfnamefont {Bing}\ \bibnamefont
  {{Cheng}}}, \bibinfo {author} {\bibfnamefont {T.}~\bibnamefont {{Ohtsuki}}},
  \bibinfo {author} {\bibfnamefont {Dipanjan}\ \bibnamefont {{Chaudhuri}}},
  \bibinfo {author} {\bibfnamefont {S.}~\bibnamefont {{Nakatsuji}}}, \bibinfo
  {author} {\bibfnamefont {Mikk}\ \bibnamefont {{Lippmaa}}}, \ and\ \bibinfo
  {author} {\bibfnamefont {N.~P.}\ \bibnamefont {{Armitage}}},\ }\bibfield
  {title} {\enquote {\bibinfo {title} {{Dielectric anomalies and interactions
  in the three-dimensional quadratic band touching Luttinger semimetal
  Pr$_{2}$Ir$_{2}$O$_{7}$}},}\ }\href {\doibase 10.1038/s41467-017-02121-y}
  {\bibfield  {journal} {\bibinfo  {journal} {Nat. Commun.}\ }\textbf {\bibinfo
  {volume} {8}},\ \bibinfo {eid} {2097} (\bibinfo {year} {2017})}\BibitemShut
  {NoStop}%
\bibitem [{\citenamefont {Kim}\ \emph {et~al.}(2018)\citenamefont {Kim},
  \citenamefont {Wang}, \citenamefont {Nakajima}, \citenamefont {Hu},
  \citenamefont {Ziemak}, \citenamefont {Syers}, \citenamefont {Wang},
  \citenamefont {Hodovanets}, \citenamefont {Denlinger}, \citenamefont
  {Brydon}, \citenamefont {Agterberg}, \citenamefont {Tanatar}, \citenamefont
  {Prozorov},\ and\ \citenamefont {Paglione}}]{Kimeaao4513}%
  \BibitemOpen
  \bibfield  {author} {\bibinfo {author} {\bibfnamefont {Hyunsoo}\ \bibnamefont
  {Kim}}, \bibinfo {author} {\bibfnamefont {Kefeng}\ \bibnamefont {Wang}},
  \bibinfo {author} {\bibfnamefont {Yasuyuki}\ \bibnamefont {Nakajima}},
  \bibinfo {author} {\bibfnamefont {Rongwei}\ \bibnamefont {Hu}}, \bibinfo
  {author} {\bibfnamefont {Steven}\ \bibnamefont {Ziemak}}, \bibinfo {author}
  {\bibfnamefont {Paul}\ \bibnamefont {Syers}}, \bibinfo {author}
  {\bibfnamefont {Limin}\ \bibnamefont {Wang}}, \bibinfo {author}
  {\bibfnamefont {Halyna}\ \bibnamefont {Hodovanets}}, \bibinfo {author}
  {\bibfnamefont {Jonathan~D.}\ \bibnamefont {Denlinger}}, \bibinfo {author}
  {\bibfnamefont {Philip M.~R.}\ \bibnamefont {Brydon}}, \bibinfo {author}
  {\bibfnamefont {Daniel~F.}\ \bibnamefont {Agterberg}}, \bibinfo {author}
  {\bibfnamefont {Makariy~A.}\ \bibnamefont {Tanatar}}, \bibinfo {author}
  {\bibfnamefont {Ruslan}\ \bibnamefont {Prozorov}}, \ and\ \bibinfo {author}
  {\bibfnamefont {Johnpierre}\ \bibnamefont {Paglione}},\ }\bibfield  {title}
  {\enquote {\bibinfo {title} {{Beyond triplet: Unconventional
  superconductivity in a spin-3/2 topological semimetal}},}\ }\href {\doibase
  10.1126/sciadv.aao4513} {\bibfield  {journal} {\bibinfo  {journal} {Science
  Adv.}\ }\textbf {\bibinfo {volume} {4}},\ \bibinfo {pages} {eaao4513}
  (\bibinfo {year} {2018})}\BibitemShut {NoStop}%
\bibitem [{\citenamefont {Abrikosov}(1974)}]{abrikosov}%
  \BibitemOpen
  \bibfield  {author} {\bibinfo {author} {\bibfnamefont {A.~A.}\ \bibnamefont
  {Abrikosov}},\ }\bibfield  {title} {\enquote {\bibinfo {title} {{Calculation
  of critical indices fo zero-gap semiconductors}},}\ }\href@noop {} {\bibfield
   {journal} {\bibinfo  {journal} {Sov. Phys. JETP}\ }\textbf {\bibinfo
  {volume} {39}},\ \bibinfo {pages} {709} (\bibinfo {year} {1974})}\BibitemShut
  {NoStop}%
\bibitem [{\citenamefont {Abrikosov}\ and\ \citenamefont
  {Beneslavskii}(1971)}]{abrben}%
  \BibitemOpen
  \bibfield  {author} {\bibinfo {author} {\bibfnamefont {A.~A.}\ \bibnamefont
  {Abrikosov}}\ and\ \bibinfo {author} {\bibfnamefont {S.~D.}\ \bibnamefont
  {Beneslavskii}},\ }\bibfield  {title} {\enquote {\bibinfo {title} {{Possible
  existence of substances intermediate between metals and dielectrics}},}\
  }\href@noop {} {\bibfield  {journal} {\bibinfo  {journal} {Sov. Phys. JETP}\
  }\textbf {\bibinfo {volume} {32}},\ \bibinfo {pages} {699} (\bibinfo {year}
  {1971})}\BibitemShut {NoStop}%
\bibitem [{\citenamefont {Moon}\ \emph {et~al.}(2013)\citenamefont {Moon},
  \citenamefont {Xu}, \citenamefont {Kim},\ and\ \citenamefont
  {Balents}}]{moon}%
  \BibitemOpen
  \bibfield  {author} {\bibinfo {author} {\bibfnamefont {Eun-Gook}\
  \bibnamefont {Moon}}, \bibinfo {author} {\bibfnamefont {Cenke}\ \bibnamefont
  {Xu}}, \bibinfo {author} {\bibfnamefont {Yong~Baek}\ \bibnamefont {Kim}}, \
  and\ \bibinfo {author} {\bibfnamefont {Leon}\ \bibnamefont {Balents}},\
  }\bibfield  {title} {\enquote {\bibinfo {title} {{Non-Fermi-Liquid and
  Topological States with Strong Spin-Orbit Coupling}},}\ }\href {\doibase
  10.1103/PhysRevLett.111.206401} {\bibfield  {journal} {\bibinfo  {journal}
  {Phys. Rev. Lett.}\ }\textbf {\bibinfo {volume} {111}},\ \bibinfo {pages}
  {206401} (\bibinfo {year} {2013})}\BibitemShut {NoStop}%
\bibitem [{\citenamefont {Herbut}\ and\ \citenamefont
  {Janssen}(2014)}]{PhysRevLett.113.106401}%
  \BibitemOpen
  \bibfield  {author} {\bibinfo {author} {\bibfnamefont {Igor~F.}\ \bibnamefont
  {Herbut}}\ and\ \bibinfo {author} {\bibfnamefont {Lukas}\ \bibnamefont
  {Janssen}},\ }\bibfield  {title} {\enquote {\bibinfo {title} {{Topological
  Mott Insulator in Three-Dimensional Systems with Quadratic Band Touching}},}\
  }\href {\doibase 10.1103/PhysRevLett.113.106401} {\bibfield  {journal}
  {\bibinfo  {journal} {Phys. Rev. Lett.}\ }\textbf {\bibinfo {volume} {113}},\
  \bibinfo {pages} {106401} (\bibinfo {year} {2014})}\BibitemShut {NoStop}%
\bibitem [{\citenamefont {Janssen}\ and\ \citenamefont
  {Herbut}(2016)}]{PhysRevB.93.165109}%
  \BibitemOpen
  \bibfield  {author} {\bibinfo {author} {\bibfnamefont {Lukas}\ \bibnamefont
  {Janssen}}\ and\ \bibinfo {author} {\bibfnamefont {Igor~F.}\ \bibnamefont
  {Herbut}},\ }\bibfield  {title} {\enquote {\bibinfo {title} {{Excitonic
  instability of three-dimensional gapless semiconductors: Large-$N$
  theory}},}\ }\href {\doibase 10.1103/PhysRevB.93.165109} {\bibfield
  {journal} {\bibinfo  {journal} {Phys. Rev. B}\ }\textbf {\bibinfo {volume}
  {93}},\ \bibinfo {pages} {165109} (\bibinfo {year} {2016})}\BibitemShut
  {NoStop}%
\bibitem [{\citenamefont {Janssen}\ and\ \citenamefont
  {Herbut}(2017)}]{PhysRevB.95.075101}%
  \BibitemOpen
  \bibfield  {author} {\bibinfo {author} {\bibfnamefont {Lukas}\ \bibnamefont
  {Janssen}}\ and\ \bibinfo {author} {\bibfnamefont {Igor~F.}\ \bibnamefont
  {Herbut}},\ }\bibfield  {title} {\enquote {\bibinfo {title} {{Phase diagram
  of electronic systems with quadratic Fermi nodes in $2<d<4$: $2+\epsilon$
  expansion, $4-\epsilon$ expansion, and functional renormalization group}},}\
  }\href {\doibase 10.1103/PhysRevB.95.075101} {\bibfield  {journal} {\bibinfo
  {journal} {Phys. Rev. B}\ }\textbf {\bibinfo {volume} {95}},\ \bibinfo
  {pages} {075101} (\bibinfo {year} {2017})}\BibitemShut {NoStop}%
\bibitem [{\citenamefont {Mandal}\ and\ \citenamefont
  {Nandkishore}(2018)}]{PhysRevB.97.125121}%
  \BibitemOpen
  \bibfield  {author} {\bibinfo {author} {\bibfnamefont {Ipsita}\ \bibnamefont
  {Mandal}}\ and\ \bibinfo {author} {\bibfnamefont {Rahul~M.}\ \bibnamefont
  {Nandkishore}},\ }\bibfield  {title} {\enquote {\bibinfo {title} {Interplay
  of coulomb interactions and disorder in three-dimensional quadratic band
  crossings without time-reversal symmetry and with unequal masses for
  conduction and valence bands},}\ }\href {\doibase 10.1103/PhysRevB.97.125121}
  {\bibfield  {journal} {\bibinfo  {journal} {Phys. Rev. B}\ }\textbf {\bibinfo
  {volume} {97}},\ \bibinfo {pages} {125121} (\bibinfo {year}
  {2018})}\BibitemShut {NoStop}%
\bibitem [{\citenamefont {Witczak-Krempa}\ and\ \citenamefont
  {Kim}(2012)}]{PhysRevB.85.045124}%
  \BibitemOpen
  \bibfield  {author} {\bibinfo {author} {\bibfnamefont {William}\ \bibnamefont
  {Witczak-Krempa}}\ and\ \bibinfo {author} {\bibfnamefont {Yong~Baek}\
  \bibnamefont {Kim}},\ }\bibfield  {title} {\enquote {\bibinfo {title}
  {{Topological and magnetic phases of interacting electrons in the pyrochlore
  iridates}},}\ }\href {\doibase 10.1103/PhysRevB.85.045124} {\bibfield
  {journal} {\bibinfo  {journal} {Phys. Rev. B}\ }\textbf {\bibinfo {volume}
  {85}},\ \bibinfo {pages} {045124} (\bibinfo {year} {2012})}\BibitemShut
  {NoStop}%
\bibitem [{\citenamefont {Savary}\ \emph {et~al.}(2014)\citenamefont {Savary},
  \citenamefont {Moon},\ and\ \citenamefont {Balents}}]{PhysRevX.4.041027}%
  \BibitemOpen
  \bibfield  {author} {\bibinfo {author} {\bibfnamefont {Lucile}\ \bibnamefont
  {Savary}}, \bibinfo {author} {\bibfnamefont {Eun-Gook}\ \bibnamefont {Moon}},
  \ and\ \bibinfo {author} {\bibfnamefont {Leon}\ \bibnamefont {Balents}},\
  }\bibfield  {title} {\enquote {\bibinfo {title} {{New Type of Quantum
  Criticality in the Pyrochlore Iridates}},}\ }\href {\doibase
  10.1103/PhysRevX.4.041027} {\bibfield  {journal} {\bibinfo  {journal} {Phys.
  Rev. X}\ }\textbf {\bibinfo {volume} {4}},\ \bibinfo {pages} {041027}
  (\bibinfo {year} {2014})}\BibitemShut {NoStop}%
\bibitem [{\citenamefont {Janssen}\ and\ \citenamefont
  {Herbut}(2015)}]{PhysRevB.92.045117}%
  \BibitemOpen
  \bibfield  {author} {\bibinfo {author} {\bibfnamefont {Lukas}\ \bibnamefont
  {Janssen}}\ and\ \bibinfo {author} {\bibfnamefont {Igor~F.}\ \bibnamefont
  {Herbut}},\ }\bibfield  {title} {\enquote {\bibinfo {title} {{Nematic quantum
  criticality in three-dimensional Fermi system with quadratic band
  touching}},}\ }\href {\doibase 10.1103/PhysRevB.92.045117} {\bibfield
  {journal} {\bibinfo  {journal} {Phys. Rev. B}\ }\textbf {\bibinfo {volume}
  {92}},\ \bibinfo {pages} {045117} (\bibinfo {year} {2015})}\BibitemShut
  {NoStop}%
\bibitem [{\citenamefont {Murray}\ \emph {et~al.}(2015)\citenamefont {Murray},
  \citenamefont {Vafek},\ and\ \citenamefont {Balents}}]{PhysRevB.92.035137}%
  \BibitemOpen
  \bibfield  {author} {\bibinfo {author} {\bibfnamefont {James~M.}\
  \bibnamefont {Murray}}, \bibinfo {author} {\bibfnamefont {Oskar}\
  \bibnamefont {Vafek}}, \ and\ \bibinfo {author} {\bibfnamefont {Leon}\
  \bibnamefont {Balents}},\ }\bibfield  {title} {\enquote {\bibinfo {title}
  {{Incommensurate spin density wave at a ferromagnetic quantum critical point
  in a three-dimensional parabolic semimetal}},}\ }\href {\doibase
  10.1103/PhysRevB.92.035137} {\bibfield  {journal} {\bibinfo  {journal} {Phys.
  Rev. B}\ }\textbf {\bibinfo {volume} {92}},\ \bibinfo {pages} {035137}
  (\bibinfo {year} {2015})}\BibitemShut {NoStop}%
\bibitem [{\citenamefont {Boettcher}\ and\ \citenamefont
  {Herbut}(2016)}]{PhysRevB.93.205138}%
  \BibitemOpen
  \bibfield  {author} {\bibinfo {author} {\bibfnamefont {Igor}\ \bibnamefont
  {Boettcher}}\ and\ \bibinfo {author} {\bibfnamefont {Igor~F.}\ \bibnamefont
  {Herbut}},\ }\bibfield  {title} {\enquote {\bibinfo {title} {{Superconducting
  quantum criticality in three-dimensional Luttinger semimetals}},}\ }\href
  {\doibase 10.1103/PhysRevB.93.205138} {\bibfield  {journal} {\bibinfo
  {journal} {Phys. Rev. B}\ }\textbf {\bibinfo {volume} {93}},\ \bibinfo
  {pages} {205138} (\bibinfo {year} {2016})}\BibitemShut {NoStop}%
\bibitem [{\citenamefont {Meinert}(2016)}]{PhysRevLett.116.137001}%
  \BibitemOpen
  \bibfield  {author} {\bibinfo {author} {\bibfnamefont {Markus}\ \bibnamefont
  {Meinert}},\ }\bibfield  {title} {\enquote {\bibinfo {title} {{Unconventional
  Superconductivity in YPtBi and Related Topological Semimetals}},}\ }\href
  {\doibase 10.1103/PhysRevLett.116.137001} {\bibfield  {journal} {\bibinfo
  {journal} {Phys. Rev. Lett.}\ }\textbf {\bibinfo {volume} {116}},\ \bibinfo
  {pages} {137001} (\bibinfo {year} {2016})}\BibitemShut {NoStop}%
\bibitem [{\citenamefont {Brydon}\ \emph {et~al.}(2016)\citenamefont {Brydon},
  \citenamefont {Wang}, \citenamefont {Weinert},\ and\ \citenamefont
  {Agterberg}}]{PhysRevLett.116.177001}%
  \BibitemOpen
  \bibfield  {author} {\bibinfo {author} {\bibfnamefont {P.~M.~R.}\
  \bibnamefont {Brydon}}, \bibinfo {author} {\bibfnamefont {Limin}\
  \bibnamefont {Wang}}, \bibinfo {author} {\bibfnamefont {M.}~\bibnamefont
  {Weinert}}, \ and\ \bibinfo {author} {\bibfnamefont {D.~F.}\ \bibnamefont
  {Agterberg}},\ }\bibfield  {title} {\enquote {\bibinfo {title} {{Pairing of
  $j=3/2$ Fermions in Half-Heusler Superconductors}},}\ }\href {\doibase
  10.1103/PhysRevLett.116.177001} {\bibfield  {journal} {\bibinfo  {journal}
  {Phys. Rev. Lett.}\ }\textbf {\bibinfo {volume} {116}},\ \bibinfo {pages}
  {177001} (\bibinfo {year} {2016})}\BibitemShut {NoStop}%
\bibitem [{\citenamefont {Goswami}\ \emph {et~al.}(2017)\citenamefont
  {Goswami}, \citenamefont {Roy},\ and\ \citenamefont
  {Das~Sarma}}]{PhysRevB.95.085120}%
  \BibitemOpen
  \bibfield  {author} {\bibinfo {author} {\bibfnamefont {Pallab}\ \bibnamefont
  {Goswami}}, \bibinfo {author} {\bibfnamefont {Bitan}\ \bibnamefont {Roy}}, \
  and\ \bibinfo {author} {\bibfnamefont {Sankar}\ \bibnamefont {Das~Sarma}},\
  }\bibfield  {title} {\enquote {\bibinfo {title} {{Competing orders and
  topology in the global phase diagram of pyrochlore iridates}},}\ }\href
  {\doibase 10.1103/PhysRevB.95.085120} {\bibfield  {journal} {\bibinfo
  {journal} {Phys. Rev. B}\ }\textbf {\bibinfo {volume} {95}},\ \bibinfo
  {pages} {085120} (\bibinfo {year} {2017})}\BibitemShut {NoStop}%
\bibitem [{\citenamefont {Boettcher}\ and\ \citenamefont
  {Herbut}(2017)}]{PhysRevB.95.075149}%
  \BibitemOpen
  \bibfield  {author} {\bibinfo {author} {\bibfnamefont {Igor}\ \bibnamefont
  {Boettcher}}\ and\ \bibinfo {author} {\bibfnamefont {Igor~F.}\ \bibnamefont
  {Herbut}},\ }\bibfield  {title} {\enquote {\bibinfo {title} {{Anisotropy
  induces non-Fermi-liquid behavior and nematic magnetic order in
  three-dimensional Luttinger semimetals}},}\ }\href {\doibase
  10.1103/PhysRevB.95.075149} {\bibfield  {journal} {\bibinfo  {journal} {Phys.
  Rev. B}\ }\textbf {\bibinfo {volume} {95}},\ \bibinfo {pages} {075149}
  (\bibinfo {year} {2017})}\BibitemShut {NoStop}%
\bibitem [{\citenamefont {Agterberg}\ \emph {et~al.}(2017)\citenamefont
  {Agterberg}, \citenamefont {Brydon},\ and\ \citenamefont
  {Timm}}]{PhysRevLett.118.127001}%
  \BibitemOpen
  \bibfield  {author} {\bibinfo {author} {\bibfnamefont {D.~F.}\ \bibnamefont
  {Agterberg}}, \bibinfo {author} {\bibfnamefont {P.~M.~R.}\ \bibnamefont
  {Brydon}}, \ and\ \bibinfo {author} {\bibfnamefont {C.}~\bibnamefont
  {Timm}},\ }\bibfield  {title} {\enquote {\bibinfo {title} {{Bogoliubov Fermi
  Surfaces in Superconductors with Broken Time-Reversal Symmetry}},}\ }\href
  {\doibase 10.1103/PhysRevLett.118.127001} {\bibfield  {journal} {\bibinfo
  {journal} {Phys. Rev. Lett.}\ }\textbf {\bibinfo {volume} {118}},\ \bibinfo
  {pages} {127001} (\bibinfo {year} {2017})}\BibitemShut {NoStop}%
\bibitem [{\citenamefont {Ghorashi}\ \emph {et~al.}(2017)\citenamefont
  {Ghorashi}, \citenamefont {Davis},\ and\ \citenamefont
  {Foster}}]{PhysRevB.95.144503}%
  \BibitemOpen
  \bibfield  {author} {\bibinfo {author} {\bibfnamefont {Sayed Ali~Akbar}\
  \bibnamefont {Ghorashi}}, \bibinfo {author} {\bibfnamefont {Seth}\
  \bibnamefont {Davis}}, \ and\ \bibinfo {author} {\bibfnamefont {Matthew~S.}\
  \bibnamefont {Foster}},\ }\bibfield  {title} {\enquote {\bibinfo {title}
  {Disorder-enhanced topological protection and universal quantum criticality
  in a spin-$\frac{3}{2}$ topological superconductor},}\ }\href {\doibase
  10.1103/PhysRevB.95.144503} {\bibfield  {journal} {\bibinfo  {journal} {Phys.
  Rev. B}\ }\textbf {\bibinfo {volume} {95}},\ \bibinfo {pages} {144503}
  (\bibinfo {year} {2017})}\BibitemShut {NoStop}%
\bibitem [{\citenamefont {Yang}\ \emph {et~al.}(2017)\citenamefont {Yang},
  \citenamefont {Xiang},\ and\ \citenamefont {Wu}}]{PhysRevB.96.144514}%
  \BibitemOpen
  \bibfield  {author} {\bibinfo {author} {\bibfnamefont {Wang}\ \bibnamefont
  {Yang}}, \bibinfo {author} {\bibfnamefont {Tao}\ \bibnamefont {Xiang}}, \
  and\ \bibinfo {author} {\bibfnamefont {Congjun}\ \bibnamefont {Wu}},\
  }\bibfield  {title} {\enquote {\bibinfo {title} {Majorana surface modes of
  nodal topological pairings in spin-$\frac{3}{2}$ semimetals},}\ }\href
  {\doibase 10.1103/PhysRevB.96.144514} {\bibfield  {journal} {\bibinfo
  {journal} {Phys. Rev. B}\ }\textbf {\bibinfo {volume} {96}},\ \bibinfo
  {pages} {144514} (\bibinfo {year} {2017})}\BibitemShut {NoStop}%
\bibitem [{\citenamefont {Savary}\ \emph {et~al.}(2017)\citenamefont {Savary},
  \citenamefont {Ruhman}, \citenamefont {Venderbos}, \citenamefont {Fu},\ and\
  \citenamefont {Lee}}]{PhysRevB.96.214514}%
  \BibitemOpen
  \bibfield  {author} {\bibinfo {author} {\bibfnamefont {Lucile}\ \bibnamefont
  {Savary}}, \bibinfo {author} {\bibfnamefont {Jonathan}\ \bibnamefont
  {Ruhman}}, \bibinfo {author} {\bibfnamefont {J\"orn W.~F.}\ \bibnamefont
  {Venderbos}}, \bibinfo {author} {\bibfnamefont {Liang}\ \bibnamefont {Fu}}, \
  and\ \bibinfo {author} {\bibfnamefont {Patrick~A.}\ \bibnamefont {Lee}},\
  }\bibfield  {title} {\enquote {\bibinfo {title} {Superconductivity in
  three-dimensional spin-orbit coupled semimetals},}\ }\href {\doibase
  10.1103/PhysRevB.96.214514} {\bibfield  {journal} {\bibinfo  {journal} {Phys.
  Rev. B}\ }\textbf {\bibinfo {volume} {96}},\ \bibinfo {pages} {214514}
  (\bibinfo {year} {2017})}\BibitemShut {NoStop}%
\bibitem [{\citenamefont {Boettcher}\ and\ \citenamefont
  {Herbut}(2018{\natexlab{a}})}]{PhysRevLett.120.057002}%
  \BibitemOpen
  \bibfield  {author} {\bibinfo {author} {\bibfnamefont {Igor}\ \bibnamefont
  {Boettcher}}\ and\ \bibinfo {author} {\bibfnamefont {Igor~F.}\ \bibnamefont
  {Herbut}},\ }\bibfield  {title} {\enquote {\bibinfo {title} {{Unconventional
  Superconductivity in Luttinger Semimetals: Theory of Complex Tensor Order and
  the Emergence of the Uniaxial Nematic State}},}\ }\href {\doibase
  10.1103/PhysRevLett.120.057002} {\bibfield  {journal} {\bibinfo  {journal}
  {Phys. Rev. Lett.}\ }\textbf {\bibinfo {volume} {120}},\ \bibinfo {pages}
  {057002} (\bibinfo {year} {2018}{\natexlab{a}})}\BibitemShut {NoStop}%
\bibitem [{\citenamefont {Venderbos}\ \emph {et~al.}(2018)\citenamefont
  {Venderbos}, \citenamefont {Savary}, \citenamefont {Ruhman}, \citenamefont
  {Lee},\ and\ \citenamefont {Fu}}]{PhysRevX.8.011029}%
  \BibitemOpen
  \bibfield  {author} {\bibinfo {author} {\bibfnamefont {J\"orn W.~F.}\
  \bibnamefont {Venderbos}}, \bibinfo {author} {\bibfnamefont {Lucile}\
  \bibnamefont {Savary}}, \bibinfo {author} {\bibfnamefont {Jonathan}\
  \bibnamefont {Ruhman}}, \bibinfo {author} {\bibfnamefont {Patrick~A.}\
  \bibnamefont {Lee}}, \ and\ \bibinfo {author} {\bibfnamefont {Liang}\
  \bibnamefont {Fu}},\ }\bibfield  {title} {\enquote {\bibinfo {title}
  {{Pairing States of Spin-$\frac{3}{2}$ Fermions: Symmetry-Enforced
  Topological Gap Functions}},}\ }\href {\doibase 10.1103/PhysRevX.8.011029}
  {\bibfield  {journal} {\bibinfo  {journal} {Phys. Rev. X}\ }\textbf {\bibinfo
  {volume} {8}},\ \bibinfo {pages} {011029} (\bibinfo {year}
  {2018})}\BibitemShut {NoStop}%
\bibitem [{\citenamefont {Ghorashi}\ \emph {et~al.}(2018)\citenamefont
  {Ghorashi}, \citenamefont {Hosur},\ and\ \citenamefont
  {Ting}}]{PhysRevB.97.205402}%
  \BibitemOpen
  \bibfield  {author} {\bibinfo {author} {\bibfnamefont {Sayed Ali~Akbar}\
  \bibnamefont {Ghorashi}}, \bibinfo {author} {\bibfnamefont {Pavan}\
  \bibnamefont {Hosur}}, \ and\ \bibinfo {author} {\bibfnamefont {Chin-Sen}\
  \bibnamefont {Ting}},\ }\bibfield  {title} {\enquote {\bibinfo {title}
  {Irradiated three-dimensional luttinger semimetal: A factory for engineering
  weyl semimetals},}\ }\href {\doibase 10.1103/PhysRevB.97.205402} {\bibfield
  {journal} {\bibinfo  {journal} {Phys. Rev. B}\ }\textbf {\bibinfo {volume}
  {97}},\ \bibinfo {pages} {205402} (\bibinfo {year} {2018})}\BibitemShut
  {NoStop}%
\bibitem [{\citenamefont {{Mandal}}(2018{\natexlab{a}})}]{Mandal}%
  \BibitemOpen
  \bibfield  {author} {\bibinfo {author} {\bibfnamefont {I.}~\bibnamefont
  {{Mandal}}},\ }\bibfield  {title} {\enquote {\bibinfo {title} {{Fate of
  superconductivity in three-dimensional disordered Luttinger semimetals}},}\
  }\href {\doibase 10.1016/j.aop.2018.03.004} {\bibfield  {journal} {\bibinfo
  {journal} {Ann. Physics}\ }\textbf {\bibinfo {volume} {392}},\ \bibinfo
  {pages} {179--195} (\bibinfo {year} {2018}{\natexlab{a}})}\BibitemShut
  {NoStop}%
\bibitem [{\citenamefont {Yu}\ and\ \citenamefont
  {Liu}(2018)}]{PhysRevB.98.104514}%
  \BibitemOpen
  \bibfield  {author} {\bibinfo {author} {\bibfnamefont {Jiabin}\ \bibnamefont
  {Yu}}\ and\ \bibinfo {author} {\bibfnamefont {Chao-Xing}\ \bibnamefont
  {Liu}},\ }\bibfield  {title} {\enquote {\bibinfo {title} {Singlet-quintet
  mixing in spin-orbit coupled superconductors with $j=\frac{3}{2}$
  fermions},}\ }\href {\doibase 10.1103/PhysRevB.98.104514} {\bibfield
  {journal} {\bibinfo  {journal} {Phys. Rev. B}\ }\textbf {\bibinfo {volume}
  {98}},\ \bibinfo {pages} {104514} (\bibinfo {year} {2018})}\BibitemShut
  {NoStop}%
\bibitem [{\citenamefont {Yao}\ and\ \citenamefont
  {Chen}(2018)}]{PhysRevX.8.041039}%
  \BibitemOpen
  \bibfield  {author} {\bibinfo {author} {\bibfnamefont {Xu-Ping}\ \bibnamefont
  {Yao}}\ and\ \bibinfo {author} {\bibfnamefont {Gang}\ \bibnamefont {Chen}},\
  }\bibfield  {title} {\enquote {\bibinfo {title}
  {{${\mathrm{Pr}}_{2}{\mathrm{Ir}}_{2}{\mathrm{O}}_{7}$: When Luttinger
  Semimetal Meets Melko-Hertog-Gingras Spin Ice State}},}\ }\href {\doibase
  10.1103/PhysRevX.8.041039} {\bibfield  {journal} {\bibinfo  {journal} {Phys.
  Rev. X}\ }\textbf {\bibinfo {volume} {8}},\ \bibinfo {pages} {041039}
  (\bibinfo {year} {2018})}\BibitemShut {NoStop}%
\bibitem [{\citenamefont {Roy}\ \emph {et~al.}(2019)\citenamefont {Roy},
  \citenamefont {Ghorashi}, \citenamefont {Foster},\ and\ \citenamefont
  {Nevidomskyy}}]{2017arXiv170807825R}%
  \BibitemOpen
  \bibfield  {author} {\bibinfo {author} {\bibfnamefont {Bitan}\ \bibnamefont
  {Roy}}, \bibinfo {author} {\bibfnamefont {Sayed Ali~Akbar}\ \bibnamefont
  {Ghorashi}}, \bibinfo {author} {\bibfnamefont {Matthew~S.}\ \bibnamefont
  {Foster}}, \ and\ \bibinfo {author} {\bibfnamefont {Andriy~H.}\ \bibnamefont
  {Nevidomskyy}},\ }\bibfield  {title} {\enquote {\bibinfo {title} {Topological
  superconductivity of spin-$3/2$ carriers in a three-dimensional doped
  luttinger semimetal},}\ }\href {\doibase 10.1103/PhysRevB.99.054505}
  {\bibfield  {journal} {\bibinfo  {journal} {Phys. Rev. B}\ }\textbf {\bibinfo
  {volume} {99}},\ \bibinfo {pages} {054505} (\bibinfo {year}
  {2019})}\BibitemShut {NoStop}%
\bibitem [{\citenamefont {{Sim}}\ \emph {et~al.}(2018)\citenamefont {{Sim}},
  \citenamefont {{Mishra}}, \citenamefont {{Jip Park}}, \citenamefont {{Kim}},
  \citenamefont {{Cho}},\ and\ \citenamefont {{Lee}}}]{2018arXiv181104046S}%
  \BibitemOpen
  \bibfield  {author} {\bibinfo {author} {\bibfnamefont {G.}~\bibnamefont
  {{Sim}}}, \bibinfo {author} {\bibfnamefont {A.}~\bibnamefont {{Mishra}}},
  \bibinfo {author} {\bibfnamefont {M.}~\bibnamefont {{Jip Park}}}, \bibinfo
  {author} {\bibfnamefont {Y.~B.}\ \bibnamefont {{Kim}}}, \bibinfo {author}
  {\bibfnamefont {G.~Y.}\ \bibnamefont {{Cho}}}, \ and\ \bibinfo {author}
  {\bibfnamefont {S.}~\bibnamefont {{Lee}}},\ }\bibfield  {title} {\enquote
  {\bibinfo {title} {{Topological d+s wave superconductors in a multi-orbital
  quadratic band touching system}},}\ }\href@noop {} {\  (\bibinfo {year}
  {2018})},\ \Eprint {http://arxiv.org/abs/1811.04046} {arXiv:1811.04046}
  \BibitemShut {NoStop}%
\bibitem [{\citenamefont {{Szabo}}\ \emph {et~al.}(2018)\citenamefont
  {{Szabo}}, \citenamefont {{Moessner}},\ and\ \citenamefont
  {{Roy}}}]{2018arXiv181112415S}%
  \BibitemOpen
  \bibfield  {author} {\bibinfo {author} {\bibfnamefont {Andras}\ \bibnamefont
  {{Szabo}}}, \bibinfo {author} {\bibfnamefont {Roderich}\ \bibnamefont
  {{Moessner}}}, \ and\ \bibinfo {author} {\bibfnamefont {Bitan}\ \bibnamefont
  {{Roy}}},\ }\bibfield  {title} {\enquote {\bibinfo {title} {{Interacting
  spin-3/2 fermions in a Luttinger (semi)metal: competing phases and their
  selection in the global phase diagram}},}\ }\href@noop {} {\  (\bibinfo
  {year} {2018})},\ \Eprint {http://arxiv.org/abs/1811.12415}
  {arXiv:1811.12415} \BibitemShut {NoStop}%
\bibitem [{\citenamefont {{Kondo}}\ \emph {et~al.}(2015)\citenamefont
  {{Kondo}}, \citenamefont {{Nakayama}}, \citenamefont {{Chen}}, \citenamefont
  {{Ishikawa}}, \citenamefont {{Moon}}, \citenamefont {{Yamamoto}},
  \citenamefont {{Ota}}, \citenamefont {{Malaeb}}, \citenamefont {{Kanai}},
  \citenamefont {{Nakashima}}, \citenamefont {{Ishida}}, \citenamefont
  {{Yoshida}}, \citenamefont {{Yamamoto}}, \citenamefont {{Matsunami}},
  \citenamefont {{Kimura}}, \citenamefont {{Inami}}, \citenamefont {{Ono}},
  \citenamefont {{Kumigashira}}, \citenamefont {{Nakatsuji}}, \citenamefont
  {{Balents}},\ and\ \citenamefont {{Shin}}}]{kondo}%
  \BibitemOpen
  \bibfield  {author} {\bibinfo {author} {\bibfnamefont {T.}~\bibnamefont
  {{Kondo}}}, \bibinfo {author} {\bibfnamefont {M.}~\bibnamefont {{Nakayama}}},
  \bibinfo {author} {\bibfnamefont {R.}~\bibnamefont {{Chen}}}, \bibinfo
  {author} {\bibfnamefont {J.~J.}\ \bibnamefont {{Ishikawa}}}, \bibinfo
  {author} {\bibfnamefont {E.-G.}\ \bibnamefont {{Moon}}}, \bibinfo {author}
  {\bibfnamefont {T.}~\bibnamefont {{Yamamoto}}}, \bibinfo {author}
  {\bibfnamefont {Y.}~\bibnamefont {{Ota}}}, \bibinfo {author} {\bibfnamefont
  {W.}~\bibnamefont {{Malaeb}}}, \bibinfo {author} {\bibfnamefont
  {H.}~\bibnamefont {{Kanai}}}, \bibinfo {author} {\bibfnamefont
  {Y.}~\bibnamefont {{Nakashima}}}, \bibinfo {author} {\bibfnamefont
  {Y.}~\bibnamefont {{Ishida}}}, \bibinfo {author} {\bibfnamefont
  {R.}~\bibnamefont {{Yoshida}}}, \bibinfo {author} {\bibfnamefont
  {H.}~\bibnamefont {{Yamamoto}}}, \bibinfo {author} {\bibfnamefont
  {M.}~\bibnamefont {{Matsunami}}}, \bibinfo {author} {\bibfnamefont
  {S.}~\bibnamefont {{Kimura}}}, \bibinfo {author} {\bibfnamefont
  {N.}~\bibnamefont {{Inami}}}, \bibinfo {author} {\bibfnamefont
  {K.}~\bibnamefont {{Ono}}}, \bibinfo {author} {\bibfnamefont
  {H.}~\bibnamefont {{Kumigashira}}}, \bibinfo {author} {\bibfnamefont
  {S.}~\bibnamefont {{Nakatsuji}}}, \bibinfo {author} {\bibfnamefont
  {L.}~\bibnamefont {{Balents}}}, \ and\ \bibinfo {author} {\bibfnamefont
  {S.}~\bibnamefont {{Shin}}},\ }\bibfield  {title} {\enquote {\bibinfo {title}
  {{Quadratic Fermi node in a 3D strongly correlated semimetal}},}\ }\href
  {\doibase 10.1038/ncomms10042} {\bibfield  {journal} {\bibinfo  {journal}
  {Nat. Commun.}\ }\textbf {\bibinfo {volume} {6}},\ \bibinfo {eid} {10042}
  (\bibinfo {year} {2015})}\BibitemShut {NoStop}%
\bibitem [{\citenamefont {Nakayama}\ \emph {et~al.}(2016)\citenamefont
  {Nakayama}, \citenamefont {Kondo}, \citenamefont {Tian}, \citenamefont
  {Ishikawa}, \citenamefont {Halim}, \citenamefont {Bareille}, \citenamefont
  {Malaeb}, \citenamefont {Kuroda}, \citenamefont {Tomita}, \citenamefont
  {Ideta}, \citenamefont {Tanaka}, \citenamefont {Matsunami}, \citenamefont
  {Kimura}, \citenamefont {Inami}, \citenamefont {Ono}, \citenamefont
  {Kumigashira}, \citenamefont {Balents}, \citenamefont {Nakatsuji},\ and\
  \citenamefont {Shin}}]{PhysRevLett.117.056403}%
  \BibitemOpen
  \bibfield  {author} {\bibinfo {author} {\bibfnamefont {M.}~\bibnamefont
  {Nakayama}}, \bibinfo {author} {\bibfnamefont {Takeshi}\ \bibnamefont
  {Kondo}}, \bibinfo {author} {\bibfnamefont {Z.}~\bibnamefont {Tian}},
  \bibinfo {author} {\bibfnamefont {J.~J.}\ \bibnamefont {Ishikawa}}, \bibinfo
  {author} {\bibfnamefont {M.}~\bibnamefont {Halim}}, \bibinfo {author}
  {\bibfnamefont {C.}~\bibnamefont {Bareille}}, \bibinfo {author}
  {\bibfnamefont {W.}~\bibnamefont {Malaeb}}, \bibinfo {author} {\bibfnamefont
  {K.}~\bibnamefont {Kuroda}}, \bibinfo {author} {\bibfnamefont
  {T.}~\bibnamefont {Tomita}}, \bibinfo {author} {\bibfnamefont
  {S.}~\bibnamefont {Ideta}}, \bibinfo {author} {\bibfnamefont
  {K.}~\bibnamefont {Tanaka}}, \bibinfo {author} {\bibfnamefont
  {M.}~\bibnamefont {Matsunami}}, \bibinfo {author} {\bibfnamefont
  {S.}~\bibnamefont {Kimura}}, \bibinfo {author} {\bibfnamefont
  {N.}~\bibnamefont {Inami}}, \bibinfo {author} {\bibfnamefont
  {K.}~\bibnamefont {Ono}}, \bibinfo {author} {\bibfnamefont {H.}~\bibnamefont
  {Kumigashira}}, \bibinfo {author} {\bibfnamefont {L.}~\bibnamefont
  {Balents}}, \bibinfo {author} {\bibfnamefont {S.}~\bibnamefont {Nakatsuji}},
  \ and\ \bibinfo {author} {\bibfnamefont {S.}~\bibnamefont {Shin}},\
  }\bibfield  {title} {\enquote {\bibinfo {title} {{Slater to Mott Crossover in
  the Metal to Insulator Transition of
  ${\mathrm{Nd}}_{2}{\mathrm{Ir}}_{2}{\mathrm{O}}_{7}$}},}\ }\href {\doibase
  10.1103/PhysRevLett.117.056403} {\bibfield  {journal} {\bibinfo  {journal}
  {Phys. Rev. Lett.}\ }\textbf {\bibinfo {volume} {117}},\ \bibinfo {pages}
  {056403} (\bibinfo {year} {2016})}\BibitemShut {NoStop}%
\bibitem [{\citenamefont {{Liang}}\ \emph {et~al.}(2017)\citenamefont
  {{Liang}}, \citenamefont {{Hsieh}}, \citenamefont {{Ishikawa}}, \citenamefont
  {{Nakatsuji}}, \citenamefont {{Fu}},\ and\ \citenamefont
  {{Ong}}}]{OngNature}%
  \BibitemOpen
  \bibfield  {author} {\bibinfo {author} {\bibfnamefont {T.}~\bibnamefont
  {{Liang}}}, \bibinfo {author} {\bibfnamefont {T.~H.}\ \bibnamefont
  {{Hsieh}}}, \bibinfo {author} {\bibfnamefont {J.~J.}\ \bibnamefont
  {{Ishikawa}}}, \bibinfo {author} {\bibfnamefont {S.}~\bibnamefont
  {{Nakatsuji}}}, \bibinfo {author} {\bibfnamefont {L.}~\bibnamefont {{Fu}}}, \
  and\ \bibinfo {author} {\bibfnamefont {N.~P.}\ \bibnamefont {{Ong}}},\
  }\bibfield  {title} {\enquote {\bibinfo {title} {{Orthogonal magnetization
  and symmetry breaking in pyrochlore iridate Eu$_{2}$Ir$_{2}$O$_{7}$}},}\
  }\href {\doibase 10.1038/nphys4051} {\bibfield  {journal} {\bibinfo
  {journal} {Nat. Phys.}\ }\textbf {\bibinfo {volume} {13}},\ \bibinfo {pages}
  {599--603} (\bibinfo {year} {2017})}\BibitemShut {NoStop}%
\bibitem [{\citenamefont {Broerman}(1972)}]{PhysRevB.5.397}%
  \BibitemOpen
  \bibfield  {author} {\bibinfo {author} {\bibfnamefont {J.~G.}\ \bibnamefont
  {Broerman}},\ }\bibfield  {title} {\enquote {\bibinfo {title}
  {{Random-Phase-Approximation Dielectric Function of $a$-Sn in the Far
  Infrared}},}\ }\href {\doibase 10.1103/PhysRevB.5.397} {\bibfield  {journal}
  {\bibinfo  {journal} {Phys. Rev. B}\ }\textbf {\bibinfo {volume} {5}},\
  \bibinfo {pages} {397--408} (\bibinfo {year} {1972})}\BibitemShut {NoStop}%
\bibitem [{\citenamefont {{Mandal}}(2018{\natexlab{b}})}]{2018arXiv181006574M}%
  \BibitemOpen
  \bibfield  {author} {\bibinfo {author} {\bibfnamefont {I.}~\bibnamefont
  {{Mandal}}},\ }\bibfield  {title} {\enquote {\bibinfo {title} {{Search for
  plasmons in isotropic Luttinger semimetals}},}\ }\href@noop {} {\  (\bibinfo
  {year} {2018}{\natexlab{b}})},\ \Eprint {http://arxiv.org/abs/1810.06574}
  {arXiv:1810.06574} \BibitemShut {NoStop}%
\bibitem [{\citenamefont {Butch}\ \emph {et~al.}(2011)\citenamefont {Butch},
  \citenamefont {Syers}, \citenamefont {Kirshenbaum}, \citenamefont {Hope},\
  and\ \citenamefont {Paglione}}]{PhysRevB.84.220504}%
  \BibitemOpen
  \bibfield  {author} {\bibinfo {author} {\bibfnamefont {N.~P.}\ \bibnamefont
  {Butch}}, \bibinfo {author} {\bibfnamefont {P.}~\bibnamefont {Syers}},
  \bibinfo {author} {\bibfnamefont {K.}~\bibnamefont {Kirshenbaum}}, \bibinfo
  {author} {\bibfnamefont {A.~P.}\ \bibnamefont {Hope}}, \ and\ \bibinfo
  {author} {\bibfnamefont {J.}~\bibnamefont {Paglione}},\ }\bibfield  {title}
  {\enquote {\bibinfo {title} {{Superconductivity in the topological semimetal
  YPtBi}},}\ }\href {\doibase 10.1103/PhysRevB.84.220504} {\bibfield  {journal}
  {\bibinfo  {journal} {Phys. Rev. B}\ }\textbf {\bibinfo {volume} {84}},\
  \bibinfo {pages} {220504} (\bibinfo {year} {2011})}\BibitemShut {NoStop}%
\bibitem [{\citenamefont {Bay}\ \emph {et~al.}(2012)\citenamefont {Bay},
  \citenamefont {Naka}, \citenamefont {Huang},\ and\ \citenamefont
  {de~Visser}}]{PhysRevB.86.064515}%
  \BibitemOpen
  \bibfield  {author} {\bibinfo {author} {\bibfnamefont {T.~V.}\ \bibnamefont
  {Bay}}, \bibinfo {author} {\bibfnamefont {T.}~\bibnamefont {Naka}}, \bibinfo
  {author} {\bibfnamefont {Y.~K.}\ \bibnamefont {Huang}}, \ and\ \bibinfo
  {author} {\bibfnamefont {A.}~\bibnamefont {de~Visser}},\ }\bibfield  {title}
  {\enquote {\bibinfo {title} {{Superconductivity in noncentrosymmetric YPtBi
  under pressure}},}\ }\href {\doibase 10.1103/PhysRevB.86.064515} {\bibfield
  {journal} {\bibinfo  {journal} {Phys. Rev. B}\ }\textbf {\bibinfo {volume}
  {86}},\ \bibinfo {pages} {064515} (\bibinfo {year} {2012})}\BibitemShut
  {NoStop}%
\bibitem [{\citenamefont {Bay}\ \emph {et~al.}(2014)\citenamefont {Bay},
  \citenamefont {Jackson}, \citenamefont {Paulsen}, \citenamefont {Baines},
  \citenamefont {Amato}, \citenamefont {Orvis}, \citenamefont {Aronson},
  \citenamefont {Huang},\ and\ \citenamefont {de~Visser}}]{BayLowT}%
  \BibitemOpen
  \bibfield  {author} {\bibinfo {author} {\bibfnamefont {T.V.}\ \bibnamefont
  {Bay}}, \bibinfo {author} {\bibfnamefont {M.}~\bibnamefont {Jackson}},
  \bibinfo {author} {\bibfnamefont {C.}~\bibnamefont {Paulsen}}, \bibinfo
  {author} {\bibfnamefont {C.}~\bibnamefont {Baines}}, \bibinfo {author}
  {\bibfnamefont {A.}~\bibnamefont {Amato}}, \bibinfo {author} {\bibfnamefont
  {T.}~\bibnamefont {Orvis}}, \bibinfo {author} {\bibfnamefont {M.C.}\
  \bibnamefont {Aronson}}, \bibinfo {author} {\bibfnamefont {Y.K.}\
  \bibnamefont {Huang}}, \ and\ \bibinfo {author} {\bibfnamefont
  {A.}~\bibnamefont {de~Visser}},\ }\bibfield  {title} {\enquote {\bibinfo
  {title} {{Low field magnetic response of the non-centrosymmetric
  superconductor YPtBi}},}\ }\href {\doibase 10.1016/j.ssc.2013.12.010}
  {\bibfield  {journal} {\bibinfo  {journal} {Solid State Commun.}\ }\textbf
  {\bibinfo {volume} {183}},\ \bibinfo {pages} {47--50} (\bibinfo {year}
  {2014})}\BibitemShut {NoStop}%
\bibitem [{\citenamefont {Luttinger}(1956)}]{luttinger}%
  \BibitemOpen
  \bibfield  {author} {\bibinfo {author} {\bibfnamefont {J.~M.}\ \bibnamefont
  {Luttinger}},\ }\bibfield  {title} {\enquote {\bibinfo {title} {{Quantum
  Theory of Cyclotron Resonance in Semiconductors: General Theory}},}\ }\href
  {\doibase 10.1103/PhysRev.102.1030} {\bibfield  {journal} {\bibinfo
  {journal} {Phys. Rev.}\ }\textbf {\bibinfo {volume} {102}},\ \bibinfo {pages}
  {1030--1041} (\bibinfo {year} {1956})}\BibitemShut {NoStop}%
\bibitem [{\citenamefont {Timm}\ \emph {et~al.}(2017)\citenamefont {Timm},
  \citenamefont {Schnyder}, \citenamefont {Agterberg},\ and\ \citenamefont
  {Brydon}}]{PhysRevB.96.094526}%
  \BibitemOpen
  \bibfield  {author} {\bibinfo {author} {\bibfnamefont {C.}~\bibnamefont
  {Timm}}, \bibinfo {author} {\bibfnamefont {A.~P.}\ \bibnamefont {Schnyder}},
  \bibinfo {author} {\bibfnamefont {D.~F.}\ \bibnamefont {Agterberg}}, \ and\
  \bibinfo {author} {\bibfnamefont {P.~M.~R.}\ \bibnamefont {Brydon}},\
  }\bibfield  {title} {\enquote {\bibinfo {title} {{Inflated nodes and surface
  states in superconducting half-Heusler compounds}},}\ }\href {\doibase
  10.1103/PhysRevB.96.094526} {\bibfield  {journal} {\bibinfo  {journal} {Phys.
  Rev. B}\ }\textbf {\bibinfo {volume} {96}},\ \bibinfo {pages} {094526}
  (\bibinfo {year} {2017})}\BibitemShut {NoStop}%
\bibitem [{\citenamefont {Brydon}\ \emph {et~al.}(2018)\citenamefont {Brydon},
  \citenamefont {Agterberg}, \citenamefont {Menke},\ and\ \citenamefont
  {Timm}}]{2018arXiv180603773B}%
  \BibitemOpen
  \bibfield  {author} {\bibinfo {author} {\bibfnamefont {P.~M.~R.}\
  \bibnamefont {Brydon}}, \bibinfo {author} {\bibfnamefont {D.~F.}\
  \bibnamefont {Agterberg}}, \bibinfo {author} {\bibfnamefont {Henri}\
  \bibnamefont {Menke}}, \ and\ \bibinfo {author} {\bibfnamefont
  {C.}~\bibnamefont {Timm}},\ }\bibfield  {title} {\enquote {\bibinfo {title}
  {Bogoliubov fermi surfaces: General theory, magnetic order, and topology},}\
  }\href {\doibase 10.1103/PhysRevB.98.224509} {\bibfield  {journal} {\bibinfo
  {journal} {Phys. Rev. B}\ }\textbf {\bibinfo {volume} {98}},\ \bibinfo
  {pages} {224509} (\bibinfo {year} {2018})}\BibitemShut {NoStop}%
\bibitem [{\citenamefont {Ill'inski}\ and\ \citenamefont
  {Keldysh}(1994)}]{BookKeldysh}%
  \BibitemOpen
  \bibfield  {author} {\bibinfo {author} {\bibfnamefont {Yu.~A.}\ \bibnamefont
  {Ill'inski}}\ and\ \bibinfo {author} {\bibfnamefont {L.~V.}\ \bibnamefont
  {Keldysh}},\ }\href@noop {} {\emph {\bibinfo {title} {{Electromagnetic
  reponse of material media}}}}\ (\bibinfo  {publisher} {Plenum press, New
  York},\ \bibinfo {year} {1994})\BibitemShut {NoStop}%
\bibitem [{\citenamefont {Dressel}\ and\ \citenamefont
  {Gr\"uner}(2002)}]{BookDressel}%
  \BibitemOpen
  \bibfield  {author} {\bibinfo {author} {\bibfnamefont {M.}~\bibnamefont
  {Dressel}}\ and\ \bibinfo {author} {\bibfnamefont {G.}~\bibnamefont
  {Gr\"uner}},\ }\href@noop {} {\emph {\bibinfo {title} {{Electrodynamics of
  Solids: Optical Properties of Electrons in Matter}}}}\ (\bibinfo  {publisher}
  {Cambridge University press, Cambridge},\ \bibinfo {year} {2002})\BibitemShut
  {NoStop}%
\bibitem [{\citenamefont {Altland}\ and\ \citenamefont
  {Simons}(2010)}]{BookAltland}%
  \BibitemOpen
  \bibfield  {author} {\bibinfo {author} {\bibfnamefont {A.}~\bibnamefont
  {Altland}}\ and\ \bibinfo {author} {\bibfnamefont {B.}~\bibnamefont
  {Simons}},\ }\href@noop {} {\emph {\bibinfo {title} {{Condensed matter field
  theory}}}}\ (\bibinfo  {publisher} {Cambridge University press, Cambridge},\
  \bibinfo {year} {2010})\BibitemShut {NoStop}%
\bibitem [{\citenamefont {Tinkham}(1996)}]{BookTinkham}%
  \BibitemOpen
  \bibfield  {author} {\bibinfo {author} {\bibfnamefont {M.}~\bibnamefont
  {Tinkham}},\ }\href@noop {} {\emph {\bibinfo {title} {{Introduction to
  Superconductivity}}}}\ (\bibinfo  {publisher} {McGraw Hill, New York},\
  \bibinfo {year} {1996})\BibitemShut {NoStop}%
\bibitem [{\citenamefont {Boettcher}\ and\ \citenamefont
  {Herbut}(2018{\natexlab{b}})}]{PhysRevB.97.064504}%
  \BibitemOpen
  \bibfield  {author} {\bibinfo {author} {\bibfnamefont {Igor}\ \bibnamefont
  {Boettcher}}\ and\ \bibinfo {author} {\bibfnamefont {Igor~F.}\ \bibnamefont
  {Herbut}},\ }\bibfield  {title} {\enquote {\bibinfo {title} {{Critical
  phenomena at the complex tensor ordering phase transition}},}\ }\href
  {\doibase 10.1103/PhysRevB.97.064504} {\bibfield  {journal} {\bibinfo
  {journal} {Phys. Rev. B}\ }\textbf {\bibinfo {volume} {97}},\ \bibinfo
  {pages} {064504} (\bibinfo {year} {2018}{\natexlab{b}})}\BibitemShut
  {NoStop}%
\bibitem [{\citenamefont {Seibold}\ \emph {et~al.}(2017)\citenamefont
  {Seibold}, \citenamefont {Benfatto},\ and\ \citenamefont
  {Castellani}}]{PhysRevB.96.144507}%
  \BibitemOpen
  \bibfield  {author} {\bibinfo {author} {\bibfnamefont {G.}~\bibnamefont
  {Seibold}}, \bibinfo {author} {\bibfnamefont {L.}~\bibnamefont {Benfatto}}, \
  and\ \bibinfo {author} {\bibfnamefont {C.}~\bibnamefont {Castellani}},\
  }\bibfield  {title} {\enquote {\bibinfo {title} {{Application of the
  Mattis-Bardeen theory in strongly disordered superconductors}},}\ }\href
  {\doibase 10.1103/PhysRevB.96.144507} {\bibfield  {journal} {\bibinfo
  {journal} {Phys. Rev. B}\ }\textbf {\bibinfo {volume} {96}},\ \bibinfo
  {pages} {144507} (\bibinfo {year} {2017})}\BibitemShut {NoStop}%
\bibitem [{\citenamefont {Dressel}(2013)}]{DresselReview}%
  \BibitemOpen
  \bibfield  {author} {\bibinfo {author} {\bibfnamefont {M.}~\bibnamefont
  {Dressel}},\ }\bibfield  {title} {\enquote {\bibinfo {title}
  {{Electrodynamics of Metallic Superconductors}},}\ }\href {\doibase
  10.1155/2013/104379} {\bibfield  {journal} {\bibinfo  {journal} {Adv. Cond.
  Matter Phys.}\ }\textbf {\bibinfo {volume} {2013}},\ \bibinfo {pages}
  {104379} (\bibinfo {year} {2013})}\BibitemShut {NoStop}%
\bibitem [{\citenamefont {Diehl}\ and\ \citenamefont
  {Wetterich}(2006)}]{PhysRevA.73.033615}%
  \BibitemOpen
  \bibfield  {author} {\bibinfo {author} {\bibfnamefont {S.}~\bibnamefont
  {Diehl}}\ and\ \bibinfo {author} {\bibfnamefont {C.}~\bibnamefont
  {Wetterich}},\ }\bibfield  {title} {\enquote {\bibinfo {title} {{Universality
  in phase transitions for ultracold fermionic atoms}},}\ }\href {\doibase
  10.1103/PhysRevA.73.033615} {\bibfield  {journal} {\bibinfo  {journal} {Phys.
  Rev. A}\ }\textbf {\bibinfo {volume} {73}},\ \bibinfo {pages} {033615}
  (\bibinfo {year} {2006})}\BibitemShut {NoStop}%
\bibitem [{\citenamefont {Nikolic}\ and\ \citenamefont
  {Sachdev}(2007)}]{PhysRevA.75.033608}%
  \BibitemOpen
  \bibfield  {author} {\bibinfo {author} {\bibfnamefont {Predrag}\ \bibnamefont
  {Nikolic}}\ and\ \bibinfo {author} {\bibfnamefont {Subir}\ \bibnamefont
  {Sachdev}},\ }\bibfield  {title} {\enquote {\bibinfo {title}
  {{Renormalization-group fixed points, universal phase diagram, and $1/N$
  expansion for quantum liquids with interactions near the unitarity limit}},}\
  }\href {\doibase 10.1103/PhysRevA.75.033608} {\bibfield  {journal} {\bibinfo
  {journal} {Phys. Rev. A}\ }\textbf {\bibinfo {volume} {75}},\ \bibinfo
  {pages} {033608} (\bibinfo {year} {2007})}\BibitemShut {NoStop}%
\bibitem [{\citenamefont {{Zwerger, W.}}(2012)}]{Zwerger}%
  \BibitemOpen
  \bibinfo {editor} {\bibnamefont {{Zwerger, W.}}},\ ed.,\ \href@noop {} {\emph
  {\bibinfo {title} {{The BCS-BEC Crossover and the Unitary Fermi Gas}}}}\
  (\bibinfo  {publisher} {Springer, Berlin},\ \bibinfo {year}
  {2012})\BibitemShut {NoStop}%
\bibitem [{\citenamefont {Boettcher}\ \emph {et~al.}(2012)\citenamefont
  {Boettcher}, \citenamefont {Pawlowski},\ and\ \citenamefont
  {Diehl}}]{Boettcher:2012cm}%
  \BibitemOpen
  \bibfield  {author} {\bibinfo {author} {\bibfnamefont {Igor}\ \bibnamefont
  {Boettcher}}, \bibinfo {author} {\bibfnamefont {Jan~M.}\ \bibnamefont
  {Pawlowski}}, \ and\ \bibinfo {author} {\bibfnamefont {Sebastian}\
  \bibnamefont {Diehl}},\ }\bibfield  {title} {\enquote {\bibinfo {title}
  {{Ultracold atoms and the Functional Renormalization Group}},}\ }\href
  {\doibase 10.1016/j.nuclphysbps.2012.06.004} {\bibfield  {journal} {\bibinfo
  {journal} {Nucl. Phys. B Proc. Suppl.}\ }\textbf {\bibinfo {volume} {228}},\
  \bibinfo {pages} {63--135} (\bibinfo {year} {2012})}\BibitemShut {NoStop}%
\bibitem [{\citenamefont {Boettcher}\ \emph {et~al.}(2014)\citenamefont
  {Boettcher}, \citenamefont {Pawlowski},\ and\ \citenamefont
  {Wetterich}}]{PhysRevA.89.053630}%
  \BibitemOpen
  \bibfield  {author} {\bibinfo {author} {\bibfnamefont {Igor}\ \bibnamefont
  {Boettcher}}, \bibinfo {author} {\bibfnamefont {Jan~M.}\ \bibnamefont
  {Pawlowski}}, \ and\ \bibinfo {author} {\bibfnamefont {Christof}\
  \bibnamefont {Wetterich}},\ }\bibfield  {title} {\enquote {\bibinfo {title}
  {{Critical temperature and superfluid gap of the unitary Fermi gas from
  functional renormalization}},}\ }\href {\doibase 10.1103/PhysRevA.89.053630}
  {\bibfield  {journal} {\bibinfo  {journal} {Phys. Rev. A}\ }\textbf {\bibinfo
  {volume} {89}},\ \bibinfo {pages} {053630} (\bibinfo {year}
  {2014})}\BibitemShut {NoStop}%
\bibitem [{\citenamefont {Liu}\ and\ \citenamefont
  {Brust}(1968)}]{PhysRevLett.20.651}%
  \BibitemOpen
  \bibfield  {author} {\bibinfo {author} {\bibfnamefont {L.}~\bibnamefont
  {Liu}}\ and\ \bibinfo {author} {\bibfnamefont {David}\ \bibnamefont
  {Brust}},\ }\bibfield  {title} {\enquote {\bibinfo {title} {{Static
  Dielectric Function of a Zero-Gap Semiconductor}},}\ }\href {\doibase
  10.1103/PhysRevLett.20.651} {\bibfield  {journal} {\bibinfo  {journal} {Phys.
  Rev. Lett.}\ }\textbf {\bibinfo {volume} {20}},\ \bibinfo {pages} {651--653}
  (\bibinfo {year} {1968})}\BibitemShut {NoStop}%
\bibitem [{\citenamefont {Broerman}(1970)}]{PhysRevLett.25.1658}%
  \BibitemOpen
  \bibfield  {author} {\bibinfo {author} {\bibfnamefont {J.~G.}\ \bibnamefont
  {Broerman}},\ }\bibfield  {title} {\enquote {\bibinfo {title} {{Temperature
  Dependence of the Static Dielectric Constant of a Symmetry-Induced Zero-Gap
  Semiconductor}},}\ }\href {\doibase 10.1103/PhysRevLett.25.1658} {\bibfield
  {journal} {\bibinfo  {journal} {Phys. Rev. Lett.}\ }\textbf {\bibinfo
  {volume} {25}},\ \bibinfo {pages} {1658--1660} (\bibinfo {year}
  {1970})}\BibitemShut {NoStop}%
\bibitem [{\citenamefont {Mattis}\ and\ \citenamefont
  {Bardeen}(1958)}]{PhysRev.111.412}%
  \BibitemOpen
  \bibfield  {author} {\bibinfo {author} {\bibfnamefont {D.~C.}\ \bibnamefont
  {Mattis}}\ and\ \bibinfo {author} {\bibfnamefont {J.}~\bibnamefont
  {Bardeen}},\ }\bibfield  {title} {\enquote {\bibinfo {title} {Theory of the
  anomalous skin effect in normal and superconducting metals},}\ }\href
  {\doibase 10.1103/PhysRev.111.412} {\bibfield  {journal} {\bibinfo  {journal}
  {Phys. Rev.}\ }\textbf {\bibinfo {volume} {111}},\ \bibinfo {pages}
  {412--417} (\bibinfo {year} {1958})}\BibitemShut {NoStop}%
\bibitem [{\citenamefont {Abrikosov}\ \emph {et~al.}(1959)\citenamefont
  {Abrikosov}, \citenamefont {Gor'kov},\ and\ \citenamefont
  {Khalatnikov}}]{AbrikosovGorkov}%
  \BibitemOpen
  \bibfield  {author} {\bibinfo {author} {\bibfnamefont {A.~A.}\ \bibnamefont
  {Abrikosov}}, \bibinfo {author} {\bibfnamefont {L.~P.}\ \bibnamefont
  {Gor'kov}}, \ and\ \bibinfo {author} {\bibfnamefont {I.~M.}\ \bibnamefont
  {Khalatnikov}},\ }\bibfield  {title} {\enquote {\bibinfo {title} {{A
  superconductor in a high frequency field}},}\ }\href@noop {} {\bibfield
  {journal} {\bibinfo  {journal} {Sov. Phys. JETP}\ }\textbf {\bibinfo {volume}
  {35}},\ \bibinfo {pages} {182} (\bibinfo {year} {1959})}\BibitemShut
  {NoStop}%
\bibitem [{\citenamefont {Herbut}(2013)}]{PhysRevD.87.085002}%
  \BibitemOpen
  \bibfield  {author} {\bibinfo {author} {\bibfnamefont {Igor~F.}\ \bibnamefont
  {Herbut}},\ }\bibfield  {title} {\enquote {\bibinfo {title} {{Majorana mass,
  time reversal symmetry, and the dimension of space}},}\ }\href {\doibase
  10.1103/PhysRevD.87.085002} {\bibfield  {journal} {\bibinfo  {journal} {Phys.
  Rev. D}\ }\textbf {\bibinfo {volume} {87}},\ \bibinfo {pages} {085002}
  (\bibinfo {year} {2013})}\BibitemShut {NoStop}%
\end{thebibliography}%

\end{document}